\documentclass[letterpaper,12pt]{article}

\pdfoutput=1  

\usepackage[top=1.0in,bottom=1.0in,left=1.0in,right=1.0in]{geometry}

\usepackage{amssymb}
\usepackage{amsmath}
\usepackage{graphicx}
\usepackage{url}
\usepackage{xspace}

\newcommand\lhood{{\cal L}}
\newcommand\accreg{{\cal A}}
\newcommand\non{n_\textnormal{\scriptsize on}}       
\newcommand\ntot{n_\textnormal{\scriptsize tot}}     
\newcommand\binp{\rho}            

\newcommand\bi{\textnormal{Bi}}
\newcommand\nbi{\textnormal{NBi}}

\newcommand\cls{$CL_s$}
\newcommand\cl{\textnormal{C.L.}}

\newcommand\events{\textnormal{events}}

\newcommand\nota{\textnormal{not~a}}
\newcommand\Not{\textnormal{not~}}
\newcommand\true{\textnormal{true}}
\newcommand\false{\textnormal{false}}
\newcommand\model{\textnormal{model}}
\newcommand\btag{\textnormal{b-tag}}
\newcommand\nobtag{\textnormal{no~b-tag}}
\newcommand\bjet{\textnormal{b-jet}}

\newcommand\mut{\mu_\textnormal{t}}

\begin{document}

\title{\bf \Large Lectures on Statistics in Theory:
\\ Prelude to Statistics in Practice}
\author{Robert D. Cousins\thanks{cousins@physics.ucla.edu}\\
Dept.\ of Physics and Astronomy\\ 
University of California, Los Angeles\\
Los Angeles, California 90095}
 
\date{June 26, 2024}

\maketitle

\begin{abstract}
This is a writeup of lectures on ``statistics'' that have evolved from the initial version for
the 2009 Hadron Collider Physics Summer School at CERN to versions for other venues and, most recently, for the African School of Fundamental Physics and Applications in 2024. The emphasis is on foundations,
using simple examples to illustrate the points that are still debated
in the professional statistics literature.  The three main approaches
to interval estimation (Neyman confidence, Bayesian, likelihood ratio)
are discussed and compared in detail, with and without nuisance
parameters.  Hypothesis testing is discussed mainly from the
frequentist point of view, with pointers to the Bayesian literature.
Various foundational issues are emphasized, including the
conditionality principle and the likelihood principle.
\end{abstract}

\clearpage
\tableofcontents

\clearpage
\section{Introduction}
This is a writeup of slides that I first prepared for two hours of lectures at the Hadron Collider Physics Summer School (HCPSS) at CERN in 2009 (and again in 2018 at Fermilab), and which eventually grew to about four hours of lectures in various venues, most recently for the 8th African School of Fundamental Physics and Applications in 2024~\cite{asp}.  Unlike commonly available lectures by many of my colleagues
on practical statistics for data analysis, my lectures focus on
discussion of the foundational aspects, for which there is much less
secondary literature written by physicists. 

I first got interested in the foundations of statistics in the late
1980's, when I learned that deep issues of great importance to
science, such as ``Is there a statistically significant departure from
expectations in my data?''  were not at all settled.  The issue is not
merely, ``How many sigma is a discovery?'', but rather, ``Is the
(equivalent) number of sigma even the right figure of merit for
inferring the presence of a discovery?''  The more I read about the
Bayesian--frequentist debates in the primary statistics literature
(which has the wonderful practice of including commentary and a
rejoinder in many major papers), the more it became a ``hobby'' to
browse this literature and (thanks to the PhyStat series organized by
Louis Lyons and colleagues) to discuss these issues with preeminent
statisticians and other interested physicists.

The style here is rather terse, reflecting the origin in the
slides~\cite{cousins-slides}, with each subsection drawing attention
to a point of interest or controversy, including pointers to more
literature.  I do not assume much advanced statistics knowledge, but
the reader may find the subject to be surprisingly {\em
  difficult}.  That is correct (!), and is one of the main
points of my lectures.  Of course, a familiarity with examples of
plots from various HEP analyses will be helpful.

By concentrating on the ``theoretical'' underpinnings, I hope to
provide the reader with {\em what one must know in order to choose
  appropriate methods} from the many possibilities.  This includes the
hope that by studying these topics, one will learn to avoid common
pitfalls (and even silly statements) that can trip up
professionals in the field.

This is a dense writeup, and I do not expect one to pick it all up in
a quick read-through. It should however be extremely useful to study
the topics, referring to the references.  I have also tried to put in
enough sub-headings so that one can use it as a reference on specific
topics. After some preliminaries, I begin with definitions and the
Bayesian approach.  That should help to understand what the
frequentist approach (described next) is {\it not}\/!  The frequentist
discussion includes interval estimation, hypothesis testing,
conditional frequentist estimation, and the much-debated issue of
downward fluctuations in a search for an excess.  After discussing
likelihood-ratio intervals, I compare the three approaches, including
major foundational issues such as the likelihood principle
(Sections~\ref{likelihoodprin}, \ref{intervalsummary}).  Finally, I add
nuisance parameters in the context of each of the approaches.  I
conclude with a word about current practice at the LHC.  

The appendices have a detailed worked example of a hypothesis test for
two simple hypotheses; further discussion of goodness of fit; a brief discussion of the look-elsewhere
effect; some further notes on Bayesian model selection; and some
remarks on point estimation.  All of these are important topics
that did not get sufficient coverage in my lectures.

\section {Preliminaries}
\subsection{Why foundations matter}

In the ``final analysis'', we often make approximations, take a
pragmatic approach, or follow a convention.  To inform such actions,
it is important to understand some foundational aspects of statistical
inference.  In Quantum Mechanics, we are used to the fact that for all
of our practical work, one's philosophical interpretation (e.g., of
collapse of the wave function) does not matter.  In statistical
inference, however, {\em foundational differences result in different
  answers}\/: one cannot ignore them!

The professional statistics community went through the topics of many
of our discussions starting in the 1920's, and revisited them in the
resurgence of Bayesian methods in recent decades.  I attempt to
summarize some of the things that we should understand from that
debate.  {\em Most importantly}: One needs to understand both
frequentist and Bayesian methods!

\subsection{Definitions are important}

As in physics, much confusion can be avoided by being precise about
definitions, and much confusion can be generated by being imprecise,
or (especially) by assuming everyday definitions in a technical
context.  You have learned in physics to see confusion in the statement,
``I did a lot of {\em  work} today by carrying this big stone around the building and then
putting it back in its original place.'' 
By the end of these lectures, you should see just as much confusion in these two statements:
\begin{enumerate}
\item ``The confidence level tells you how much confidence
one has that the true value is in the confidence interval,'' 
\item ``A noninformative prior probability density contains
no information.''
\end{enumerate}

Confusion is also possible because the statistics literature uses some
words differently than does the HEP literature. A few examples are in
Table~\ref{jargon}, adapted from James~\cite{james2006}.  Here I tend
to use words from both columns, with nearly exclusive use of the
statisticians' definition of ``estimation'', as discussed below.

\begin{table}[h]
\begin{center}
\caption{Potential for confusion}
\begin{tabular}{lll}  
\\
\hline 
Physicists say\dots &&Statisticians say\dots \\ \hline
Determine, Measure && Estimate \\
Estimate && (Informed) Guess \\
Gaussian &&  Normal    \\ 
Breit-Wigner, Lorentzian &~~& Cauchy \\ \hline
\end{tabular}
\label{jargon}
\end{center}
\end{table}

\subsection{Key tasks: Important to distinguish}

The most common tasks to be performed in statistical inference are
typically classified as follows.
\begin{itemize}
\item {\em Point estimation}: What single ``measured'' value of a
  parameter do you report? While much is written about point
  estimation, in the end it is not clear what the criteria are for a
  ``best'' estimator. Decision Theory can be used to specify criteria
  and choose among point estimators.  However, in HEP this is only
  implicitly done, and point estimation is usually not a contentious
  issue: typically the maximum-likelihood (point) estimator (MLE)
  serves our needs rather well, sometimes with a small correction for
  bias if desired. (See Appendix~\ref{pointest}.)  We generally put a
  ``hat'' (accent circumflex) over a variable to denote a point
  estimate, e.g., $\hat\mu$.

\item {\em Interval estimation} : What interval (giving a measure of
  uncertainty of the parameter inference) do you report?  This is
  crucial in HEP (and in introductory physics laboratory courses), and
  as discussed below, is deeply connected to frequentist hypothesis
  testing.  In HEP it is fairly mandatory that there is a {\em
    confidence level} that gives the {\em frequentist coverage}
  probability (Section~\ref{sec-coverage}) of a method, even if it is
  a Bayesian-inspired recipe.

Point estimation and interval estimation can be approached
consistently by insisting that the interval estimate contain the point
estimate; in that case, one can construct the point estimate by taking
the limit of interval estimates as intervals get smaller (limit of
confidence level going to zero).

For many problems in HEP, there is reasonable hope of approximate
reconciliation between Bayesian and frequentist methods for point and
interval estimation, especially with large sample sizes.

\item{\em Hypothesis testing}: There are many special cases, 
including a test of:
\begin{description}
\item{(a)}
A given functional form (``model'') vs another functional form.  
Also known as ``model selection'';
\item{(b)}
A single value of a parameter (say 0 or 1) vs all other values;
\item{(c)}
{\em Goodness of Fit}: A given functional form against all other
(unspecified) functional forms (also known as ``model checking'')
\end{description}
Bayesian methods for hypothesis testing generally attempt to calculate
the probability that a hypothesis is true.  Frequentist methods cannot
do this, and often lead to results expressed as $p$-values
(Section~\ref{pvalues}).  There is a large literature bashing
$p$-values, but they are still deemed essential in HEP.

\item {\em Decision making}: What action should I take (tell no one,
  issue press release, propose a new experiment, \dots) based on the
  observed data?  Decision making is rarely performed formally in HEP,
  but it is important to understand the outline of the formal theory,
  in order to avoid confusion with statistical inference that stops
  short of a decision, and to inform informal application.
\end{itemize}

In frequentist statistics, the above hypothesis testing case (b) maps
identically onto interval estimation.  This is called the duality of
``inversion of a hypothesis test to get confidence intervals'', and
vice versa.  I discuss this in more detail in Section~\ref{duality}.

In contrast, in Bayesian statistics, hypothesis testing case (b) is an
especially controversial form of case (a), model selection.  The model
with fixed value of the parameter is considered to be a
lower-dimensional model in parameter space (one fewer parameter) than
the model with parameter not fixed. I just mention this here to
foreshadow a very deep issue.  Because of the completely different
structure of the approaches to testing, there can be dramatic differences
between frequentist and Bayesian hypothesis testing methods, with
conclusions that apparently disagree, even in the limit of large data
sets. Beware!  See Appendix~\ref{modelselection} and my paper on the
Jeffreys-Lindley paradox~\cite{cousinsJL}.

\section{Probability}

\subsection{Definitions, Bayes's theorem}
\begin{figure}
\begin{center}
\includegraphics[width=0.95\textwidth]{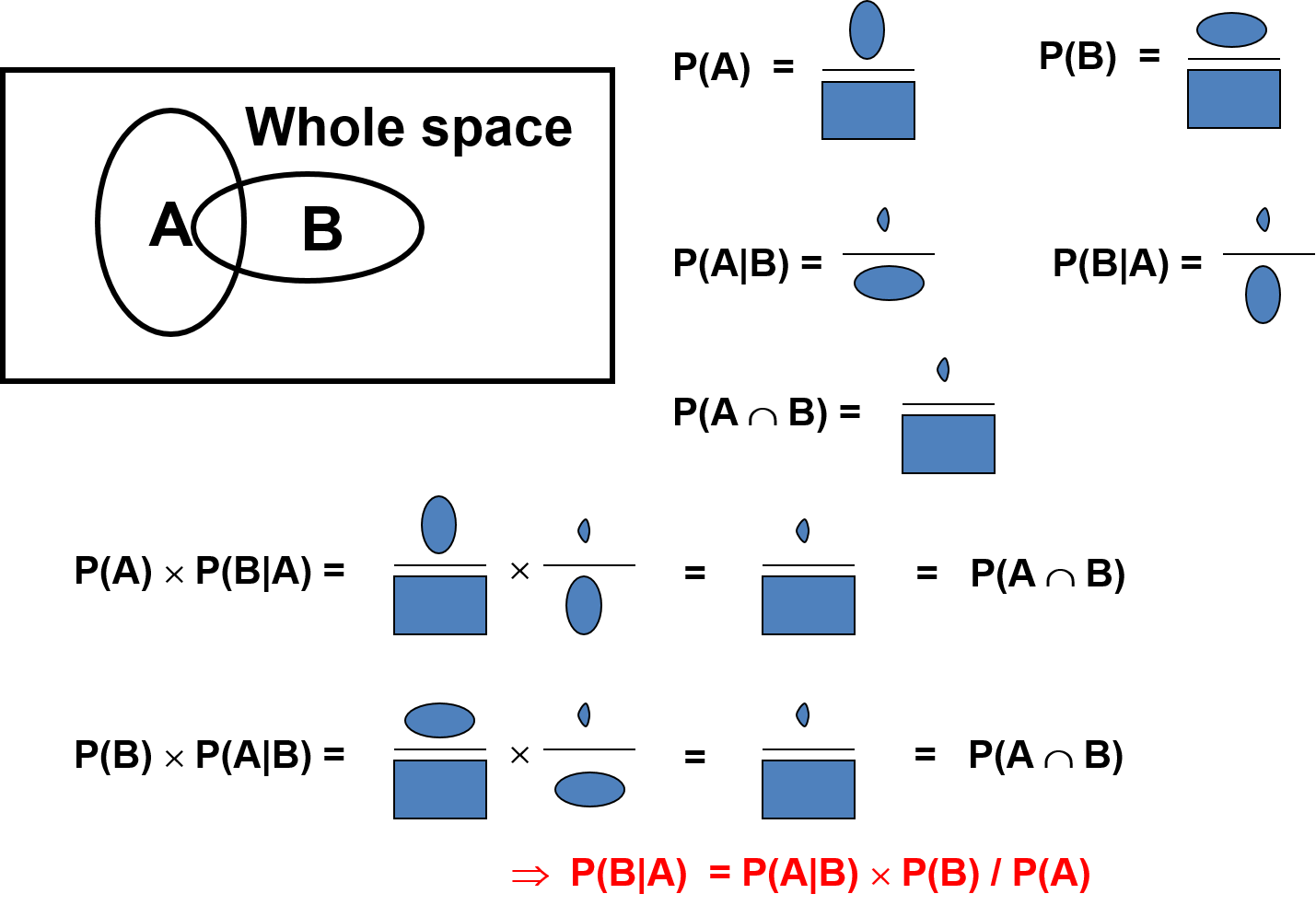}
\caption{A cartoon explanation of probability, conditional
  probability, and Bayes's Theorem, using picture arithmetic.}
\label{bayes_in_pix}
\end{center}
\end{figure}

Abstract mathematical probability $P$ can be defined in terms of sets
and axioms that $P$ obeys, as outlined in Ref.~\cite{cowan} (Chapter
1) and discussed in much more detail in Ref.~\cite{persi} (Chapter
5). {\em Conditional probabilities} $P(B|A)$ (read ``$P$ of $B$ given
$A$'') and $P(A|B)$ are related by Bayes's Theorem (or ``Bayes's
Rule''):
\begin{equation} 
\label{eqn-bayes}
P(B|A) = P(A|B) P(B) / P(A).
\end{equation}
A cartoon illustration of conditional probabilities and a
``derivation'' of Bayes's theorem is in Fig.~\ref{bayes_in_pix}.

Two established incarnations of $P$ (still argued about, but to less
practical effect) are:
\begin{itemize}
\item {\em Frequentist} $P$: the limiting frequency in an ensemble of
  imagined repeated samples (as usually taught in Q.M.).
  $P(\textnormal{constant of nature})$ and $P(\textnormal{SUSY is
    true})$ do not exist (in a useful way) for this definition of $P$
  (at least in one universe).

\item {\em (Subjective) Bayesian $P$}: subjective (personalistic) {\em
  degree of belief} (as advocated by de Finetti~\cite{definetti},
  Savage~\cite{savagefoundations}, and
  others). $P(\textnormal{constant of nature})$ and
  $P(\textnormal{SUSY is true})$ exist for You.  (In the literature,
  ``You'' is often capitalized to emphasize the personalistic aspect.)
  This has been argued to be a basis for coherent personal
  decision-making (where ``coherent'' has a technical meaning).

\end{itemize}

{\em It is important to be able to work with either definition of $P$,
  and to know which one you are using!}

The descriptive word ``Bayesian'' applies to the use of the above
definition of probability as degree of belief! The adjective
``Bayesian'' thus also normally applies to the use of a probability
density for any parameter (such as a constant of nature) whose true
value is {\em fixed but unknown}, even in a context where the
practitioner is not really aware of the degree-of-belief definition.

In contrast, Bayes's theorem applies to any definition of probability
that obeys the axioms and for which the probabilities are defined in
the relevant context!  The distinction was noted by statistician
Bradley Efron in his keynote talk at the 2003 PhyStat meeting at
SLAC~\cite{efronslac}:

\begin{description}
\item{} ``Bayes' rule is satisfying, convincing, and fun to use. But
  using Bayes' rule does not make one a Bayesian; {\em always} using
  it does, and that's where difficulties begin.''  (Emphasis in
  original.)
\end{description}

Clearly, Bayes's theorem has more applications for Bayesian $P$ than
for frequentist $P$, since Bayesian $P$ can be used in more contexts.
But one of the sillier things one sometimes sees in HEP is the use of
a frequentist example of Bayes's Theorem as a foundational argument
for ``Bayesian'' statistics!  Below I give a simple example for each
definition of $P$.

\subsection[Example of Bayes's theorem using frequentist 
$P$]{\boldmath Example of Bayes's theorem using frequentist $P$}

\label{btagsec}
A b-tagging method is developed, and one measures:
\begin{description}
\item{} $P(\btag | \bjet)$,    i.e., efficiency for tagging b-jets
\item{} $P(\btag | \nota~\bjet)$,      i.e., efficiency for background
\item{} $P(\nobtag | \bjet)     = 1 - P(\btag | \bjet)$
\item{} $P(\nobtag | \nota~\bjet) = 1 - P(\btag | \nota~\bjet)$.
\end{description}

\noindent
{\bf Question:} Given a selection of jets tagged as b-jets, what
fraction of them is b-jets?  I.e., what is $P(\bjet | \btag)$ ?

\medskip\noindent {\bf Answer:} {\em Cannot be determined from the
  given information!}

\medskip\noindent One needs in addition: $P(\bjet)$, the true fraction
of {\em all} jets that are b-jets.  Then Bayes's Theorem inverts the
conditionality:
\begin{equation}
P(\bjet | \btag) \propto P(\btag |\bjet) P(\bjet),
\end{equation}
(where I suppress the normalization denominator).

As noted, in HEP $P(\btag | \bjet)$ is called the {\em efficiency} for
tagging b-jets.  Meanwhile $P(\bjet | \btag)$ is often called the {\em
  purity} of a sample of b-tagged jets.  As this should be a
conceptually easy distinction for experienced data analysts in HEP, it
is helpful to keep it in mind when one encounters cases where it is
perhaps tempting to make the logical error of equating $P(A|B)$ and
$P(B|A)$.

Note: Looking ahead, when we talk about frequentist hypothesis testing later in the lectures, we will mention names for analogous probabilities in other fields of science. 
E.g., in medicine, $P$(Covid test is positive $|$ patient has Covid) is called the ``sensitivity" of the Covid test, or (unfortunately in my opinion), the ``true positive rate".

\subsection[Example of Bayes's theorem using Bayesian 
$P$]{\boldmath Example of Bayes's theorem using Bayesian $P$}
\label{bayesbayes}
In a {\em background-free} experiment, a theorist uses a ``model'' to
predict a signal with Poisson mean of 3.0 events. From the Poisson
formula (Eqn.~\ref{eqn-poisson}) we know:
\begin{description}
\item{} $P(0~\events | \model~\true) = 3.0^0e^{-3.0}/0! = 0.05$
\item{} $P(0~\events | \model~\false) = 1.0$
\item{} $P(>0~\events | \model~\true) = 0.95$
\item{} $P(>0~\events | \model~\false) = 0.0$.
\end{description}
The experiment is performed, and {\em zero events are observed}.

\medskip\noindent 
{\bf Question:} Given the result of the experiment, what is the
probability that the model is true? I.e., What is $P(\model~\true |
0~\events)$?

\medskip\noindent {\bf Answer:} {\em Cannot be determined from the
  given information!}

\medskip\noindent One needs in addition: $P(\model~\true)$, the degree
of belief in the model prior to the experiment.  Then Bayes's Theorem
inverts the conditionality:
\begin{equation}
P(\model~\true | 0~\events) \propto P(0~\events | \model~\true) P(\model~\true)
\end{equation}
(again suppressing the normalization).  It is instructive to apply
Bayes's Theorem in a little more detail, with the normalization. In
Eqn.~\ref{eqn-bayes}, let ``$A$'' correspond to ``0~\events'' and
``$B$'' correspond to ``\model~\true''.  Similarly, with $P(\Not B) =
1 - P(B)$, we can write a version of Bayes's Theorem (replacing $B$
with ``$\Not B$'') as
\begin{equation}
\label{bayesnotb}
P(\Not B|A)  = P(A| \Not B) \times P(\Not B) / P(A).
\end{equation}
(As a check, we can add Eqns.~\ref{eqn-bayes} and \ref{bayesnotb} and
get unity, confirming that $P(A)$ is the correct normalization.)
Solving Eqn.~\ref{bayesnotb} for $P(A)$ and substituting into
Eqn.~\ref{eqn-bayes}, and inserting numerical values from above,
yields $P(B|A) = 0.05 P(B) / (1- 0.95 P(B))$, i.e.,
\begin{equation}
P(\model~\true | 0~\events) 
    = \frac{0.05 \times P(\model~\true)}{(1- 0.95 P(\model~\true))}.
\end{equation}

We can examine the limiting cases of strong prior belief in the model
and very low prior belief. If we let the ``model'' be the Standard
Model (SM), then we could express our high prior belief as $P(\model~\true)
= 1 - \epsilon_1$, where $\epsilon_1 \ll 1$. Plugging in gives, to lowest
order,
\begin{equation}
P(\model~\true | 0~\events) \approx 1 - 20\epsilon_1.
\end{equation}  
This is still very high degree of belief in the SM.  Unfortunately,
one still finds (in the press and even among scientists) the fallacy
that is analogous to people saying, ``$P(0~\events | \model~\true) =
5$\%, with 0 events observed, means there is 5\% chance the SM is
true.'' (UGH!)

In contrast, let the ``model'' be large extra dimensions, so that for
a skeptic, the prior belief can be expressed as $P(\model~\true) =
\epsilon_2$, for some other small $\epsilon_2$.  Then to lowest order we
have,
\begin{equation}
P(\model~\true | 0~\events) \approx 0.05 \epsilon.
\end{equation}  
Low prior belief becomes even lower.

More realistic examples are of course more complex.  But this example
is good preparation for avoiding misinterpretation of $p$-values in
Section~\ref{pvalues}.

\subsection{A note re {\em Decisions}}

Suppose that as a result of the previous experiment, your degree of
belief in the model is $P(\model~\true | 0~\events) = 1$\%, and you
need to {\em decide} on an action, e.g., announcing in a press release that the model is false, or making no announcement while taking more data. 

\medskip\noindent
{\bf Question:} What should you {\em decide}?

\medskip\noindent
{\bf Answer:} {\em Cannot be determined from the given information}!    

\medskip\noindent One needs in addition: The {\em utility} function
(or its negative, the {\em loss} function), which quantifies the
relative costs (to You) of
\begin{description}
\item{} {\em Type I error:} announcing that the model is false, when it is true (thus eventually harming your reputation);
 and
  of
\item{} {\em Type II error:} not announcing that the model is false when it is false, thus potentially allowing another experiment to make the announcement first.
\end{description}
Thus, Your {\em decision} requires two subjective inputs: Your prior
probabilities, and the relative costs (or benefits) to You of
outcomes.

Statisticians often focus on decision-making.  In HEP, the tradition
thus far is to communicate experimental results (well) short of formal
decision calculations.  It is important to realize that frequentist
(classical) ``hypothesis testing'' as discussed in
Section~\ref{hypotest} below (especially with conventions like 95\%
\cl\ or 5$\sigma$) is {\em not} a complete theory of decision-making!
One must always keep this in mind, since the traditional
``accept/reject'' language of frequentist hypothesis testing is too
simplistic for ``deciding''.

\subsection{Aside: What is the ``whole space''?}
\label{aside}
For probabilities to be well-defined, the ``whole space'' needs to be
defined. This can be difficult or impossible for both frequentists and
Bayesians. For frequentists, specification of the whole space may
require listing the experimental protocol in detail, including the
experimenters' reaction to potentially unexpected results that did not
occur!  For Bayesians, normalization of probabilities of hypotheses
requires enumerating all possible hypotheses and assigning degree of
belief to them, including hypotheses not yet formulated!

Thus, the ``whole space'' itself is more properly thought of as a
conditional space, conditional on the assumptions going into the model
(Poisson process, whether or not the total number of events was fixed,
etc.), and simplifying assumptions or approximations.

Furthermore, it is widely accepted that restricting the ``whole
space'' to a relevant (``conditional'') subspace can sometimes improve
the quality of statistical inference.  The important topic of such
``conditioning'' in frequentist inference is discussed in detail in
Section~\ref{conditioning}.  

In general, I do not clutter the notation with
explicit mention of the assumptions defining the ``whole space'', but
some prefer to do so. In any case, it is important to keep them in mind 
and to be aware of their effect on the results obtained.

\section{Probability, probability density, likelihood}
\label{probpdf}
These are key building blocks in both frequentist and Bayesian
statistics, and it is crucial to distinguish among them.  In the
following, we let $x$ be an observed quantity; sometimes we use $n$ if
the observation is integer-valued and we want to emphasize that (to
aging Fortran programmers).  A {\em ``(statistical) model''} is an expression
specifying probabilities or probability densities for observing $x$.
Here we use $\mu$ for parameters (sometimes vector-valued) in the model. (In the
statistical literature, $\theta$ is more common.) 

In Bayesian statistics, the parameters themselves are considered to be
``random variables''. The notation for such a general model is
$p(x|\mu)$, where the vertical line (read ``given'') means conditional
probability, conditional on a particular value of $\mu$. In frequentist
statistics, typically the dependence on $\mu$ is not a proper
conditional probability, and thus many experts advocate using notation
with a semi-colon: $p(x\/;\mu)$. The modern text by George Casella and
Roger Berger~\cite{casellabergerbook} (p.\ 86) however uses the
vertical line for ``given'' in the context where the parameter is not
a random variable being conditioned on.  I do not know of any examples
where this causes trouble, so at the risk of offending some, I use a
vertical line throughout for both Bayesian and frequentist models.
Then the most common examples in HEP are:

\begin{itemize}
\item Binomial probability for $\non$ successes out of $\ntot$ trials
  (Section~\ref{sec-binomial}):
\begin{equation}
\label{eqn-binomial}
\bi(\non|\ntot,\binp) = \frac{\ntot!}{\non!(\ntot-\non)!}\, 
\binp^{\non}\,(1-\binp)^{(\ntot-\non)}
\end{equation}

\item Poisson probability for $n$ events to be observed:
\begin{equation}
\label{eqn-poisson} 
P(n|\mu) = \frac{\mu^n \textnormal{e}^{-\mu}}{n!}
\end{equation} 
\item Gaussian probability {\em density} function (pdf):
\begin{equation}
\label{eqn-gaussian} 
p(x| \mu,\sigma) = 
   \frac{1}{\sqrt{2\pi\sigma^2}} \textnormal{e}^{-(x-\mu)^2/2\sigma^2},
\end{equation} 
so that $p(x|\mu,\sigma)dx$ is a differential of probability $dP$.
\end{itemize}
A typical course in statistical physics shows how the latter two can
be viewed as limiting cases of the first.  The binomial and Poisson
formulas are sometimes called probability {\em mass} functions in the
statistics literature.

\begin{figure}
\begin{center}
\includegraphics[width=0.49\textwidth]{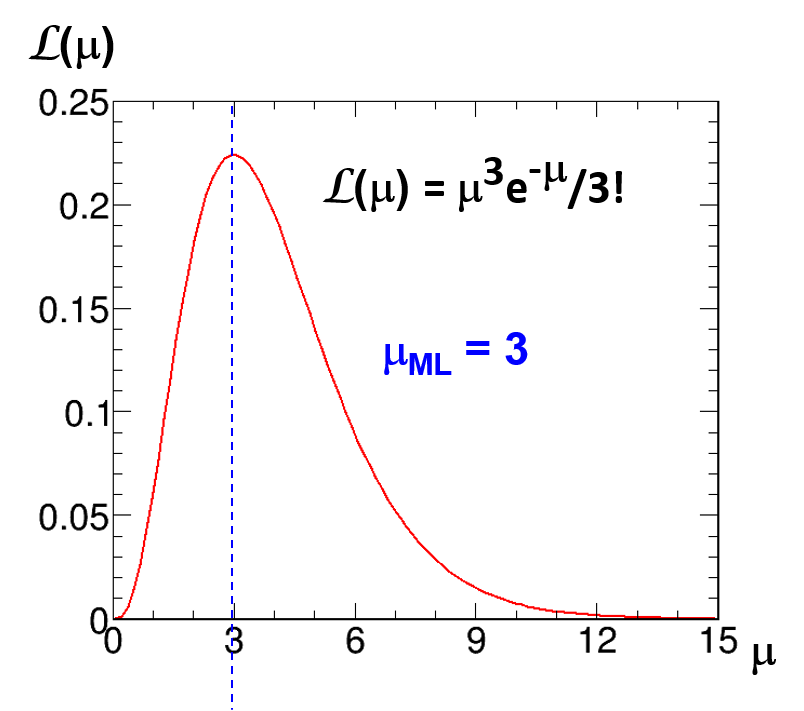}
\caption{The likelihood function $\lhood(\mu)$ 
after observing $n=3$ in the Poisson model of Eqn.~\ref{eqn-poisson}.
The likelihood obtains its maximum (ML) at $\mu=3$.}
\label{pois_likli_3obs}
\end{center}
\end{figure}

In the Poisson case, suppose that $n=3$ is observed. Substituting this
{\em observed value} $n=3$ into $P(n|\mu)$ yields the {\em likelihood
  function}, $\lhood(\mu) = \mu^3 \exp(-\mu)/3!$, plotted in
Fig.~\ref{pois_likli_3obs}. 

It is tempting to consider the area under $\lhood$ as meaningful, but
$\lhood(\mu)$ is {\em not} a probability density in $\mu$. The area
under $\lhood$ (or parts of it) is meaningless!  The Poisson example
makes this particularly clear, since the definition of $\lhood(\mu)$
starts with a probability (not a probability density); it makes no
sense to multiply $\lhood(\mu)$ by $d\mu$ and integrate.

As we shall see, likelihood {\em ratios} $\lhood(\mu_1)/\lhood(\mu_2)$
are useful and frequently used. In fact, in the statistics literature, the likelihood 
is only defined up to an overall constant factor.

\subsection[Change of observable variable (``metric'') $x$ 
in pdf $p(x|\mu)$]{\boldmath Change of observable variable
  (``metric'') $x$ in pdf $p(x|\mu)$}
\label{metricchange}
For pdf $p(x|\mu)$ and a 1-to-1 change of variable (metric) from
(vector) $x$ to (vector) $y(x)$, the volume element is modified by the
Jacobian. For a 1D function $y(x)$, we have $p(y) |dy| = p(x) |dx|$,
so that
\begin{equation}
\label{jacobian}
p(y(x)|\mu) = p(x|\mu)\ /\ |dy/dx|.
\end{equation}
The Jacobian modifies the probability density in such a way to
guarantee that
\begin{equation}
P( y(x_1)< y < y(x_2) ) = P(x_1 < x < x_2 ),
\end{equation}
(or equivalent with decreasing $y(x)$).  That is, the Jacobian
guarantees that {\em probabilities are invariant under change of
  variable} $x$.

Because of the Jacobian in Eqn.~\ref{jacobian}, the {\em mode} of
probability {\em density} is in general {\em not} invariant under
change of metric.  That is, the value of $y$ for which $p(y(x))$ is a
maximum is not trivially related to the value of $x$ for which $p(x)$
is a maximum.  

Another consequence of the Jacobian in Eqn.~\ref{jacobian} is that the
likelihood function $\lhood(\mu)$ differs for different choices of the
data variable.  However, likelihood {\em ratios} such as
$\lhood(\mu_1) /\lhood(\mu_2)$ {\em are} invariant under change of
variable $x$ to $y(x)$. The Jacobian in the denominator cancels that
in the numerator.  Thus, for example, the value of $\mu$ that
maximizes $\lhood(\mu)$ will be independent of the choice of data
variable, but the value of $\lhood(\mu)$ at that maximum is different.
The latter point also explains why the value of $\lhood(\mu)$ at its
maximum is not an appropriate test statistic for goodness of fit;
useful tests based on the likelihood function use likelihood {\em
  ratios}, as discussed in Appendix B.

\subsection[Change of parameter $\mu$ in pdf 
$p(x|\mu)$]{\boldmath Change of parameter $\mu$ in pdf $p(x|\mu)$}
\label{invariantl}

The pdf for $x$ given parameter $\mu=3$ is the {\em same} as the pdf
for $x$ given $1/\mu=1/3$, or given $\mu^2=9$, or given any specified
function of $\mu$. They all imply the same $\mu$, and hence the same
pdf for $x$.

In slightly confusing notation, that is what we mean by changing the
parameter from $\mu$ to $f(\mu)$, and saying that
\begin{equation}
p(x|f(\mu)) = p(x|\mu).
\end{equation}
    
Inserting an observed value of $x$, we see the important result that
the likelihood function $\lhood(\mu)$ is {\em invariant} (!) under
reparameterization from parameter $\mu$ to $f(\mu)$: $\lhood( f(\mu) )
= \lhood(\mu)$.  The absence of a Jacobian reinforces the fact that
$\lhood(\mu)$ is {\em not} a pdf in $\mu$.  Furthermore, the mode of
$\lhood(\mu)$ is thus also invariant: if we let $\hat\mu$ be the value
of $\mu$ for which $\lhood(\mu)$ is a maximum, and if we let $\hat f$
be the value of $f(\mu)$ for which $\lhood(f)$ is a maximum, then
\begin{equation}
\hat f = f(\hat\mu). 
\end{equation}
This is an important property of the popular ML technique of point
estimation. (See discussion in Ref.~\cite{barlow}, Section 5.3.1.)

\subsection{Probability integral transform}
\label{PIT}
In 1938 Egon Pearson commented on a paper of his father Karl (of
$\chi^2$ fame), noting that the {\em probability integral transform}
``\dots seems likely to be one of the most fruitful conceptions
introduced into statistical theory in the last few
years''~\cite{pearson1938}.  Indeed this simple transformation makes many
issues more clear (or trivial).  Given continuous $x \in (a,b)$, and
its pdf $p(x)$, one transforms to $y$ given by
\begin{equation}
\label{eqpit}
      y(x) = \int_a^x  p(x^\prime) dx^\prime.
\end{equation}
Then trivially $y \in (0,1)$, and with the chain rule and the
fundamental theorem of calculus, it follows that the pdf for $y$ is
$p(y) = 1$ (uniform) for all $y$ !  (If $x$ is discrete, there are
complications.)

{\em So, for continuous variables, there always exists a metric y in
  which the pdf is uniform}.  As an aside, the inverse transformation
can provide for efficient Monte Carlo (MC) generation of $p(x)$
starting from a pseudo-random number generator uniform on (0,1). See
Section 42.2 in Ref.~\cite{pdg2024}.

Looking ahead to Section~\ref{bayesintro}, I mention here that the
specification of a Bayesian prior pdf $p(\mu)$ for parameter $\mu$ is
thus equivalent to the choice of the metric $g(\mu)$ in which the pdf
is uniform.  This is a deep issue, not always recognized by users of
uniform prior pdf's in HEP!

\subsection{Bayes's theorem generalized to probability {\em densities}}

Recall (Eqn.~\ref{eqn-bayes}) that $P(B|A) \propto P(A|B) P(B)$. For
Bayesian $P$, continuous parameters such as $\mu$ are random variables
with pdf's.

Let pdf $p(\mu|x)$ be the conditional pdf for parameter $\mu$, given
data $x$.  As usual, $p(x|\mu)$ is the conditional pdf for data $x$,
given parameter $\mu$. Then Bayes's Theorem becomes $p(\mu|x) \propto
p(x|\mu) p(\mu)$.  Substituting in a particular set of observed data
$x_0$, we have $p(\mu|x_0) \propto p(x_0|\mu) p(\mu)$.
 
Recognizing the likelihood (variously written as $\lhood(x_0|\mu)$,
$\lhood(\mu)$, or unfortunately even $\lhood(\mu|x_0)$), then
\begin{equation}
\boxed{ p(\mu|x_0) \propto \lhood(x_0|\mu) p(\mu),}
\end{equation}
 where:

$p(\mu|x_0)  =$ the {\em posterior pdf} for $\mu$, given the results of 
this experiment,

$\lhood(x_0|\mu) =$ the likelihood function of $\mu$ from this experiment,

$p(\mu) =$ the {\em prior pdf} for $\mu$, before applying the results of
this experiment.

\medskip\noindent
Note that there is one (and only one) probability {\em density in}
$\mu$ on each side of the equation, consistent with $\lhood(x_0|\mu)$
{\em not} being a density in $\mu$!

(Aside: occasionally someone in HEP refers to the prior pdf as
``{\em a priori}''.  This is incorrect, as is obvious when one
considers that the posterior pdf from one experiment can serve as the
prior pdf for the next experiment.)

\section{Bayesian analysis}
\label{bayesintro}

All equations up until now are true for {\em any} definition of
probability $P$ that obeys the axioms, including frequentist $P$, as
long as the probabilities exist. For example, if $\mu$ is sampled from
an ensemble with known ``prior'' pdf, then it has a frequentist
interpretation. (This is however unusual.) The word ``Bayesian''
refers {\em not} to these equations, but to the choice of definition
of $P$ as {\em personal subjective degree of belief}. For example, if
$\mu$ is a constant of nature, its Bayesian pdf expresses one's
relative belief in different values.

Bayesian $P$ applies to hypotheses and constants of nature (while
frequentist $P$ does not), so there are many Bayesian-only
applications of Bayes's Theorem.  Bayesian analysis is based on the
Bayesian posterior pdf $p(\mu|x_0)$, as sketched here.

\begin{itemize}
\item{\em Point estimation}: Some Bayesians use the posterior mode
  (maximum posterior density) as the point estimate of $\mu$.
  This has the problem that it is metric dependent: the point estimate
  of the mean lifetime $\tau$ will not be the inverse of the point
  estimate of the decay rate $\Gamma$.  This is because the Jacobian
  moves the mode around under change of parameter from lifetime $\tau$
  to decay rate $\Gamma=1/\tau$. (Recall Section~\ref{metricchange}.)
  The posterior median can be used in 1D and is metric-independent.
  There are also Bayesians (as well as frequentists) who think that
  emphasis on point estimation is misguided.
\item{\em Interval estimation}: The {\em credibility} of $\mu$ being
  in any interval $[\mu_1,\mu_2]$ can be calculated by integrating
  $p(\mu|x_0)$ over the interval.  For reporting a default interval
  with, say, 68\% credibility, one needs in addition a convention for
  which 68\% to use (lower, upper, or central quantiles are common
  choices).  It is preferable to refer to such intervals as ``{\em
    credible intervals}'', as opposed to ``confidence intervals'',
  unless the Bayesian machinery is used just as a technical device to
  obtain valid (at least approximate) frequentist confidence intervals
  (as is often the case in HEP).
\item{\em Hypothesis testing}: Unlike frequentist statistics, testing
  credibility of whether or not $\mu$ equals a particular value
  $\mu_0$ is {\it not} performed by examining interval estimates (at
  least assuming a regular posterior pdf).  One starts over with
  Bayesian model selection, as discussed in
  Appendix~\ref{modelselection}.  (Dirac $\delta$-functions in the
  prior and posterior pdfs can however connect interval estimation to
  model selection, with its issues.)
\item{\em Decision making}: All {\em decisions} about $\mu$ require
  not only $p(\mu|x_0)$ but also further input: the utility function
  (or its negative, the loss function). See, e.g.,
  Ref.~\cite{james2006} (Chapter 6) and Ref.~\cite{bergerdecision1985}
  (Chapter 2).
\end{itemize}
Since Bayesian analysis {\em requires} a prior pdf, big issues in
Bayesian estimation include:

\begin{itemize}
\item What prior pdf to use, and how sensitive is the result to the
  prior?
\item How to interpret posterior probability if the prior pdf is not
  Your personal subjective belief?
\end{itemize}
{\em Frequentist tools can be highly relevant to both questions!}

\subsection{Can ``subjective'' be taken out of ``degree of belief''?}

There are compelling arguments (Savage~\cite{savagefoundations}, De
Finetti~\cite{definetti} and others) that Bayesian reasoning with {\em
  personal subjective} $P$ is the uniquely ``coherent'' way (with
technical definition of coherent) of updating {\em personal} beliefs
upon obtaining new data and making decisions based on them. However, these
foundational works can be very heavy going.  For a more accessible review
by an outspoken subjectivist, with more complete references, see
Lindley's 2000 review~\cite{lindley2000}. A 2018 book by Diaconis
and Skyrms~\cite{persi} is also very detailed and at a deep level
that seems mostly comprehensible to physicists.

A huge question is: {\em Can the Bayesian formalism be used by
  scientists to report the results of their experiments in an
  ``objective'' way (however one defines ``objective''), and does any
  of the glow of coherence remain when subjective $P$ is replaced by
  something else?}

An idea vigorously pursued by physicist Harold
Jeffreys~\cite{jeffreys1961} in the mid-20th century is: {\em Can one
  define a prior $p(\mu)$ that contains as little information as
  possible?}

The really {\em really} thoughtless idea (despite having a fancy name,
``Laplace's Principle of Insufficient Reason''), recognized by
Jeffreys as such, but dismayingly common historically in HEP is: just choose prior
$p(\mu)$ uniform in whatever metric you happen to be using!

\subsection{``Uniform prior'' requires a choice of metric}

Recall that the probability integral transform {\em always} allows one
to find a metric in which $p$ is uniform (for continuous $\mu$).  Thus
the question, ``What is the prior pdf $p(\mu)$?'' is equivalent to the
question, ``For what function $g(\mu)$ is $p(g)$ uniform?''  There is
usually {\it no reason} to choose $g$ arbitrarily as $g(\mu) = \mu$
(!).

\subsection{Jeffreys's choice of metric in which prior is uniform}
\label{jeffprior}

The modern foundation of the vast literature on prior pdfs that one
may hope (in vain) to be uninformative is the monograph by Harold
Jeffreys~\cite{jeffreys1961}.  He proposes more than one approach, but
the one that is commonly referred to as the ``Jeffreys prior'' (and
considered the default ``noninformative'' prior by statisticians for
estimation in 1-parameter problems) is {\em derived from the
  statistical model} $p(x|\mu)$.

{\em This means that the prior pdf depends on the measurement
  apparatus!}  For example, if the measurement apparatus has a
resolution function that is Gaussian for mass $m$ (with $\sigma$
independent of $m$), then the Jeffreys prior pdf $p(m)$ for the mass
is uniform in $m$.  If a different measurement apparatus has a
resolution function that is Gaussian for $m^2$, then the Jeffreys
prior pdf $p(m^2)$ is uniform in $m^2$. In the latter case, by the
rules of probability (Eqn.~\ref{jacobian}), the prior pdf $p(m)$ is
not uniform, but rather proportional to $m$ (!).

Jeffreys's derivation of his eponymous prior is based on the idea that
the prior should be uniform in a metric related to the Fisher
information, calculated from the curvature of the log-likelihood function
averaged over sample space.  Some examples are:
\begin{description}
\item{} Poisson signal mean $\mu$, no background:   
$p(\mu) = 1/\sqrt{\mu}$
\item{} Poisson signal mean $\mu$, known mean background $b$: 
$p(\mu) = 1/\sqrt{\mu+b}$
\item{} Mean $\mu$ of Gaussian with fixed $\sigma$ 
(unbounded or bounded $\mu$): $p(\mu) = 1$
\item{} rms deviation $\sigma$ of a Gaussian when mean fixed: 
$p(\sigma) = 1/\sigma$
\item{} Binomial parameter $\rho$, $0 \le \rho \le 1$ : 
$p(\rho) = \rho^{-1/2}(1 - \rho)^{-1/2} = \textnormal{Beta}(1/2,1/2)$.
\end{description}

If parameter $\mu$ is changed to $f(\mu)$, the recipe for obtaining
the Jeffreys prior for $f(\mu)$ yields a different-looking prior that
corresponds to the {\em same choice of uniform metric}. So if you use
Jeffreys's recipe to obtain a prior pdf for $\mu$, and your friend
uses Jeffreys's recipe to obtain a prior pdf for $f(\mu)$, then those
pdfs will be correctly related by the appropriate Jacobian.  (This is
not true for some other rules, in particular if each of you takes a
uniform prior in the metric you are using.) Thus {\em probabilities}
(integrals of pdfs over equivalent endpoints) using Jeffreys prior are
invariant under choices of different parameterizations.

For a detailed modern review of Jeffreys's entire book, including his
prior, with discussion by six prominent statisticians (including
outspoken subjectivist Lindley), and rejoinder, see
Ref.~\cite{robert2009}.

\subsubsection{Reference priors of Bernardo and Berger}
As Jeffreys noted, his recipe encounters difficulties with models
having more than one parameter.  Jos\'e Bernardo
\cite{bernardoRSS1979} and
J.O. Berger~\cite{bergerbern89,bergerbern92} advocate an approach that
they argue works well in higher dimensions, with the crucial
observation that one must choose an ordering of the parameters in order to
well-define the multi-dimensional prior pdf. Their so-called
``Reference prior'' reduces to the Jeffreys prior in 1D, with a
different rationale, namely the prior that leads to a posterior pdf
that is most dominated by the likelihood.  Earlier this year, Berger and Bernardo (with Donchu Sun) published a new book~\cite{berger2024} discussing in detail these issues of estimation
(though not hypothesis testing).

There are many subtleties. Beware!  See also Bernardo's talk at
PhyStat-2011, which includes hypothesis testing, and discussion
following~\cite{bernardo2011phystat}.  Demortier, Jain, and Prosper
pioneered the use of Bernardo/Berger reference priors in a case of
interest in HEP~\cite{demortier2010}, with about 40 citations thus far (mid-2024)
at inspirehep.net.

\subsection{What to call such non-subjective priors?}

\begin{itemize}
\item ``{\em Non}informative priors''? Traditional among
  statisticians, even though {\em they know it is a misnomer}. (You
  should too!)
\item ``Vague priors''? 
\item ``Ignorance priors''? 
\item ``Default priors''?
\item ``Reference priors''? (Unfortunately also refers to the specific
  recipe of Bernardo and Berger)
\item ``Objective priors''?  Despite the highly questionable use of
  the word, Jeffreys prior and its generalization by Bernardo and
  Berger are now widely referred to as ``objective priors''.
\end{itemize}

Kass and Wasserman~\cite{kasswasserman} give the best (neutral) name
in my opinion: Priors selected by ``formal rules''.  Their article is
required reading for anyone using Bayesian methods!

Whatever the name, the prior pdf in one metric determines it in all
other metrics: be careful in the choice of metric in which it is
uniform!

For one-parameter models, the ``Jeffreys prior'' is the most common
choice among statisticians for a non-subjective ``default'' prior---so
common that statisticians can be referring to the Jeffreys prior when
they say ``flat prior'' (e.g., D.R. Cox in discussion at PhyStat
2005~\cite{coxdiscussion}, p.\ 297).

A key point: priors such as the Jeffreys prior are based on the
likelihood function and thus inherently derived from the {\em
  measurement apparatus and procedure}, not from thinking about the
parameter!  This may seem strange, but does give advantages,
particularly for frequentist (!) coverage
(Section~\ref{sec-coverage}), as mentioned in Section~\ref{pragmatism}
and emphasized to us by Jim Berger at the Confidence Limits Workshop
at Fermilab in 2000~\cite{bergerclk}.

\subsection{Whatever you call them, non-subjective priors {\em cannot} 
represent ignorance!}

Although some authors have claimed that various invariance principles
can be invoked to yield priors that represent complete ``ignorance'',
I do not know of any modern statistician who thinks that this is
possible.  On the contrary, (subjectivist) Dennis Lindley
wrote,~\cite{lindley1990}, ``the mistake is to think of them [Jeffreys
  priors or Bernardo/Berger's reference priors] as representing
ignorance.''  This Lindley quote is emphasized in the monograph by
prominent Bayesian Christian Robert~\cite{robert2007} (p.\ 29).

Objectivist Jose Bernardo says, regarding his reference priors,
``[With non-subjective priors,] The contribution of the data in
constructing the posterior of interest should be `dominant'. Note that
this does not mean that a non-subjective prior is a mathematical
description of `ignorance'. Any prior reflects some form of
knowledge.''

Nonetheless, Berger~\cite{bergerdecision1985} (p.\ 90) argues that
Bayesian analysis with noninformative priors (older name for objective
priors), such as those of Jeffreys and Bernardo/Berger, ``{\em is the
  single most powerful method of statistical analysis}, in the sense
of being the {\em ad hoc} method most likely to yield a sensible
answer for a given investment of effort.'' [emphasis in original].

\subsection{Priors in high dimensions}

Is there a sort of informational ``phase space'' that can lead us to a
sort of probability Dalitz plot? I.e., the desire is that structure in
the posterior pdf represents information in the data, not the effect
of Jacobians in the priors.  This is a {\em notoriously hard problem!}

Be careful: Uniform priors push the probability away from the origin
to the boundary! (The volume element goes as $r^2dr$.)  The state of
the art for ``objective'' priors may be the ``reference priors'' of
Bernardo and Berger, but multi-D tools have been lacking.  Subjective
priors are also very difficult to construct in high dimensions: human
intuition is poor.

Subjective Bayesian Michael Goldstein~\cite{goldsteinphystat} told us
in 2002 at Durham, ``\dots meaningful prior specification of beliefs
in probabilistic form over very large possibility spaces is very
difficult and may lead to a lot of arbitrariness in the specification
\dots''.

Bradley Efron at PhyStat-2003~\cite{efronslac} concluded: ``Perhaps
the most important general lesson is that the facile use of what
appear to be uninformative priors is a dangerous practice in high
dimensions.''

Sir David Cox~\cite{cox2006}, p.\ 46: ``With multi-dimensional
parameters\dots naive use of flat priors can lead to procedures that
are very poor from all perspectives\dots''.  Also, on p.\ 83: ``\dots the
notion of a flat or indifferent prior in a many-dimensional problem is
untenable.''

\subsection{Types of Bayesians}
\label{fivefaces}
The broad distinction between the subjective and objective Bayesians
is far from the complete story.  At PhyStat-LHC in 2007, Sir David Cox
described ``Five faces of Bayesian statistics''~\cite{coxphystat}:
\begin{itemize}
\item empirical Bayes: number of similar parameters with a frequency
  distribution
\item neutral (reference) priors: Laplace, Jeffreys, Jaynes, Berger
  and Bernardo
\item information-inserting priors (evidence-based)
\item personalistic priors
\item technical device for generating frequentist inference
\end{itemize}
Currently in HEP, the main application is the last ``face'' on his
list: we typically desire good frequentist properties for point or
interval estimation when using a nominally ``Bayesian'' recipe.  In
particular, for {\em upper limits} on a Poisson mean, we use a uniform
prior (i.e., not what objective statisticians recommend), for {\it
  frequentist} reasons. (See Section~\ref{intervalsummary} and
Ref.~\cite{cousinsajp1995}.)  Cox's third ``face'' also arises in HEP
when likelihoods are incorporated from subsidiary measurements and
are used to provide evidence-based priors.  Unfortunately, some
people in HEP have in effect also added to Cox's list a 6th ``face'':
\begin{itemize}
\item Priors uniform in arbitrary variables, or in ``the parameter of
  interest''.  
\end{itemize}
I know of no justification for this in modern subjective or objective
Bayesian theory.  It is an ``ignorance'' prior only in the sense that
it betrays ignorance of the modern Bayesian literature!

\subsection{Do ``Bayesians'' care about frequentist properties of 
their results?}

Another claim that is dismaying to see among some physicists is the
blanket statement that ``Bayesians do not care about frequentist
properties''.  While that may be true for pure subjective Bayesians,
most of the pragmatic Bayesians that we have met at PhyStat meetings
do use frequentist methods to help calibrate Bayesian statements.
That seems to be essential when using ``objective'' priors to obtain
results that are used to communicate inferences from experiments.

A variety of opinions on this topic are in the published comments in
the inaugural issue of the journal {\em Bayesian Analysis}, following
the papers by Jim Berger~\cite{berger2006} and Michael
Goldstein~\cite{goldstein2006} that advocate the objective and
subjective points of view, respectively.  In particular, Robert
Kass~\cite{kass2006} provided a list of nine questions for which
answers ``may be used to classify various kinds of Bayesians''.

The first question is, ``Is it important for Bayesian inferences to
have good frequentist operating characteristics?''  The questions, the
explicit answers from Kass, Berger, and Goldstein, and other
commentary (from which answers can be gleaned) from other prominent
statisticians, are part of my list of ``required reading'' for
physicists.  They should go a long way toward broadening the view of
any physicists who have swallowed some of the extreme polemics about
Bayesian analysis (as found for example in Jaynes~\cite{jaynes2003},
which unfortunately seems to have been read uncritically by too many
scientists).

\subsection{Analysis of sensitivity to the prior}

Since a Bayesian result depends on the prior probabilities, which are
either personalistic or with elements of arbitrariness, it is widely
recommended by Bayesian statisticians to study the {\em sensitivity}
of the result to varying the prior.
\begin{itemize}
\item ``Objective'' Bayesian Jos\'e Bernardo, quoted in 
Ref.~\cite{bernardodialog}: ``Non-subjective
  Bayesian analysis is just a part---an important part, I believe
 ---of a healthy {\em sensitivity analysis} to the prior
  choice\dots''.

\item ``Subjective'' Bayesian Michael Goldstein, from the
  Proceedings~\cite{goldsteinphystat}: ``\dots Again, different
  individuals may react differently, and the sensitivity analysis for
  the effect of the prior on the posterior is the analysis of the
  scientific community\dots''. In his transparencies at the conference,
  he put it simply: ``Sensitivity Analysis is at the heart of
  scientific Bayesianism.''
\end{itemize}
I think that historically, too little emphasis was given to this important point by
Bayesian advocates in HEP.

\subsection{Bayesian must-read list for HEP/Astro/Cosmo 
(including discussion!)}

In my experience, some high energy physicists and astrophysicists
appear to be overly influenced by the polemical book by
E.T. Jaynes~\cite{jaynes2003}, (which, for example, argues for the
existence priors representing ignorance).  I strongly urge anyone
diving into Bayesian statistics to read as well the following {\em
  minimal} set of papers by Bayesian subjectivists and objectivists,
and the associated discussion, and rejoinders.  (If there is one thing
that HEP journals could learn from statisticians, it is to publish
such discussion and rejoinder accompanying major papers and reviews!)
As I made this list over ten years ago, I would welcome suggestions for
more recent additions.

\begin{description}
\item Robert E. Kass and Larry Wasserman, ``The Selection of Prior
  Distributions by Formal Rules,'' \cite{kasswasserman}

\item Telba Z. Irony and Nozer D. Singpurwalla, ``Non-informative
  priors do not exist: A dialogue with Jose M. Bernardo,''
  \cite{bernardodialog}

\item James Berger, ``The Case for Objective Bayesian Analysis,''
  \cite{berger2006}

\item Michael Goldstein, ``Subjective Bayesian Analysis: Principles
  and Practice,'' \cite{goldstein2006}

\item J.O. Berger and L.R. Pericchi, ``Objective Bayesian Methods for
  Model Selection: Introduction and Comparison,''
  \cite{bergerpericchi2001}

\end{description}

\subsection{Pseudo-Bayesian analyses}
\label{pseudobayes}
Jim Berger in 2006~\cite{berger2006}: 

``One of the mysteries of modern Bayesianism is the lip service that
is often paid to subjective Bayesian analysis as opposed to objective
Bayesian analysis, but then the practical analysis actually uses a
very ad-hoc version of objective Bayes, including use of constant
priors, vague proper priors, choosing priors to `span' the range of
the likelihood, and choosing priors with tuning parameters that are
adjusted until the answer `looks nice.' I call such analyses {\em
  pseudo-Bayes} because, while they utilize Bayesian machinery, they
do not carry with them any of the guarantees of good performance that
come with either true subjective analysis (with a very extensive
elicitation effort) or (well-studied) objective Bayesian analysis\dots
I do not mean to discourage this approach. It simply must be realized
that pseudo-Bayes techniques do not carry the guarantees of proper
subjective or objective Bayesian analysis, and hence must be validated
by some other route.''

Berger goes on to give examples of pseudo-Bayes analyses, with the
first being (what else?), ``Use of the constant prior density''.

Pseudo-Bayes analyses pop up from time to time in HEP, for example
by those using priors ``uniform in the parameter of
interest''.  Here I mention three examples of ``pseudo-Bayes'' in HEP
that have been criticized by me and others.  I previously discussed
them at the PhyStat-nu workshop in Tokyo in 2016~\cite{cousinstokyo}
(slides 62--68).

The first example was exposed in Luc Demortier's talk in 2002 at
Durham~\cite{demortierdurham}.  In a method of Bayesian upper limit
calculation that was common at the Tevatron at the time, the use of a
uniform prior for a Poisson mean, along with a Gaussian truncated at
the origin for a systematic uncertainty in efficiency, led to an
integral that Luc showed ``by hand'' to diverge!  The integral was
typically evaluated numerically, without first checking that it
exists.  The answer thus depended on the choice of cutoff that was
used in the numerical evaluation.  Alternatives to the truncated
Gaussian prior are mentioned in Section~\ref{nuisprior}.

The second example comes from Joel Heinrich at the Oxford PhyStat in
2005~\cite{heinrich2005}.  It had been known for a long time that a
uniform prior for a Poisson mean of a {\em signal} yields good
frequentist properties for {\it upper limits} (but not lower
limits). (See Section~\ref{intervalsummary}.)  Joel showed the dangers
of naively using uniform priors for the means of several background
processes.

The third example is in the category that I find has some of the worst
pseudo-Bayes examples, namely Bayesian model selection
(Appendix~\ref{modelselection}), which is however not attempted as
often as Bayesian estimation in HEP.  Practitioners are sometimes
unaware that:
\begin{enumerate}
\item In model selection, unlike estimation, the dependence on some
  prior pdfs for parameters does {\em not} become negligible as the
  amount of data increases without bound (even for so-called Bayes
  factors that attempt to separate out the prior probabilities of the
  hypotheses).
\item Improper priors (such as uniform over a line or half-line) are a
  disaster, and if they are made proper by adding a cutoff, then the
  model selection answer is directly proportional to the (often
  arbitrary) cutoff.  Using ``1'' for the prior just hides the
  problem.
\item In fact, Jeffreys and followers use priors for model selection
  that are different from those used for estimation (!).
\end{enumerate}
Thus, Bayesian model selection should not be approached naively.  For
an example in PRL that I criticized, see my Comment in
Ref.~\cite{cousins2008}.

Harrison Prosper (an early and sustained advocate of Bayesian methods
in HEP) has provided an excellent discussion of the care needed in
Bayesian analyses in Chapter 12 of {\it Data Analysis in High Energy
  Physics}, edited by O. Behnke et al.~\cite{behnke2013}.

\section{Frequentist estimation: confidence intervals}
\label{secconfidence}

What can be computed without using a prior, with only the frequentist
definition of $P$?
\begin{itemize}
\item{\em Not} $P(\textnormal{constant of nature is in some specific
  interval} \,|\, \textnormal{data}) $
\item{\em Not}  
$P(\textnormal{SUSY is true} \,|\, \textnormal{data}) $
\item{\em Not}  
$P(\textnormal{Standard Model is false} \,|\, \textnormal{data}) $
\end{itemize}
Rather, without a prior, we can compute:
\begin{enumerate}
\item {\em Confidence Intervals} for constants of nature or other
  parameter values, as defined in the 1930's by Jerzy
  Neyman. Statements are made about probabilities in {\em ensembles}
  of intervals (fraction containing unknown true value).  Confidence
  intervals have further applications in frequentist hypothesis
  testing.
\item Likelihoods and thus {\em likelihood ratios}, the basis for a
  large set of techniques for point estimation, interval estimation,
  and hypothesis testing.
\end{enumerate}
Both can be constructed using the {\em frequentist} definition of $P$.
In this section, we introduce confidence intervals, and in
Section~\ref{likelihood}, introduce likelihood ratios for interval
estimation.

``Confidence intervals'', and this phrase to describe them, were
invented by Jerzy Neyman in 1934-37~\cite{neyman1937}.  Statisticians
typically mean Neyman's intervals (or an approximation thereof) when
they say ``confidence interval''.  In HEP the language is a little
loose.  I highly recommend using ``confidence interval'' (and
``confidence regions'' when multi-D) only to describe intervals and
regions corresponding to Neyman's construction, described below, or by
recipes of other origins (including Bayesian recipes) only if they
yield good approximations thereof.

The following subsections use upper/lower limits and closely related
central confidence intervals to introduce and illustrate the basic
notions, and then discuss Neyman's more general construction (used
e.g. by Feldman and Cousins).  Then, after introducing frequentist
hypothesis testing (Section~\ref{hypotest}), we return to make the
connection between confidence intervals and hypothesis testing of a
particular value of parameter vs other values (Section~\ref{duality}).

\subsection{Notation}

\label{notation}
\begin{description}
\item{} $x$ denotes observable(s). More generally, $x$ is any
  convenient or useful function of the observable(s), and is called a
  ``statistic'' or ``test statistic''.
\item{} $\mu$ denotes parameter(s).  (Statisticians often use
  $\theta$.)
\item{} $p(x| \mu)$ is the probability or pdf (from context)
  characterizing everything that determines the
  probabilities/densities of the observations, from laws of physics to
  experimental setup and protocol. The function $p(x| \mu)$ is called
  ``the statistical model'', or simply ``the model'', by
  statisticians.
\end{description}

\subsection{The conceptual idea of confidence intervals in two 
sentences}
\label{twosentences}
Given the model $p(x| \mu)$ and the observed value $x_0$, we ask: For
what values of $\mu$ is $x_0$ an ``extreme'' value of $x$?  Then we
include in the confidence interval $[\mu_1, \mu_2]$ those values of
$\mu$ for which $x_0$ is {\em not} ``extreme''.

(Note that this basic idea sticks strictly to the frequentist
probability of obtaining $x$, and makes no mention of probability (or
density) for $\mu$.)

\subsubsection[Ordering principle is required for possible values of 
$x$]{\boldmath Ordering principle is required for possible values of
  $x$}
\label{ordering}

In order to define ``extreme'', one needs to choose an {\em ordering
  principle} that ranks the possible values of $x$ applicable to each
$\mu$.  By convention {\em high rank means not extreme}.

Some common ordering choices in 1D (when $p(x| \mu)$ is such that
higher $\mu$ implies higher average $x$) are:

\begin{itemize}
\item Order $x$ from largest to smallest: the smallest values of $x$
  are the most extreme.  Given $x_0$, the confidence interval that
  contains $\mu$ for which $x_0$ is not extreme will typically not
  contain the largest values of $\mu$.  This leads to confidence
  intervals known as {\em upper limits} on $\mu$.
\item Order $x$ from smallest to largest.  This leads to {\em lower
  limits} on $\mu$.
\item Order $x$ using {\em central} quantiles of $p(x| \mu)$, with 
the quantiles shorter in $x$ (least integrated probability of $x$) containing higher-ranked
$x$, with lower-ranked $x$ added as the central quantile gets longer (contains more
integrated probability of $x$).
This leads to {\em central} confidence intervals
  for $\mu$. 
\end{itemize}

These three orderings apply only when $x$ is 1D. A fourth ordering,
using a particular likelihood ratio advocated by Feldman and Cousins,
is still to come (Section~\ref{orderFC}); it is applicable in both 1D
and multi-D.

\subsubsection{A ``confidence level'' must be specified}

Given model $p(x| \mu)$ and an ordering of $x$, one chooses a
particular fraction of highest-ranked values of $x$ that are {\em not}
considered as ``extreme''.  This fraction is called the {\em
  confidence level} (\cl), say 68\% or 95\%.

(In this discussion, 68\% is more precisely 68.27\%; 84\% is 84.13\%;
etc.  Also, in the statistics literature there is a fine distinction
between confidence level and confidence coefficient, which we ignore
here.)

We also define $\alpha = 1 - $\cl, i.e., the fraction that is the
lower-ranked, ``extreme'' set of values.

\subsubsection{One-sentence summary}
Then the confidence interval $[\mu_1, \mu_2]$ contains those values of
$\mu$ for which $x_0$ is {\em not} ``extreme'' at the chosen \cl, {\em
  given the chosen ordering of $x$}.  E.g., at 68\% \cl, $[\mu_1,
  \mu_2]$ contains those $\mu$ for which $x_0$ is in the
highest-ranked (least extreme) 68\% values of $x$ for each respective
$\mu$, according probabilities obtained from the model $p(x|\mu)$.

\subsection{Correspondence between upper/lower limits and central 
confidence intervals} 

\begin{figure}
\begin{center}
\includegraphics[width=0.40\textwidth]{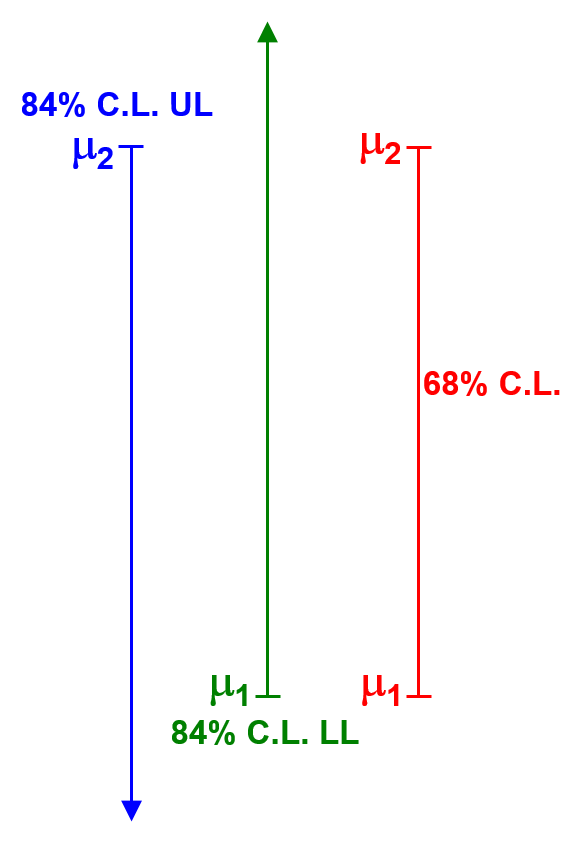}
\caption{Sketch to illustrate the relationship among 84\% C.L. upper
  limits, 84\% C.L. lower limits, and 68\% C.L. central confidence
  interval.  The two one-sided tails of 16\% compose the two-sided
  tails totaling 32\%.}
\label{limits_central}
\end{center}
\end{figure}

As illustrated in Fig.~\ref{limits_central}, the 84\% \cl\ upper limit
$\mu_2$ excludes $\mu$ for which $x_0$ is in the lowest 16\% values of
$x$.  The 84\% \cl\ lower limit $\mu_1$ excludes $\mu$ for which $x_0$
is in the highest 16\% values of $x$.  Then, the interval $[\mu_1,
  \mu_2]$ includes $\mu$ for which $x_0$ is in the central 68\%
quantile of $x$ values.  It is a 68\% \cl\ {\em central} confidence
interval (!).

For general $\cl$, the endpoints of central confidence intervals are,
by the same reasoning, the same as upper/lower limits with confidence
level given by $1-(1-\cl)/2 = (1+\cl)/2$.

Examples follow for a couple of illustrative models, first with
continuous $x$, then with discrete $x$.

\subsection[Gaussian pdf $p(x| \mu,\sigma)$ with $\sigma$ 
a function of $\mu$: $\sigma = 0.2\mu$]{\boldmath Gaussian pdf $p(x|
  \mu,\sigma)$ with $\sigma$ a function of $\mu$: $\sigma = 0.2\mu$}
\label{secgauss02sigma}

\begin{figure}
\begin{center}
\includegraphics[width=0.49\textwidth]{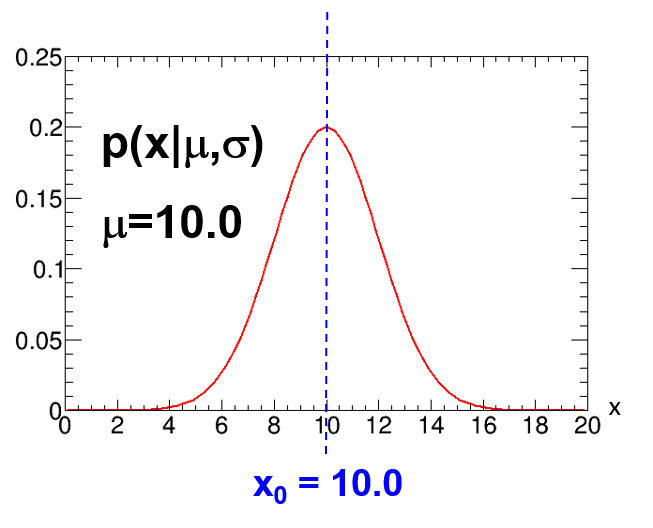}
\caption{Plot of $p(x| \mu,\sigma)$, with $\mu=10.0$, $\sigma=0.2\mu$,
  as in Eqn.~\ref{eqn-gaussian_prob}.  In the subsequent discussion,
  we suppose that the observed value is $x_0 = 10.0$.}
\label{fig-gaussian_prob}
\end{center}
\end{figure}

It is common in HEP to express uncertainties as a percentage of a mean
$\mu$ (even though situations in which this is rigorously motivated
are rare).  Thus, instead of the most trivial example of Gaussian with
unknown mean $\mu$ and {\em known} rms $\sigma$, let's assume that
$\sigma$ is 20\% of the unknown true value $\mu$:
\begin{equation}
\label{eqn-gaussian_prob}
p(x| \mu,\sigma) 
= \frac{1}{\sqrt{2\pi\sigma^2}} \textnormal{e}^{-(x-\mu)^2/2\sigma^2}
 = p(x| \mu) = \frac{1}{\sqrt{2\pi(0.2\mu)^2}} \textnormal{e}^{-(x-\mu)^2/2(0.2\mu)^2},
\end{equation}
as illustrated in Fig.~\ref{fig-gaussian_prob}.

\begin{figure}
\begin{center}
\includegraphics[width=0.49\textwidth]{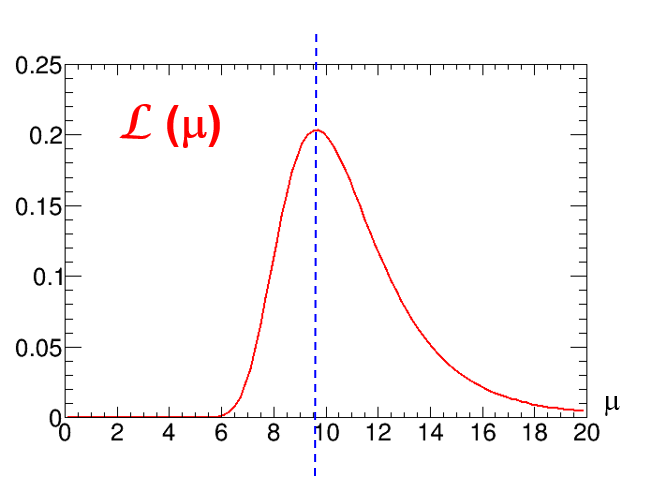}
\caption{Plot of $\lhood(\mu)$ for observed $x_0 = 10.$, for the model
  in Eqn.~\ref{eqn-gaussian_prob}.  The maximum is at $\mu_{ML}=
  9.63$}
\label{fig-gaussian_likl}
\end{center}
\end{figure}

With $\mu$ (and hence $\sigma$) unknown, suppose $x_0 = 10.0$ is
observed.  What can one say about $\mu$?  Regardless of what else is
done, it is always useful to plot the likelihood function, in this
case obtained by plugging $x_0$ into Eqn.~\ref{eqn-gaussian_prob}:
\begin{equation}
\lhood(\mu)  = 
\frac{1}{\sqrt{2\pi(0.2\mu)^2}} \textnormal{e}^{-(10.0-\mu)^2/2(0.2\mu)^2},
\end{equation}
as illustrated in Fig.~\ref{fig-gaussian_likl}.  The first thing one
notices is that the likelihood is asymmetric and obtains its maximum at
$\mu$ less than the $x_0$.  (There is food for thought here, but we
move on.)

\subsubsection[What is the {\em central confidence interval} for 
$\mu$?]{\boldmath What is the {\em central confidence interval} for
  $\mu$?}
\label{secgauss02sigmacentral}

\begin{figure}
\begin{center}
\includegraphics[width=0.49\textwidth]{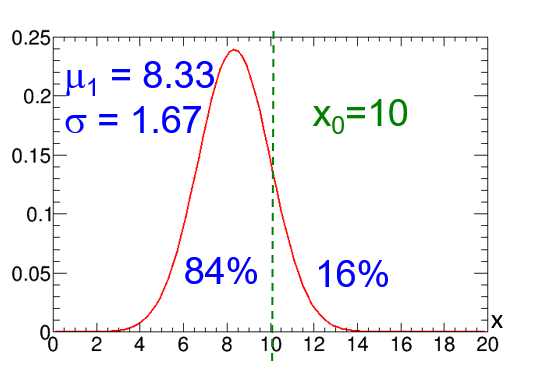}
\includegraphics[width=0.49\textwidth]{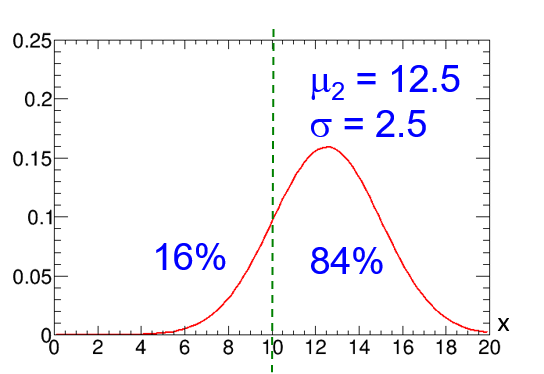}
\caption{(left) The model in Eqn.~\ref{eqn-gaussian_prob}, with $\mu$
  chosen such that 84\% of the probability is {\em below} $x_0 =
  10.0$. (right) The model in Eqn.~\ref{eqn-gaussian_prob}, with $\mu$
  chosen such that 84\% of the probability is {\em above} $x_0 =
  10.0$.}
\label{gaussian_lower_upper}
\end{center}
\end{figure}

First we find $\mu_1$ such that 84\% of $p(x|\mu_1,\sigma=0.2\mu_1)$
is {\em below} $x_0 = 10.0$; 16\% of the probability is above.  Solve:
$\mu_1 = 8.33$, as in Fig.~\ref{gaussian_lower_upper}(left).  So
$[\mu_1,\infty]$ is an 84\% \cl\ confidence interval, and $\mu_1$ is
84\% \cl\ {\em lower} limit for $\mu$.

Then we find $\mu_2$ such that 84\% of $p(x|\mu_2,\sigma=0.2\mu_2)$ is
{\em above} $x_0 = 10.0$; 16\% of the probability is below.  Solve:
$\mu_2 = 12.52$, as in Fig.~\ref{gaussian_lower_upper}(right). So
$[-\infty,\mu_2]$ is an 84\% \cl\ confidence interval, and $\mu_2$ is
84\% \cl\ {\em upper} limit for $\mu$.

Then the 68\% \cl\ central confidence interval is $[\mu_1,\mu_2] =
[8.33,12.52]$.  This is ``exact''.  In fact, the reasoning used
here was already laid out by E.B. Wilson in 1927~\cite{wilson1927} in
his discussion of intervals in the Gaussian approximation of the
binomial model (Section~\ref{wilson}).

\subsubsection{Contrast Wilson reasoning with ``Wald interval'' 
reasoning}
\label{wald}

The Wilson-inspired reasoning is crucial. Note the difference from the
superficially attractive reasoning that proceeds as follows, leading
to so-called Wald intervals:
\begin{enumerate}
\item For $x_0 = 10.0$, the ``obvious'' point estimate $\hat\mu$
  (perhaps thought to be justified by minimum-$\chi^2$) is $\hat\mu =
  10.0$.  (This makes the mistake of interpreting the sampled value of
  $x$ as being the ``measured value of $\mu$'', and hence as the point
  estimate of $\mu$, $\hat\mu$.)
\item
Then one estimates $\sigma$ by $\hat\sigma = 0.2 \times \hat\mu =
2.0$.  (This is potentially dangerous since it estimates one
parameter by plugging the point estimate of another parameter into a
relationship between {\it true} values of parameters.)
\item
Then $\hat\mu\, \pm\, \hat\sigma$ yields the interval $[8.0,12.0]$.  (Again
this is potentially dangerous, as it evaluates tail probabilities of a
model having true parameters via plugging in estimates of the
parameters.)
\end{enumerate}
Such ``Wald intervals'' are {\it at best} only approximations to exact
intervals.  For (``exact'') confidence intervals, the reasoning must
{\it always} involve probabilities for $x$ calculated considering
particular possible true values of parameters, as in the Wilson
reasoning!  Clearly, the validity of the approximate Wald-interval
reasoning (i.e., how dangerous the ``potentially dangerous'' steps
are) depends on how much $\sigma(\mu)$ changes for $\mu$ relevant to
problem at hand. The important point is that the Wald reasoning is not
the correct reasoning for confidence intervals. Beware!

\subsection[Confidence intervals for binomial parameter 
$\rho$]{\boldmath Confidence intervals for binomial parameter $\rho$}
\label{sec-binomial}

The binomial model is directly relevant to efficiency calculations in
HEP, as well as other contexts.  Recall the binomial distribution
$\bi(\non|\ntot,\binp)$ from Eqn.~\ref{eqn-binomial} for the
probability of $\non$ successes in $\ntot$ trials, each with binomial
parameter $\binp$.  In repeated trials, $\non$ has mean
\begin{equation}
\label{meannon}
\ntot\,\binp
\end{equation}
and rms deviation
\begin{equation}
\label{rmsnon}
\sqrt{\ntot\,\binp\,(1-\binp)}\,.
\end{equation}
For asymptotically large $\ntot$, $\bi(\non|\ntot,\binp)$ can be
approximated by a normal distribution with this mean and rms
deviation.

With observed number of successes $\non$, the likelihood function
$\lhood(\binp)$ follows from reading Eqn.~\ref{eqn-binomial} as a
function of $\binp$.  The maximum is at
\begin{equation}
\hat\binp = \non/\ntot.
\end{equation}

\begin{figure}
\begin{center}
\includegraphics[width=0.49\textwidth]{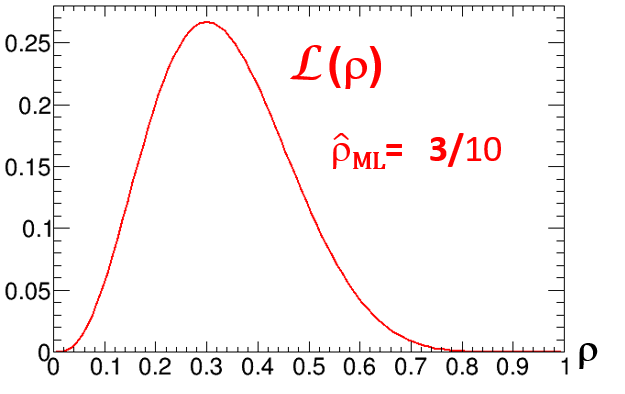}
\includegraphics[width=0.49\textwidth]{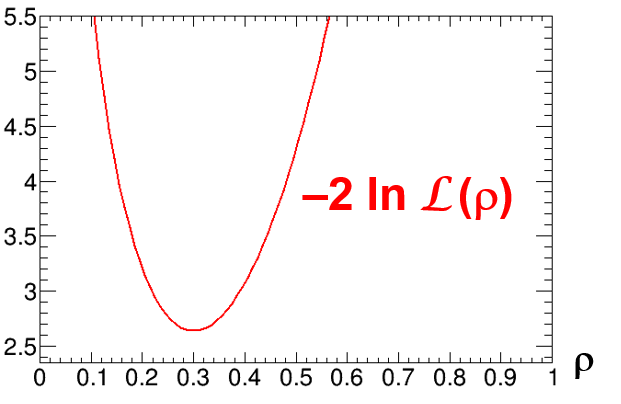}
\caption{(left) Likelihood function $\lhood(\binp)$ for $\non=3$
  successes in $\ntot=10$ trials in the binomial model of
  Eqn.~\ref{eqn-binomial}. (right) Looking ahead to
  Section~\ref{likelihood}, the plot of $-2\ln\lhood(\binp)$.}
\label{binomial_likl}
\end{center}
\end{figure}

Suppose one observes $\non=3$ successes in $\ntot=10$ trials.  The
likelihood function $\lhood(\binp)$ is plotted in
Fig.~\ref{binomial_likl}.

\subsubsection[What {\em central confidence interval} should 
we report for $\rho$?]{\boldmath What {\em central confidence
    interval} should we report for $\rho$?}

We have $\non = 3$, $\ntot=10$.  Let us find the ``exact'' 68\%
\cl\ central confidence interval $[\rho_1,\binp_2]$.  Recall the
shortcut in Section~\ref{secgauss02sigmacentral} for central
intervals: Find the {\em lower limit} $\binp_1$ with \cl\ $=
(1+0.68)/2 = 84$\%; and find {\em upper limit} $\binp_2$ with \cl\ $=
84$\%.  Then $[\binp_1,\binp_2]$ is a 68\% \cl\ central confidence
interval.

\begin{figure}
\begin{center}
\includegraphics[width=0.49\textwidth]{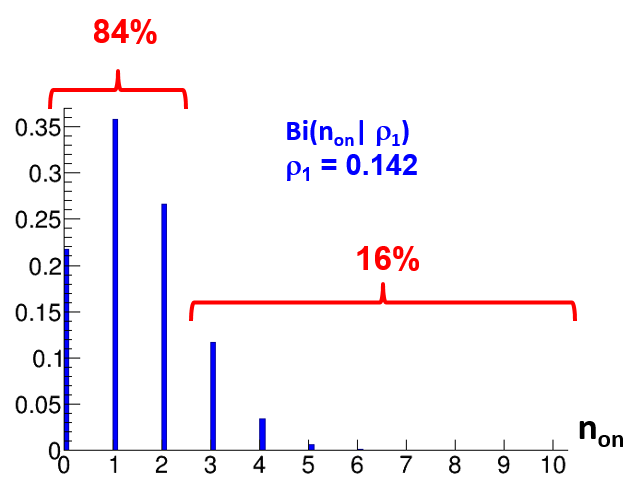}
\includegraphics[width=0.49\textwidth]{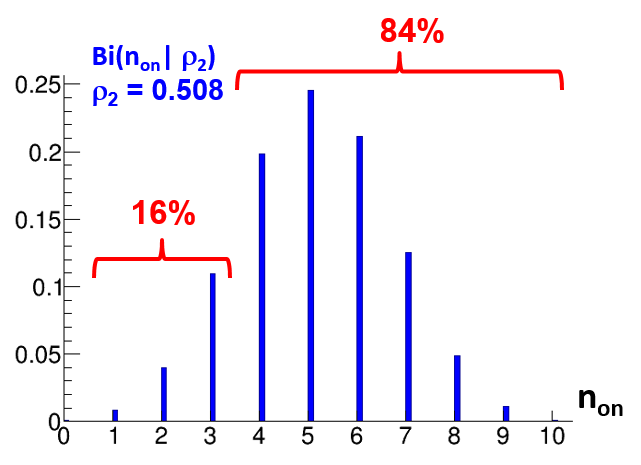}
\caption{(left) Plot of $\bi(\non|\ntot=10,\binp)$, with $\binp$ 
chosen such that  84\% of the probability is {\em below} $\non=3$.
(right) Plot of $\bi(\non|\ntot=10,\binp)$, with $\binp$ 
chosen such that  84\% of the probability is {\em above} $\non=3$.}
\label{binomial_lower}
\end{center}
\end{figure}

\begin{enumerate}
\item For lower limit at 84\% \cl, find
$\binp_1$ such that
$\bi(\non < 3 | \binp_1)    = 84$\%, i.e.,
$\bi(\non \ge 3 | \binp_1)  = 16$\%.
Solve: $\binp_1 = 0.142$ as in Fig.~\ref{binomial_lower} (left).

\item For upper limit at 84\% \cl), find
$\binp_2$ such that
$\bi(\non > 3 | \binp_2)    = 84$\%, i.e.,
$\bi(\non \le 3 | \binp_2)  = 16$\%.
Solve: $\binp_2 = 0.508$ as in Fig.~\ref{binomial_lower} (right).

\item Then $[\binp_1,\binp_2] = [0.142, 0.508]$ is a {\em central}
confidence interval with 68\% \cl\ 
\end{enumerate}

This is the same result as obtained by Clopper and Pearson (C-P) in
1934~\cite{clopper1934}, citing Fisher, Neyman, and Neyman's students.
In HEP, such C-P intervals are the standard for a binomial parameter;
they have been in PDG RPP since 2002~\cite{pdg2024} (although the
connection of the given formula with C-P is obscure).

The same method was applied to confidence intervals for a Poisson mean
by Garwood in his 1934 thesis, and published in 1936~\cite{garwood36}. (See
Ref.~\cite{cousinsajp1995}, and references therein.)  They are also
the standard in HEP when there is no background.  There is controversy
when there is background; see PDG RPP (Section 39.4.2.4).

Many tables and online calculators for C-P intervals and Garwood
intervals exist; I usually use Ref~\cite{confint}.

The use of the word ``exact'' (dating to Fisher) for intervals such as
C-P refers to the {\it construction} above. But the discreteness of
observed $x$ ($\non$ in this case) causes the frequentist coverage
(defined and discussed Section~\ref{sec-coverage} below) not to be
exact, but rather generally greater than the chosen \cl\ C-P intervals
are thus criticized by some as ``wastefully conservative''.  For a
comprehensive review of both central and non-central confidence
intervals for a binomial parameter and for the ratio of Poisson means,
see Cousins, Hymes, and Tucker~\cite{cht2010}. Many choices, including
C-P, are implemented in ROOT~\cite{tefficiency}.

For the closely related construction of upper/lower limits and central
confidence intervals for a {\em Poisson mean}, see
Ref.~\cite{cousinsajp1995}.

\subsubsection{Gaussian approximation for binomial confidence intervals: 
Wilson score interval of 1927}
\label{wilson}
As mentioned above, $\non$ has mean and rms deviation given by
Eqns.~\ref{meannon} and~\ref{rmsnon}, and for asymptotically large
$\ntot$, $\bi(\non|\ntot,\binp)$ converges to a Gaussian with mean
$\mu(\binp) = \ntot \binp$ and rms  $\sigma(\binp) =
\sqrt{\ntot\binp(1-\binp)}$.

We can thus compute {\it approximate} confidence intervals for $\rho$
while invoking this Gaussian approximation to the binomial
distribution.

The idea is {\it not} to substitute $\hat\binp$ for $\binp$ in
Eqns.~\ref{meannon} and~\ref{rmsnon} (potentially big mistake!), but
rather to follow the logic already used above from E.B. Wilson in
1927~\cite{wilson1927}. (This is in fact the example that he was
illustrating in that paper!) That is, we use the above recipe for
upper and lower limits:

\begin{enumerate}
\item Find $\binp_1$ such that Gauss$(x\ge3 |
  \textnormal{mean}\ \binp_1, \sigma(\binp_1) ) = 0.16$.
\item Find $\binp_2$ such that Gauss$(x\le3 |
  \textnormal{mean}\ \binp_2, \sigma(\binp_2) ) = 0.16$.
\end{enumerate}

This consistently uses the (different) values of $\sigma$ associated
with each $\binp$, {\em not} $\sigma(\hat\binp)$. It leads to a
quadratic equation that, for our example, has solution
$[\binp_1,\binp_2] = = [0.18, 0.46]$. This is the approximate 68\%
\cl\ confidence interval known as the {\em Wilson score
  interval}. (See Ref.~\cite{cht2010} and references therein.)

Although this Wilson score interval needs only the quadratic formula,
it is for some reason relatively unknown, as students are taught the
Wald interval (UGH) of the next subsection.

\subsubsection{Gaussian approximation for binomial confidence 
intervals: potentially disastrous Wald interval to avoid}

It is tempting instead to follow the so-called ``Wald reasoning''
mentioned in Section~\ref{wald}, and to substitute $\hat\rho =
\ntot\binp$ for $\binp$ in the expression for the rms deviation in
Eqn.~\ref{rmsnon}.  One obtains: $\hat\sigma =
\sqrt{\ntot\hat\binp(1-\hat\binp)}$ and it seems a simple step to the
potentially disastrous ``Wald interval'' $\hat\binp \pm \hat\sigma$,
i.e., $[\binp_1,\binp_2] = [\hat\binp - \hat\sigma, \hat\binp +
  \hat\sigma]$.

The Wald interval does not use the correct logic for frequentist
confidence!  In fact, $\hat\sigma=0$ when $\non = 0$ or $\non =
\ntot$.

Incredibly, the failure of the Wald interval when $\non = 0$ (or $\non
= \ntot$) has been used as a {\em foundational argument} in favor of
Bayesian intervals in at least four public HEP postings (one
retracted) and one published astrophysics paper!  Of all the misguided
things that have been written about statistics (sometimes by me) in
nominally scholarly writings, this is among the most uninformed that I
have seen. (Typically the authors were not informed about Bayesian
statistics either, and thought that a prior uniform in $\binp$ was
obvious, without having read the vast Bayesian literature on priors
for a binomial parameter.)

Beware! Avoid the Wald interval except for ``hallway estimates'' ---
there is no reason to use it.  And certainly do not make the silly
claim that problems with the Wald interval point to foundational
issues with confidence intervals.  The Wald interval does not use
``confidence'' reasoning, and was already obsolete in 1927 as the
Gaussian approximation to binomial confidence intervals; and by 1934
the exact intervals using confidence reasoning were published by C-P!
They have no problem with the cases ``fatal'' to Wald intervals, $\non
= 0$ or $\non = \ntot$, as can be easily verified.

\subsubsection{HEP applications of confidence intervals for 
binomial parameter}

As mentioned, the binomial model is directly relevant to efficiency
calculations in the usual case where $\non$ is not fixed by the
experiment design but rather sampled.

Using a famous math identity, the binomial model is also directly
applicable to confidence intervals for the {\em ratio of Poisson
  means}~\cite{cht2010,jamesroos}.  Thus, it applies to
statistical significance ($Z_\textnormal{Bi}$) of an excess in a
signal bin when a sideband is used to estimate the mean background.
(See Cousins, Linnemann, and Tucker~\cite{clt2008}.)  As discussed in
our paper, one can even stretch the use of $Z_\textnormal{Bi}$ (using
a ``rough correspondence'') to the problem of a signal bin when a
Gaussian estimate of mean background exists.

\subsection{Perceived problems with upper/lower limits and hence for 
central confidence intervals}
\label{centralproblems}

For decades, issues with upper limits and central confidence intervals
have been discussed in two prototype problems in HEP:

\begin{itemize}
\item Gaussian measurement resolution near a physical boundary (e.g.,
  neutrino mass-squared is positive, but the sample $x$ is negative);
\item Poisson signal mean measurement when observed number of events
  is less than the mean expected background (so that the naive
  ``background-subtracted'' mean is negative).
\end{itemize}

Many ideas have been put forward, and by 2002 the PDG RPP settled on
three, which remains the case~\cite{pdg2024}.  I have described some
of this interesting history in a ``virtual
talk''~\cite{cousinsvirtual} and in an arXiv
post~\cite{cousins2011}. (See Section~\ref{downward} in this paper.)
In this section, I describe just one of these three ideas, namely
using Neyman's confidence interval construction, going beyond the
common orderings of $x$ discussed thus far (Section~\ref{ordering}),
to that advocated by Feldman and Cousins (F-C)~\cite{feldman1998}.
(The other two ideas in the RPP are Bayesian upper limits and
\cls\ (Section~\ref{cls}.)

\subsection{Beyond upper/lower limits and {\em central} confidence 
intervals}
\label{orderFC}
Among the more general choices for ordering $x$ in $p(x| \mu)$, the
most common in HEP is that based on a {\em likelihood ratio} (LR):

For each $\mu$, order $x_0$ using the likelihood ratio
$\lhood(x_0|\mu) / \lhood(x_0| \mu_\textnormal{best-fit})$, where
$\mu_\textnormal{best-fit}$ respects the physical boundaries of $\mu$.
This was advocated in HEP by Feldman and Cousins in
1998~\cite{feldman1998}. In fact, as learned by F-C while their paper
was in proof, the dual hypothesis test (Section~\ref{duality})
appeared in the hypothesis test section of the frequentist classic by
``Kendall and Stuart''~\cite{kendall1999} long before and since.
Unlike the orderings described above, the LR ordering is applicable in
both 1D and multi-D for $x$.

Recall from Section~\ref{metricchange} that likelihood {\em ratios} as
in F-C are independent of metric in $x$ since the Jacobians cancel, so
there is no issue there.  In contrast, an alternative that might come
to mind, namely ordering $x$ by the probability {\em density} $p(x|
\mu)$, is {\em not} recommended. A change of metric from $x$ to $y(x)$
leads to a Jacobian $|dy/dx|$ in $p(y| \mu) = p(x| \mu) / |dy/dx|$ (as
in Section~\ref{metricchange}).  So ordering by $p(y| \mu)$ is
different than ordering by $p(x| \mu)$, and so all that would follow
from ordering by a pdf would depend on the arbitrary choice of metric.

In order to implement this more general ordering by likelihood ratios,
we must go beyond finding the confidence interval endpoints $\mu_1$
and $\mu_2$ independently, and perform the construction of the whole
interval at the same time, as in the next subsection.

\subsection{Neyman's Construction of Confidence Intervals}
\label{neymanconstruction}

The general method for constructing ``confidence intervals'', and the
name, were invented by Jerzy Neyman in 1934-37.  It takes a bit of
time to sink in---given how often confidence intervals are
misinterpreted, perhaps the argument is a bit too ingenious!  In
particular, you should understand that the confidence level does not
tell you ``how confident you are that the unknown true value is in the
specific interval you report''---only a {\em subjective} Bayesian
credible interval has that property! Rather, as stated in
Section~\ref{twosentences}, the confidence interval $[\mu_1, \mu_2]$
contains those values of $\mu$ for which the observed $x_0$ is {\em
  not} ``extreme'', according to the ordering principle chosen.  In
this section, we describe this construction in more detail and more
generality.  We begin with 1D data and one parameter $\mu$.

\begin{figure}
\begin{center}
\includegraphics[width=0.49\textwidth]{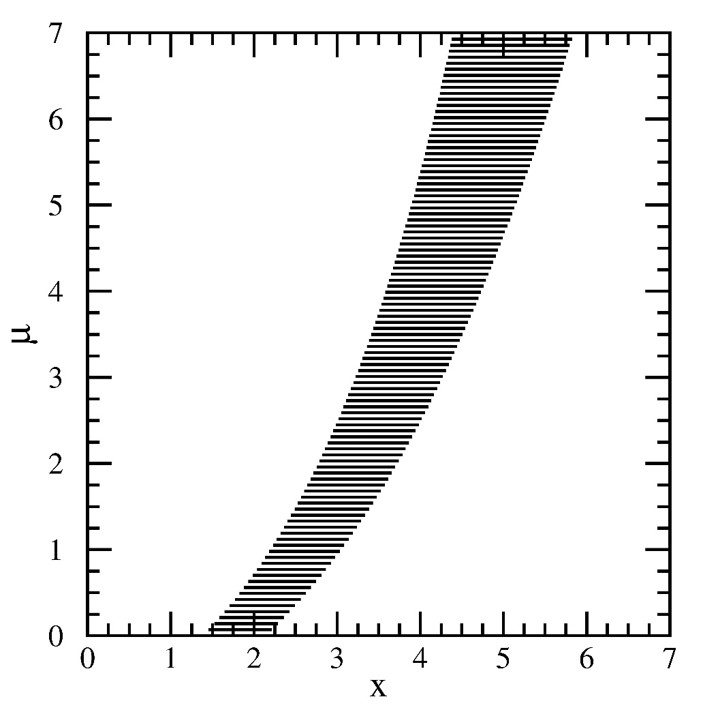}
\includegraphics[width=0.49\textwidth]{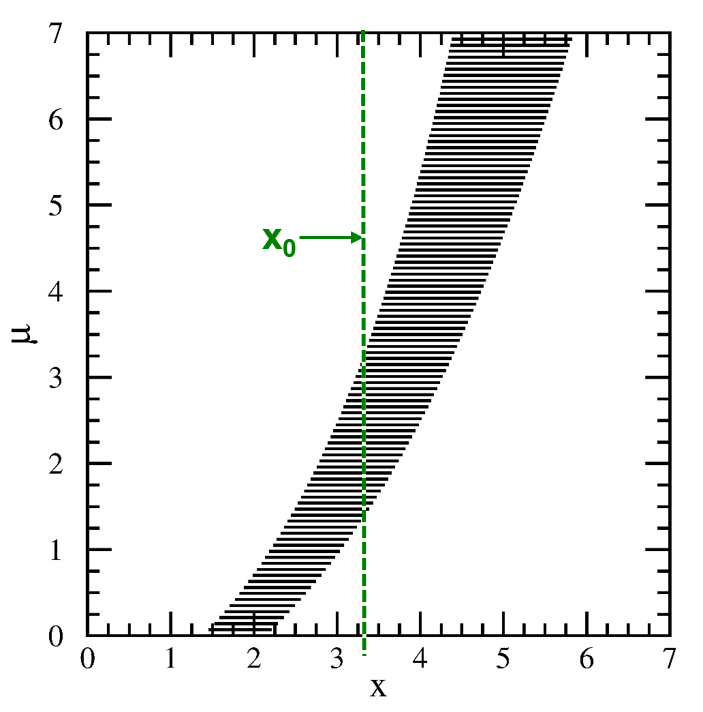}
\includegraphics[width=0.49\textwidth]{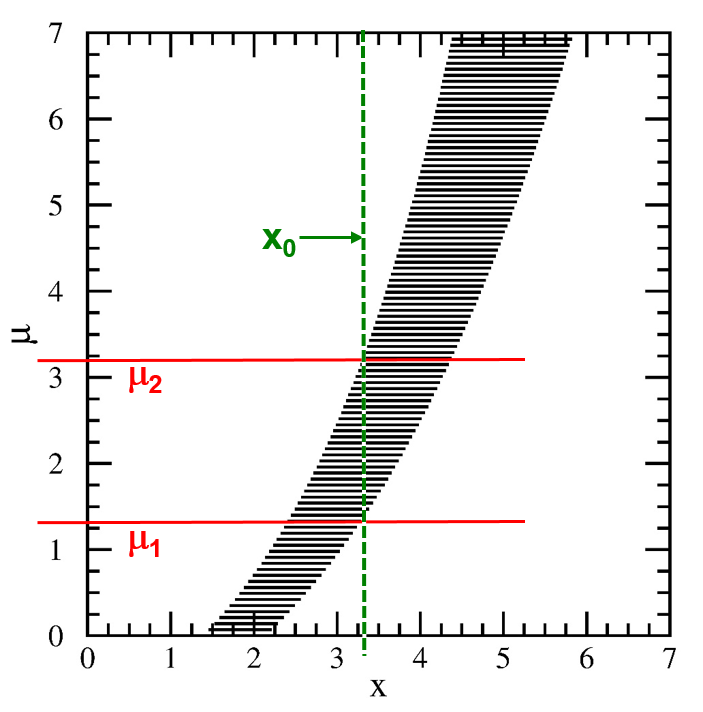}
\caption{Steps in the Neyman construction, as described in the
  text. From Ref.~\cite{feldman1998}.}
\label{neyman_const_1.png}
\end{center}
\end{figure}

Given a statistical model $p(x|\mu)$: For each value of $\mu$, one draws a
horizontal {\em acceptance interval} $[x_1,x_2]$ such that
\begin{equation}
\label{eqnacc}
p(x \in [x_1,x_2] | \mu ) = \cl = 1 - \alpha,
\end{equation} 
as in Fig.~\ref{neyman_const_1.png}(left).  The ``ordering principle''
that was chosen for $x$ is used to well-define which values of $x$ are
included.

Upon observing $x$ and obtaining the value $x_0$, one draws a vertical
line through $x_0$, as in Fig.~\ref{neyman_const_1.png}(right).  The
vertical {\em confidence interval} $[\mu_1, \mu_2]$ with confidence
level $\cl = 1 - \alpha$ is the union of all values of $\mu$ for which
the corresponding acceptance interval is intercepted by the vertical
line, as in Fig.~\ref{neyman_const_1.png}(bottom).  (It need not be
simply connected, as in highly nonlinear applications such as neutrino
oscillations.)

{\em Important note:} $x$ and $\mu$ need not have the same range,
units, or (in generalization to higher dimensions) dimensionality!

In fact, I think it is {\em much} easier to avoid confusion if $x$ and
$\mu$ are qualitatively different.  Louis Lyons gives the example
where $x$ is the flux of solar neutrinos and $\mu$ is the temperature
at the center of the sun.  I like examples where $x$ and $\mu$ have
different dimensions: Neyman's original paper~\cite{neyman1937} has a
2D observation space and 1D parameter space; his figure was crucial
for my own understanding of the construction.

\subsection[Famous confusion re Gaussian $p(x| \mu)$ where 
$\mu$ is mass $\ge0$]{\boldmath Famous confusion re Gaussian 
$p(x| \mu)$ where $\mu$ is mass $\ge0$}
\label{negativex}

A prototype problem is the Gaussian model $p(x| \mu,\sigma)$ in
Eqn.~\ref{eqn-gaussian} in the case where $\mu$ corresponds to a
quantity that is physically non-negative, e.g., a mass, a
mass-squared, or an interaction cross section.  Note that in this case
negative values of $\mu$ {\em do not exist in the model}\,!

A particle's mass-squared might be computed by measuring a particle's
energy-squared ($E^2$) with Gaussian resolution and also
(independently) its momentum-squared ($p^2$) with Gaussian resolution.
Then the observable $x$ corresponding to mass-squared could be
computed from $E^2-p^2$, with Gaussian resolution.  If the true $m^2$
is zero or small compared to the resolution, 
then such a procedure can easily obtain a value of $x$ that
is negative.  There is {\em nothing anomalous} about negative $x$.

A {\em key} point that was a source of a lot of confusion historically
in HEP, and which still trips up people, is the following: It is {\em
  crucial} to distinguish between the observed data $x$, which {\em
  can} be negative (no problem), and the model parameter $\mu$, for
which negative values {\em do not exist in the model}.  For parameter
$\mu <0$, $p(x|\mu)$ does not exist: You would not know how to
simulate the physics of your detector response to negative mass!

The constraint $\mu\ge 0$ thus has {\em nothing} to do with a Bayesian
prior pdf for $\mu$ (!!!) as sometimes mistakenly thought.  The
constraint is in the {\em model}, and hence in the likelihood
$\lhood(\mu)$, not in the prior pdf.

The confusion is encouraged since we often refer to $x$ as the
``measured value of $\mu$'', and say that $x<0$ is ``unphysical''.
This is a bad habit!  The value $x$ is an observed sample from a
Gaussian centered on $\mu$, with $\mu$ being ``physical''.

\begin{figure}
\begin{center}
\includegraphics[width=0.48\textwidth]{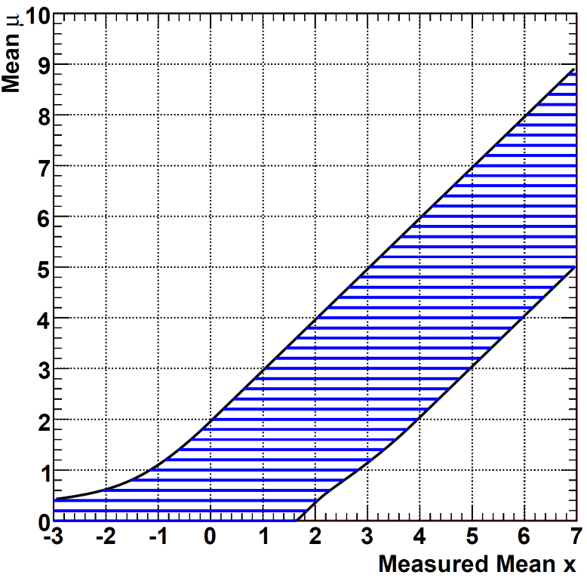}
\includegraphics[width=0.49\textwidth]{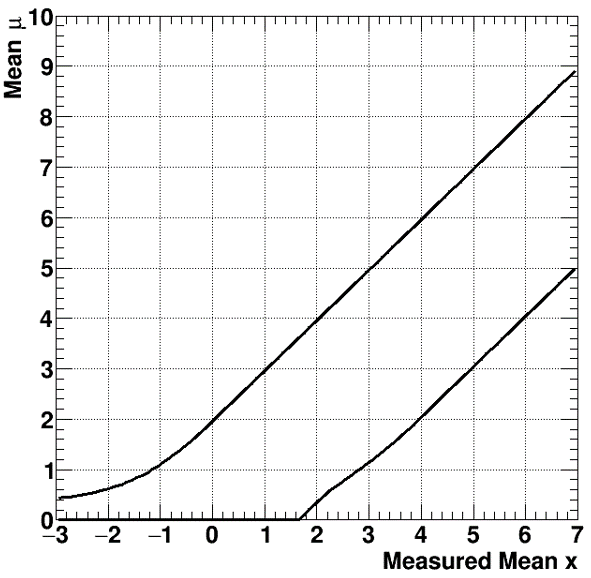}
\caption{(left) The Neyman construction for the nonnegative mean of a
  Gaussian with acceptance intervals shown in blue, as advocated in
  Ref.~\cite{feldman1998}.  (right) The ``confidence belt'' obtained
  from the construction on the left, showing the envelope of the
  acceptance intervals, which are not drawn.}
\label{neyman_const_fc.png}
\end{center}
\end{figure}

Thus, a proper Neyman construction graph for the case at hand has $x$
with {\em both arithmetic signs} but only non-negative $\mu\ge 0$.
This is the case in the plot in Fig.~\ref{neyman_const_fc.png}(left),
following that in F-C~\cite{feldman1998}.
There was a lot of confusion in the early days when the vertical axis
on this plot was extended to non-physical values of $\mu$.  Since the
model does not exist there, it is a conceptual mistake to draw the
acceptance intervals there (as some did).

For alternative methods for dealing with this situation, see
Section~\ref{downward}.

\subsection{Confidence {\em belts}}

From the earliest days (as in the 1934 Clopper-Pearson
paper~\cite{clopper1934}), the horizontal line segments of the
acceptance intervals (such as in Fig.~\ref{neyman_const_fc.png}(left))
{\em have been suppressed}, and only their envelope, as in the black
curves in Fig.~\ref{neyman_const_fc.png}(right), have been
plotted. The black curves (and interior) are called a {\em confidence
  belt}.

I added the line segments for demonstrating the construction in the
F-C paper after reading Neyman's 1937 paper~\cite{neyman1937}, in
which his Fig.~1 has a couple of planes with acceptance regions shown,
and a 1D vertical line at the observed 2D point intercepting them.
This practice in the F-C paper is fortunately spreading.

I had found other descriptions of the construction, showing only the
belt, to be too obscure when I was trying to learn the
construction. There would be statements that I found cryptic, such as
``Notice that the confidence belt is {\em constructed horizontally}
but {\em read vertically}.''  This made sense to me only {\em after} I
had understood the construction!

\subsection{Confidence intervals and coverage}
\label{sec-coverage}
Do you recall how a vector is defined in an abstract math class?  In
math, one defines a {\em vector space} as a set with certain
properties, and then the definition of a vector is ``an element of a
vector space''.  A vector is not defined in isolation!

Similarly, whether constructed in practice by Neyman's construction or
some other technique, a confidence interval is defined to be ``an
element of a confidence set'', where the confidence set is a set of
intervals defined to have the property of frequentist coverage under
repeated sampling:

Let $\mut$ be the unknown true value of $\mu$. In repeated
experiments, confidence intervals will have different endpoints $[
  \mu_1,\mu_2]$, since the endpoints are functions of the randomly
sampled $x$.

A little thought will convince you that a fraction \cl$ = 1 - \alpha$
of intervals obtained from Neyman's construction will contain
(``cover'') the fixed but unknown $\mut$. I.e.,
\begin{equation}
P(\mut \in [  \mu_1,\mu_2]) = \cl = 1 - \alpha.
\label{coverage}
\end{equation}
This equation is the definition of {\em frequentist coverage}. In this
(frequentist!) equation, $\mut$ is {\em fixed and unknown}.  The
endpoints $\mu_1, \mu_2$ are the random variables (!).  Coverage is a
property of the set of confidence intervals, not of any one interval.

[Here is the ``little thought'' that explains why Neyman's
  construction leads to coverage: For the unknown true value $\mut$,
  the probability that $x_0$ is in its {\em acceptance interval} is
  \cl, by construction.  When those $x_0$'s are obtained, the vertical
  line will intercept $\mut$'s acceptance region, and so $\mut$ be
  will be put into the confidence interval. Thus the coverage equation
  is satisfied.]

One of the complaints about confidence intervals is that the consumer
often forgets (if he or she ever knew) that the random variables in
Eqn.~\ref{coverage} are $\mu_1$ and $\mu_2$, and not $\mut$; and that
coverage is a property of the set, not of an individual interval!
Please don't forget! A lot of confusion might have been avoided if
Neyman had chosen the names ``{\em coverage intervals}'' and ``{\em
  coverage level}\,''!  (Maybe we can have a summit meeting treaty 
where frequentists stop saying ``confidence'' and Bayesians stop saying 
``noninformative''!)

Note: It {\em is} true (in precisely the sense defined by the ordering
principle used for $x$ in the Neyman construction) that the confidence
interval consists of those values of $\mu$ for which the observed
$x_0$ is among the \cl\ least extreme values to be observed.

\subsubsection[Over-coverage when observable $x$ is 
discrete]{\boldmath Over-coverage when observable $x$ is discrete}
\label{overcov}

A problem arises in Neyman's construction when the observable $x$ (or
more generally, the test statistic) is discrete.  This was already the
case in the Clopper-Pearson paper.  When constructing an acceptance
interval, typically Eqn.~\ref{eqnacc} cannot be satisfied exactly.
The traditional convention (still typically observed in HEP) is to
include enough values of $x$ so that the equality becomes ``$\ge$''.
This means that the coverage equation includes so-called
over-coverage:
\begin{equation}
\label{overcoverage}
P(\mut \in [  \mu_1,\mu_2]) \ge \cl = 1 - \alpha.
\end{equation}
For a discussion of this issue and various opinions about it, see
Ref.~\cite{cht2010} and references therein.

\section{Frequentist (classical) hypothesis testing}
\label{hypotest}

At this point, we set aside confidence intervals for the moment and
consider from the beginning the nominally different topic of
hypothesis testing.  In fact, we will soon find that in frequentist
statistics, certain hypothesis tests will take us immediately back to
confidence intervals.  But first, we consider the more general
framework.

Frequentist hypothesis testing, often called ``classical'' hypothesis
testing, was developed by J. Neyman and E. Pearson (N-P) in unfriendly
competition with R.A. Fisher. Modern testing has a mix of ideas from
both ``schools''.  Following the N-P
approach~\cite{james2006,neymanpearson1933a}, we frame the discussion
in terms of a null hypothesis H$_0$ (e.g., the Standard Model), and an
alternative H$_1$ (e.g., some Beyond-SM model).  

As in Section~\ref{notation}, $x$ is a test statistic (function of the observed
data) and $\mu$ represents parameters.  Then the model
$p(x| \mu)$ is
different for H$_0$ and H$_1$, either because parameter $\mu$ (often
called $\theta$ by statisticians) has a different value, or because
$p(x| \mu)$ itself is a different functional form, perhaps with
additional parameters.

For the null hypothesis H$_0$, we order possible observations $x$ from
least extreme to most extreme, using an ordering principle (which can
depend on H$_1$ as well). We choose a probability $\alpha$ (smallish
number).

We then ``reject'' H$_0$ if the observed $x_0$ is in the most extreme
fraction $\alpha$ of observations $x$ (generated under H$_0$). By
construction:
\begin{description}
\item{$\alpha$} $=$ probability (with $x$ generated according to
  H$_0$) of rejecting H$_0$ when it is true, i.e., false discovery
  claim (Type I error).  It is called the {\em size} or {\em
    significance level} of the test.
\end{description}
To quantity the performance of this test if H$_1$ is true, we further
define:
\begin{description}
\item{$\beta$} $=$ probability (with $x$ generated according to H$_1$)
  of not rejecting (``accepting'') H$_0$ when it is false, i.e., not
  claiming a discovery when there is one (Type II error).  
\item The {\em power} of the test is defined as $1-\beta$.
\end{description}

Note: If the alternative H$_1$ is not specified, then $\beta$ is not
known and optimality cannot be well-defined. The test is then called a
{\em goodness-of-fit} test of H$_0$, as discussed in
Section~\ref{goodness}.

\begin{figure}
\begin{center}
\includegraphics[width=0.49\textwidth]{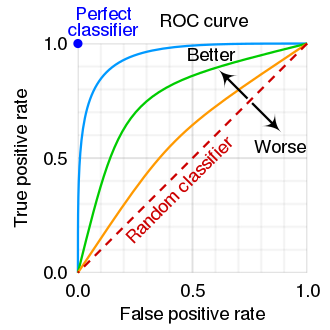}
\caption{Sketch of ROC curves for three different classifying algorithms, from Wikipedia~\cite{wikiroc}. The ``True positive rate" is another name for the power $1-\beta$, and the ``False positive rate" is another name for the Type I error probability $\alpha$.}
\label{figroc}
\end{center}
\end{figure}

There is a tradeoff between Type I and Type II errors.  Typically, internal to the data analysis, there is a threshold on a test statistic $x$ that determines the tradeoff. This tradeoff, as well as competing
analysis algorithms (resulting in different test statistics) can be
compared by looking at graphs of the power $1-\beta$ vs $\alpha$ at various $\mu$, as in
Fig.~\ref{figroc}, 
and at graphs of $1-\beta$ vs $\mu$ at various $\alpha$ (power
function). (See, e.g., Ref.~\cite{james2006}, pp.\ 258, 262.)

Testing H$_0$ vs H$_1$, including the binary classification problem of assigning items to one of two classes, is ubiquitous in science. There are several sets of words used instead of N-P’s Type I error probability $\alpha$ and Type II error probability $\beta$ and their “power” $1-\beta$.
Typically, H$_0$ is the “boring” hypothesis (background event, no disease, no signal of something new), and H$_1$ is something new (signal for new science, presence of disease, etc.) Rejecting H$_0$ is called a “positive” result of the test (even if news of a disease), while not rejecting H$_0$ is a “negative” test result. 
So test results are called true $+$, true $-$, false $+$, or false $-$.

Appendix~\ref{secjhep} illustrates the tradeoff between $\alpha$ and $\beta$
with a toy example of spin discrimination of a new resonance.
\subsection{Hypothesis testing jargon, ROC curves}
The acronym ROC (for “receiver operating characteristic”) from the early days of radar is commonly used in many fields of engineering and science. The ROC curve is a graph of power $1-\beta$ vs Type 1 error probability $\alpha$, often with new names: True positive rate vs.\ False positive rate.
This definition of ROC~\cite{wikiroc} uses “rate” as a synonym for “probability” (UGH), as follows:

False positive rate = $P$(reject H$_0\,|\,$ H$_0$ true)

= $\alpha$ = N-P Type I error probability

= $1-${\em specificity} in medical jargon, and
\medskip

True positive rate = $P$(reject H$_0\,|\,$ H$_0$ false)

= $1-\beta$ = N-P power of test

= {\em sensitivity} in medical jargon.

\medskip
Again, there is internally a threshold of a test statistic $x$ that determines the tradeoff between $\alpha$ and $1-\beta$, and hence between FPR and TPR.

The commonly used machine learning software suite scikit-learn \cite{scikit} provides ROC curves with TPR vs FPR. 

\subsubsection{Jargon in HEP}

\begin{itemize}
    \item $P$(object called signal $|$ bkgnd) $ = \alpha$  is the “mistag” probability for background. The object is “tagged” as signal even though it is background.
 \item $P$(object called signal $|$ signal) $  = 1-\beta$ is the “efficiency” for signal
   (fraction of true signal events that are tagged as signal)
 \item P(object called bkgnd $|$ signal) $  = \beta$ is the “inefficiency” for signal
\end{itemize}
As the threshold on a test statistic for tagging signal is changed, one maps out curves with axes labeled with these terms instead of $1-\beta$ vs.\ $\alpha$, or TPR vs. FPR.

Note: The assignment of H$_0$ and H$_1$ is arbitrary and can be reversed, with corresponding reversal of $\alpha$ and $\beta$. 
Typically H$_0$ is the simpler hypothesis.

Warning: In the TMVA package in ROOT, there is a nonstandard convention for “ROC curves”; the axes are labeled “background rejection” vs “efficiency”, meaning $1-\alpha$ vs $1-\beta$. Beware of the many definitions and conventions! 

\subsubsection{Jargon for medical tests}

Recall: results are true $+$, true $-$, false $+$, or true $-$.
\begin{itemize}
\item $P$(test is $+$ $|$ disease)  = $1-\beta$ is called the “sensitivity” of the test: the
   probability of correct diagnosis if actually diseased (true $+$).
\item
P(test is $-$ $|$ no disease)  = $1-\alpha$ is called the “specificity” of the test: the
   probability of correct diagnosis if no disease (true $-$).
\end{itemize}
Again, the chosen value of a threshold on a test statistic determines the tradeoff between sensitivity and specificity.

\medskip\noindent 
{\bf Question:} Suppose we know the sensitivity and specificity. Consider people with positive test results. What fraction of them have the disease?  
I.e., what is $P$(disease $|$ test is $+$) ? This is called the {\em positive predictive value} (PPV) in medicine.

\medskip\noindent {\bf Answer:} {\em Cannot be determined from the
  given information!}

\medskip\noindent 
One needs in addition: $P$(disease), the true fraction of all people that have the disease.  Then Bayes’s Theorem inverts the conditionality: 

$P$(disease $|$ test is $+$) $\propto$ $P$(test is $+$ $|$ disease) $P$(disease)


\subsection[The choice of Type I error probability 
$\alpha$]{\boldmath The choice of Type I error probability 
$\alpha$}
\label{alphachoice}
The choice of operating point on the $1-\beta$ vs $\alpha$ curve (or ROC
curve) is a long discussion.  (It is even longer when considered as the number
$N$ of events increases, so that both $\alpha$ and $\beta$ are reduced.)  The N-P
language of ``accept'' or ``reject'' H$_0$ should not be mistaken for
a complete theory of decision-making: A decision on whether or not to
declare discovery (falsifying H$_0$) requires 2 more inputs:
\begin{itemize}
\item Prior belief in H$_0$ vs H$_1$. (Can affect choice of $\alpha$)
\item Cost of Type I error (false discovery claim) vs cost of Type II
  error (missed discovery).  (Can also affect choice of $\alpha$)
\end{itemize}
A one-size-fits-all criterion of $\alpha$ corresponding to 5$\sigma$
is without foundation!  For a discussion of this point, see
Refs.~\cite{cousinsJL,lyons2013}.

Considerations such as these for the choice of $\alpha$ typically
depend on the context.  Using the result of an experiment
for a single test of a physics hypothesis is a very different context
than repeatedly selecting candidate b-jets with a b-tagging algorithm 
as in Section~\ref{btagsec}.  In the latter case, the tradeoff between 
$\beta$ vs $\alpha$ is usually determined by considerations downstream
in the analysis.

\subsection{Frequentist hypothesis testing: Simple hypotheses}
\label{simple}
In idealized cases that are sometimes reasonably well-approximated in
HEP, a hypothesis may have no floating (unfixed) parameters.  N-P
called such hypotheses {\em simple}, in contrast to {\em composite}
hypotheses that have unfixed parameters.

Examples in HEP where both H$_0$ and H$_1$ are truly simple are rare,
but we do have a few examples where the quantity of interest is simple
in both hypotheses, and the role of unfixed nuisance parameters does
not badly spoil the ``simplicity''.  For example, the hypotheses H$_0$
vs H$_1$ might be:

\begin{itemize}
\item ``jet originated from a quark'' vs ``jet originated from a gluon'', for a
  jet reconstructed in the detector
\item spin-1 vs spin-2 for a new resonance in $\mu^+\mu^-$
\item J$^{\rm P}=0^+$ vs J$^{\rm P}=0^-$ for the Higgs-like boson
\end{itemize}
A simplified, detailed illustration of the spin test is in Appendix~\ref{secjhep},
while an example of the J$^{\rm P}$ test published by CMS is in Section~\ref{higgsCP}.
Of course, framing these tests in this way makes strong assumptions
(in particular that one of the two hypotheses is true) that need to be
revisited once data are in hand.

\subsubsection{Testing Simple hypotheses: Neyman--Pearson lemma}
\label{secNP}

If the Type I error probability $\alpha$ is specified in a test of
simple hypothesis H$_0$ against simple hypothesis H$_1$, then the Type
II error probability $\beta$ is minimized by ordering $x$ according to
the likelihood ratio~\cite{neymanpearson1933a},
\begin{equation}
\label{NPLR}
\lambda = \lhood(x| {\rm H}_0) /\lhood(x| {\rm H}_1).
\end{equation} 
One finds the cutoff $\lambda_{{\rm cut,}\alpha}$ for the desired $\alpha$ and rejects
H$_0$ if $\lambda\le  \lambda_{{\rm cut,}\alpha}$.  For an outline of a proof, see
Ref.~\cite{kendall1999} (p.\ 176).  As mentioned, Appendix~\ref{secjhep}
has an example.

This ``Neyman--Pearson lemma'' applies only to a very special case: no
fitted parameters, not even undetermined parameters of interest!  However,
it has inspired many generalizations, and likelihood ratios are an
oft-used component of both frequentist and Bayesian methods.

\subsection{Nested hypothesis testing}
\label{nested}

In contrast to two disjoint simple hypotheses, it is common in HEP for
H$_0$ to be {\em nested} in H$_1$. For example, commonly H$_0$
corresponds to the parameter $\mu$ in H$_1$ being equal to a
particular value $\mu_0$. (Typical values of $\mu_0$ are 0 or 1.) So
we often consider:
\begin{description}
\item{H$_0$:} $\mu = \mu_0$ (the ``point null'', or ``sharp
  hypothesis'') vs
\item{H$_1$:} $\mu \ne \mu_0$ (the ``continuous alternative'').
\end{description}
Common examples, including a decay of the Higgs boson H$^0$, are: 
\begin{itemize}
\item Signal strength $\mu$ of new physics: null $\mu_0 = 0$,
  alternative $\mu>0$
\item H$^0$ $\rightarrow \gamma\gamma$ before discovery of this decay,
  $\mu =$ signal strength: null $\mu_0 = 0$, alternative $\mu>0$
\item H$^0$ $\rightarrow \gamma\gamma$ after discovery of this decay:
  $\mu$ is the ratio of the signal strength to the SM prediction; null
  $\mu_0 =1$ (i.e., SM prediction), alternative is any other $\mu \ne
  \mu_0$.
\end{itemize}

\subsection{Nested hypothesis testing: Duality with intervals}
\label{duality}

In the classical frequentist formalism (but not the Bayesian
formalism), the theory of these tests maps to that of confidence
intervals!  The argument is as follows.

\begin{enumerate}
\item Having observed data $x_0$, suppose the 90\% C.L. confidence
  interval for $\mu$ is $[\mu_1,\mu_2]$.  This contains all values of
  $\mu$ for which the observed $x_0$ is ranked in the {\em least}
  extreme 90\% of possible outcomes $x$ according to $p(x|\mu)$ and
  the ordering principle in use.
\item With the same data $x_0$, suppose that we wish to test H$_0$ vs
  H$_1$ (as defined in Section~\ref{nested}) at Type I error
  probability $\alpha = 10$\%.  We reject H$_0$ if $x_0$ is ranked in
  the {\em most} extreme 10\% of $x$ according to $p(x|\mu)$ and the
  ordering principle in use.
\end{enumerate}

Comparing the two procedures, we see that we reject H$_0$ at
$\alpha=10$\% if and only if $\mu_0$ is outside the 90\%
C.L. confidence interval $[\mu_1,\mu_2]$. 

(In this verbal description, I am implicitly assuming that $x$ is
continuous and that $p(x|\mu)$ is a pdf that puts zero probability on
a point $x$ with measure zero.  Thus I ignore any issues concerning
endpoints of intervals.)

We conclude: {\em Given an ordering:} a test of H$_0$ vs H$_1$ at
significance level $\alpha$ is equivalent to asking: Is $\mu_0$ outside the
confidence interval for $\mu$ with C.L. $= 1- \alpha$?

As Kendall and Stuart put it, ``There is thus no need to derive
optimum properties separately for tests and for intervals; there is a
one-to-one correspondence between the problems as in the dictionary in
Table 20.1''~\cite{kendall1999} (p.\ 175).  The table mentioned maps
the terminology that historically developed separately for intervals
and for testing, e.g.,
\begin{description}
\item  $\alpha$ $\leftrightarrow$ $1 -\cl$
\item  Most powerful $\leftrightarrow$  Uniformly most accurate
\item Equal-tailed tests $\leftrightarrow$ central confidence
  intervals
\end{description}
Use of this duality is referred to as ``inverting a test'' to obtain
confidence intervals, and vice versa.

\subsection{Feldman-Cousins}

As mentioned in Section~\ref{orderFC}, the ``new'' likelihood-ratio
(LR) ordering principle that Gary Feldman and I advocated for
confidence intervals in Ref.~\cite{feldman1998} turned out to be one
and the same as the time-honored ordering (generalized to include
nuisance parameters) described in Ref.~\cite{kendall1999} in the
chapter of {\em hypothesis testing} using likelihood ratios.  We had
scoured statistics literature for precedents of ``our'' intervals
without finding anything. But we over-looked the fact that we should
also be scouring the literature on hypothesis tests, until Gary
realized that, just in time to get a note added in proof.  In fact, it
was all on $1\frac{1}{4}$ pages of ``Kendall and Stuart'', plus
nuisance parameters! This led to rapid inclusion of the LR ordering in
the PDG RPP, since the proposal suddenly had the weight of real
statistics literature behind it.

\subsection[Post-data $p$-values and $Z$-values]{\boldmath Post-data 
$p$-values and $Z$-values}
\label{pvalues}

The above N-P theory is all a {\em pre-data} characterization of the
hypothesis test.  A deep issue is how to apply it after $x_0$ is
known, i.e., {\em post-data}.

In N-P theory, $\alpha$ is {\em specified in advance}.  Suppose after
obtaining data, you notice that with $\alpha=0.05$ previously
specified, you reject H$_0$, but with $\alpha=0.01$ previously
specified, you accept H$_0$.  In fact, you determine that with the
data set in hand, H$_0$ would be rejected for $\alpha \ge 0.023$.

This interesting value has a name: {\em After} data are obtained, the
{\em p-value} is the smallest value of $\alpha$ for which H$_0$ would
be rejected, {\em had that value been specified in advance}.

This is numerically (if not philosophically) the same as the
definition used e.g. by Fisher and often taught: ``The $p$-value is
the probability under H$_0$ of obtaining $x$ as extreme {\em or more
  extreme} than the observed $x_0$.''~\cite{cowan} (p.\ 58).  See also
Ref.~\cite{james2006} (p.\ 299) and Ref.~\cite{pdg2024} (Section
40.3.2).

In HEP, a $p$-value is typically converted to a $Z$-value
(unfortunately sometimes called ``the significance S''), which is the
equivalent number of Gaussian standard deviations.  E.g., for a
one-tailed test in a search for an {\em excess},
$p$-value${} = 2.87 \times 10^{-7}$ corresponds to $Z = 5$.  

Note that Gaussianity of the test statistic is typically {\em not}
assumed when the $p$-value is computed; this conversion to equivalent
Gaussian ``number of sigma'' is just for perceived ease of
communication.  This needs to be emphasized when communicating outside
HEP, as I hear too often statisticians wondering about Gaussian
assumptions, in effect making the conversion counter-productive (!).

Although these lectures are not ``statistics in practice'', I mention
ROOT commands for one-tailed conversions (improved version courtesy of
Igor Volobouev):
\begin{verbatim}
zvalue = -TMath::NormQuantile(pvalue)
pvalue = 0.5*TMath::Erfc(zvalue/sqrt(2.0))
\end{verbatim}

In our usual one-tailed test convention, $p$-value $>$ 0.5 corresponds
to $Z<0$.

\subsubsection[Interpreting $p$-values and 
$Z$-values]{\boldmath Interpreting $p$-values and $Z$-values}

In the example above, it is crucial to realize that that value of
$\alpha$ equal to the $p$-value (0.023 in the example) was typically
{\em not} specified in advance.  So $p$-values do {\em not} correspond
to Type I error probabilities of experiments reporting them.

The interpretation of $p$-values (and hence $Z$-values) is a long,
contentious story---beware! They are widely bashed; I discuss why in
Section~\ref{likelihoodprin}.  I also defend their use in HEP.  (See
for example my paper on the Jeffreys-Lindley
Paradox~\cite{cousinsJL}.)

Whatever they are, $p$-values are {\em not} the probability that H$_0$
is true!  This misinterpretation of $p$-values is unfortunately so
common as to be used as an argument against frequentist statistics.
Please keep in mind:
\begin{itemize}
\item That $p$-values are calculated {\em assuming that} H$_0$ {\em is
  true}, so they can hardly tell you the probability that H$_0$ is
  true!
\item That the calculation of the ``probability that H$_0$ is true''
  requires prior(s) to invert the conditional probabilities, as in
  Section~\ref{bayesbayes}.
\end{itemize}
Please help educate press officers and journalists!  (and physicists)!

\subsubsection[Early CMS Higgs boson spin-parity test of $0^+$ 
vs. $0^-$]{\boldmath Early CMS Higgs boson spin-parity test 
of $0^+$ vs. $0^-$}
\label{higgsCP}

\begin{figure}
\begin{center}
\includegraphics[width=0.6\textwidth]{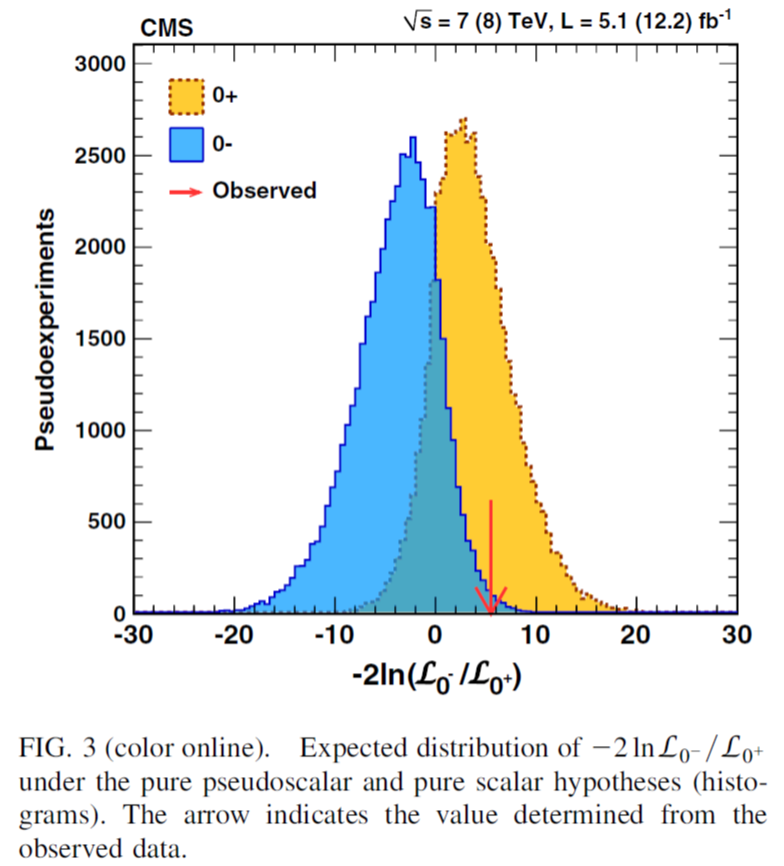}
\caption{Figure and caption from the early CMS paper on a test of
  J$^{\rm P}=0^+$ vs J$^{\rm P}=0^-$ for the Higgs-like
  boson~\cite{cmshiggsprop2012}.}
\label{cmsjp}
\end{center}
\end{figure}

As noted at the beginning of Section~\ref{simple}, the test of H$_0$:
J$^{\rm P}=0^+$ vs H$_1$: J$^{\rm P}=0^-$ for the Higgs-like boson is
a case where the parameter of interest (parity P) has two discrete
values of interest, and the role of the (many) unfixed nuisance
parameters do not badly spoil this ``simplicity'' in practice.  Thus
the test statistic used is (twice the negative log of)
the Neyman--Pearson likelihood ratio
$\lambda$ (with nuisance parameters separately optimized for each P).
An early result from CMS~\cite{cmshiggsprop2012} is shown in
Fig.~\ref{cmsjp}.  The pdf of the test statistic, obtained by
simulation, is shown for each hypothesis.

The similarity with the spin discrimination example in Appendix~\ref{secjhep}
is evident; the principles are the same, but in the Higgs boson case the
likelihood functions are more complicated and there are nuisance parameters.
As such examples are fairly rare in HEP (compared to cases of a
continuous alternative), there was a fair amount of discussion within
CMS about how best to present the results post-data.  CMS reported:

\begin{enumerate}
\item The observed value $\lambda = -2\ln(\lhood_{0^-}
  /\lhood_{0^+}) = 5.5$, favoring $0^+$;
\item For a test of H$_0$: $0^-$,  the $p$-value $= 0.0072$;
\item Reversing the role of the two hypotheses and testing H$_0$:
  $0^+$, the $p$-value $= 0.7$;
\item \cls${} = (0.0072)/(1-0.7) = 0.024$, ``a more conservative value
  for judging whether the observed data are compatible with $0^-$.''
(See Section~\ref{cls}.)
\end{enumerate}

Note that for each $p$-value calculation, the relevant tail
probability is for the tail in the direction of the other hypothesis.
The Bayes factor (Appendix~\ref{modelselection}) for this test is
similar to the observed value of $\lambda$ reported, as it differs
only in the treatment of the nuisance parameters (marginalization
rather than separate optimization).

Demortier and Lyons~\cite{lyonsp0p1} discuss the two $p$-values in
such a simple-vs-simple case, contrasting tail probabilities with the
likelihood ratio. See also Fig.~4 of Ref.~\cite{lyonswardle}.

\subsubsection{Post-data choice of the C.L.}

Section~\ref{duality} describes how for a hypothesis test dual to
intervals, the (pre-data) $\alpha$ corresponds to $1 -\cl$\ Then in
Section~\ref{pvalues}, the (post-data) $p$-value is the smallest value
of $\alpha$ for which H$_0:\mu = \mu_0$ would be rejected.  In light
of the duality, the corresponding post-data quantity for intervals is
``the largest value of the C.L. (or limiting value thereof) such that
$\mu_0$ is not in the confidence interval for $\mu$.''  This is
clearly equal to one minus the $p$-value, but (remarkably to me) it
seems not to have a standard name in the statistics literature.  Thus
I will call it the ``critical C.L.''\ A more natural way to think
about this critical value may be ``the smallest value of the C.L. such
that $\mu_0$ is {\em in} the confidence interval for $\mu$.''

For people focused more on the interval aspect of the duality, the
critical C.L. may seem a natural way to express a post-data result
when a particular value $\mu_0$ is of interest.  In fact, shortly
after the F-C paper, the CDF collaboration~\cite{cdffc} measured a
quantity called $\sin2\beta$ for which a key scientific question was
whether $\sin2\beta>0$.  They reported the critical value of
C.L. (0.93) for which 0 was (just) included in the F-C interval (so
that 0 was an endpoint of the interval).  I was in a couple of email
exchanges regarding what this 0.93 number was, and how to interpret it
(as people understood that it was not a pre-data coverage probability
for the CDF interval).  Eventually, we all agreed (I think) that it had
exactly the same status as a post-data $p$-value.  Given the LR
ordering used by F-C, a $p$-value associated with that ordering for
the test of $\sin2\beta=0$ vs $\sin2\beta>0$ was simply $1 - 0.93 =
0.07$.  Reporting the critical C.L. was dual to reporting that critical
$\alpha$.

\subsection{Classical frequentist goodness of fit (g.o.f.)}
\label{goodness}

A more extended discussion of goodness of fit is in Appendix~\ref{app-gof}.

As noted in Section~\ref{hypotest}, if H$_0$ is specified but the alternative H$_1$ is not, then only the
Type I error probability $\alpha$ can be calculated, since the Type II
error probability $\beta$ depends on H$_1$.  A test with this
feature is called a test for {\em goodness-of-fit} (to H$_0$).
(Fisher called them significance tests. I leave it to others, e.g.,
Ref.~\cite{lehmann1993}, to try to explain his seemingly pathological
opposition to explicit formulation of alternative hypotheses.)  With
no alternative specified, the question ``Which \xspace test is best?''
is thus ill-posed.  Despite the popularity of tests with universal
maps from test statistics to $\alpha$ (in particular $\chi^2$ and
Kolmogorov tests), they may be ill-suited for many problems (i.e.,
they may have poor power ($1- \beta$) against relevant alternative
H$_1$'s).

In 1D, the difficulty of the unbinned g.o.f.\ test is
exemplified by the following simple example: ``Given 3 numbers
(e.g., neutrino mixing angles) in $(0,1)$, are they consistent with
three calls to a random number generator that is uniform on $(0,1)$
?''  Have fun with that!  A plethora of possible tests in 1D are
described in the book by D'Agostino and Stephens~\cite{dagostino}
(a must-read for those wanting to invent a new test).
 
As multi-D unbinned ML fits have proliferated in recent decades, there
are increasing needs for multi-D unbinned g.o.f.\ tests.  E.g., is it
reasonable that 1000 events scattered in a 5D sample space have been
drawn from a particular pdf (which may have parameters that were fit
using an unbinned M.L. fit to those 1000 events)?  Of course, this is
an ill-posed question, but we are looking for good omnibus tests.
Then getting the null distribution of the test statistic from 
simulation (Section~\ref{toymc}) is typically doable, it seems.  One
can follow an unbinned ML fit with a binned g.o.f.\ test such as
$\chi^2$, but this brings in its own issues. At a loss of power but
increase in transparency, one can also perform tests on 1D or 2D
distributions of the marginalized densities.

Appendix B has further discussion of g.o.f., based on my informal
note~\cite{cousinsgoodness}.

\section{Likelihood (ratio) intervals for 1 parameter}
\label{likelihood}
Recall from Section~\ref{invariantl}: the likelihood function $\lhood(\mu)$ is invariant under
reparameterization from $\mu$ to $f(\mu)$: $\lhood(\mu) =
\lhood(f(\mu))$.  So {\em likelihood ratios} $\lhood(\mu_1)
/\lhood(\mu_2)$ and {\em log-likelihood differences} $\ln\lhood(\mu_1)
- \ln\lhood(\mu_2)$ are also invariant.

After using the maximum-likelihood method to obtain the estimate $\hat\mu$
that maximizes either $\lhood(\mu)$ or $\lhood(f(\mu))$, one can
obtain a likelihood interval $[ \mu_1,\mu_2]$ as the union of all
$\mu$ for which
\begin{equation}
\label{deltal}
2\ln\lhood(\hat\mu) - 2\ln\lhood(\mu) \le Z^2, 
\end{equation}
for $Z$ real.
 
As the sample size increases (under important regularity conditions) this
interval approaches a central confidence interval with
\cl\ corresponding to $\pm Z$ Gaussian standard deviations.  This is a special case of an important theorem by S.S Wilks~\cite{Wilks38}, discussed in Sections~\ref{rppregion} and~\ref{nuislikl}.
Section 9.3.2 of Ref.~\cite{james2006} has a heuristic argument
why this might work; more rigorous discussions are in the literature
on ``asymptotic'' (large sample size)
approximations. (See Ref.~\cite{igorlikl}; Chapter 6 in Ref.~\cite{cox2006};
and the discussions in Severini's monograph devoted to 
likelihood methods~\cite{severini2000likelihood}.)

But!  Regularity conditions, in particular the requirement that
$\hat\mu$ is not on the boundary (which can also cause practical problems if
it is close to a boundary), needs to be carefully checked. 
If $\mu \ge 0$ on physical grounds, then $\hat\mu = 0$ requires
care. (See, e.g., Ref.~\cite{ccgv2011} and references therein.)

\subsection[LR interval example: Gaussian pdf $p(x| \mu,\sigma)$ with 
$\sigma = 0.2\mu$]{\boldmath LR interval example: \boldmath Gaussian
  pdf $p(x| \mu,\sigma)$ with $\sigma = 0.2\mu$}

\begin{figure}
\begin{center}
\includegraphics[width=0.6\textwidth]{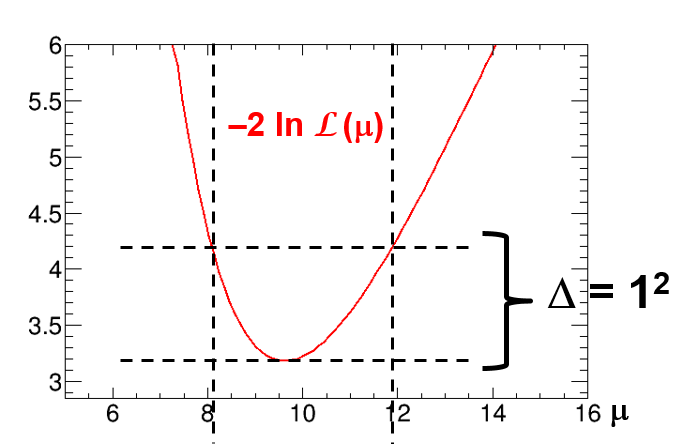}
\caption{Plot of $- 2\ln\lhood(\mu)$ corresponding to
  Fig.~\ref{fig-gaussian_likl}, the case of Gaussian pdf $p(x|
  \mu,\sigma)$ with $\sigma = 0.2\mu$.  Superimposed is the
  construction using Eqn.~\ref{deltal} with $Z=1$ to find the
  approximate 68\% C.L. likelihood ratio interval.}
\label{gaussian_like_2}
\end{center}
\end{figure}

Recall from Section~\ref{secgauss02sigma} the Gaussian pdf with
$\sigma = 0.2\mu$. The likelihood function $\lhood(\mu)$ for observed
$x_0 = 10.0$ is plotted in Fig.~\ref{fig-gaussian_likl}, with a
maximum at $\mu_{\rm ML}= 9.63$.  Fig.~\ref{gaussian_like_2} is a plot
of $- 2\ln\lhood(\mu)$, from which the likelihood ratio interval for
$\mu$ at approximate 68\% \cl\ is $[\mu_1,\mu_2] = [8.10, 11.9]$.
Compare with the exact confidence interval, [8.33,12.5].

\subsection{Binomial likelihood-ratio interval example}
\label{binomialLR}

Recall (Section~\ref{sec-binomial}) the example of $\non=3$ successes
in $\ntot=10$ trials, with $\lhood(\binp)$ and $-2\ln\lhood(\binp)$
plotted in Fig.~\ref{binomial_likl}.  The minimum value of the latter
is 2.64.  Solving for solutions to $-2\ln\lhood(\rho) = 2.64 + 1 =
3.64$, one obtains the likelihood-ratio interval $[\rho_1,\rho_2] =
[0.17, 0.45]$.  This can be compared to the Clopper-Pearson interval,
$[\rho_1,\rho_2] = [0.14, 0.51]$, and the Wilson interval,
$[\rho_1,\rho_2] = [0.18, 0.46]$.

\subsection{Poisson likelihood-ratio interval example}

\begin{figure}
\begin{center}
\includegraphics[width=0.49\textwidth]{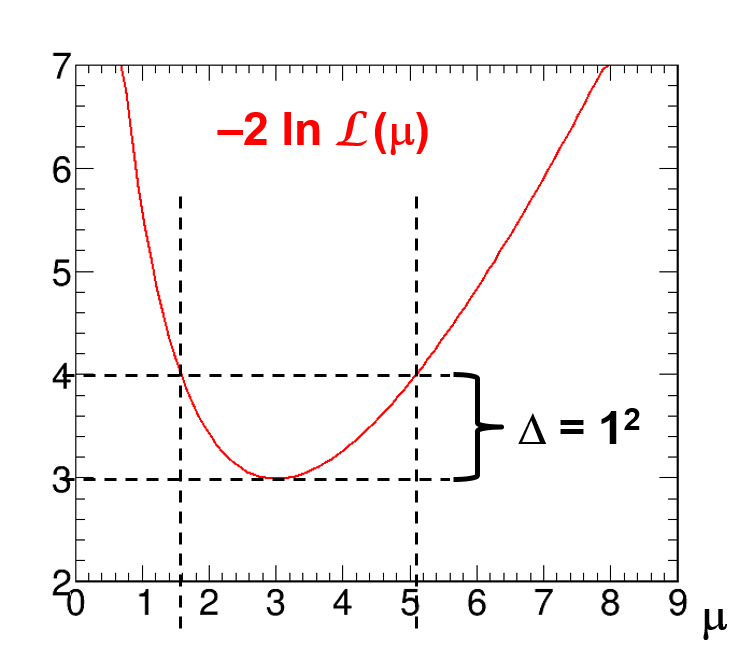}
\caption{Plot of $- 2\ln\lhood(\mu)$ corresponding to
  Fig.~\ref{pois_likli_3obs}, the case of Poisson probability in
  Eqn.~\ref{eqn-poisson} with $n=3$ observed.  Superimposed is the
  construction using Eqn.~\ref{deltal} with $Z=1$ to find the
  approximate 68\% C.L. likelihood ratio interval.  See also
Ref.~\cite{cousinsajp1995} }
\label{poisson_2negloglik}
\end{center}
\end{figure}

Recall the plot of $\lhood(\mu)$ in Fig.~\ref{pois_likli_3obs} for a
Poisson process with $n=3$ observed.  Fig.~\ref{poisson_2negloglik} is
a plot of $-2\ln\lhood(\mu)$, from which the likelihood ratio interval
at approximate 68\% C.L. can be similarly extracted, yielding $[\mu_1,
  \mu_2] = [1.58, 5.08]$.  The central confidence interval (Garwood) is $[\mu_1,
  \mu_2] = [1.37, 5.92]$.

\section{Likelihood principle}
\label{likelihoodprin}

Recall the three methods of interval construction for binomial
parameter $\binp$ upon observing $\non=3$ out of $\ntot=10$ trials:
Bayesian intervals as briefly outlined in Section~\ref{bayesintro},
confidence intervals in Section~\ref{sec-binomial}, and LR intervals
in Section~\ref{binomialLR}.  We can note that:

\begin{itemize}
\item For constructing Bayesian and likelihood intervals,
  $\bi(\non|\ntot,\binp)$ is evaluated {\em only} at the observed
  value $\non=3$.
\item For constructing confidence intervals we use, in addition, {\em
  probabilities for values of $\non$ not observed}.
\end{itemize}
This distinction turns out to be a {\em huge} deal!

In both Bayesian methods and likelihood-ratio-based methods, the
probability (density) for obtaining the {\em data at hand} is used
(via the likelihood function), {\em but probabilities for obtaining
  other data are not used!}

In contrast, in typical frequentist calculations (confidence
intervals, $p$-values), one also uses probabilities of data that could
have been observed but that were {\em not observed}.

The assertion that only the former is valid is captured by the
\begin{itemize}
\item {\em Likelihood Principle}: If two experiments yield likelihood
  functions that are proportional, then Your inferences from the two
  experiments should be identical.
\end{itemize}

There are various versions of the L.P., strong and weak forms,
etc. See Ref.~\cite{kendall1999} and the book by Berger and
Wolpert~\cite{bergerwolpert}.

The L.P. is built into Bayesian inference (except e.g., when Jeffreys
prior leads to violation).  The L.P. is violated by $p$-values and
confidence intervals.  Jeffreys~\cite{jeffreys1961} (p.\ 385) still
seems to be unsurpassed in his ironic criticism of tail probabilities,
which include probabilities of data {\em more extreme} than that
observed: ``{\em What the use of [the $p$-value] implies, therefore,
  is that a hypothesis that may be true may be rejected because it has
  not predicted observable results that have not occurred.}''

Although practical experience indicates that the L.P. may be too
restrictive, it is useful to keep in mind.  When frequentist results
``make no sense'' or ``are unphysical'', in my experience the
underlying reason can be traced to a bad violation of the L.P.

\subsection{Likelihood principle example \#1: the ``Karmen problem''}

You expect background events sampled from a Poisson distribution with
mean $b=2.8$, assumed known precisely.  For signal mean $\mu$, the
total number of events $n$ is then sampled from a Poisson distribution
with mean $\mu+b$.
So $P(n) = (\mu+b)^n \exp(- \mu-b)/n!$.

Then suppose you observe no events at all! I.e., $n=0$. (The numbers
are taken from an important neutrino experiment~\cite{karmen}.)
Plugging in,
\begin{equation}
\lhood(\mu) = (\mu+b)^0 \exp(- \mu-b)/0!  = \exp(- \mu) \exp(-b)
\end{equation}
Note that changing $b$ from 0 to 2.8 changes $\lhood(\mu)$ only by the
constant factor $\exp(-b)$.  This gets renormalized away in any
Bayesian calculation, and is irrelevant for likelihood {\em ratios}.
So for zero events observed, likelihood-based inference about signal
mean $\mu$ is {\em independent of expected $b$} when zero events are
observed. (If the prior depends on $b$, as does the Jeffreys prior
for this example in Section~\ref{jeffprior},
then there is potentially an issue. But I do not see this used in HEP.)

For essentially all frequentist {\em confidence interval}
constructions, the fact that $n=0$ is less likely for $b=2.8$ than for
$b=0$ results in {\em narrower} confidence intervals for $\mu$ as $b$
increases.  This is a clear violation of the L.P.

\subsection{Likelihood principle example \#2: binomial stopping rule}

This is a famous example among statisticians, translated to HEP. You
want to measure the efficiency $\epsilon$ of some trigger
selection. You count until reaching $\ntot=100$ zero-bias events, and
note that of these, $m=10$ passed the selection.  The probability for
$m$ is binomial with binomial parameter $\epsilon$:

\begin{equation}
\bi(m | \ntot , \epsilon) = 
\frac{\ntot!}{m! (\ntot-m)!} \epsilon^m  (1-\epsilon)^{(\ntot - m)}
\end{equation}

The point estimate is $\hat\epsilon = 10/100$, and we can compute the
binomial confidence interval (Clopper-Pearson) for $\epsilon$.  Also,
plugging in the observed data, the likelihood function is
\begin{equation}
\lhood(\epsilon) = \frac{100!}{10! 90!} \epsilon^{10} (1-\epsilon)^{90}
\end{equation}

Suppose that your colleague {\em in a different experiment} counts
zero-bias events until $m=10$ have passed her trigger selection. She
notes that this required $\ntot=100$ minimum-bias events (a
coincidence).  Intuitively, the fraction 10/100 {\em over-estimates}
her trigger's $\epsilon$ because she stopped just upon reaching 10
passed events. Indeed an unbiased estimate of $\epsilon$ and
confidence interval will be slightly different from the binomial case.

The relevant distribution here is (a version of) the {\em negative
  binomial}:
\begin{equation}
\nbi(\ntot |m , \epsilon) = 
\frac{\ntot - 1!}{(m-1)!} \epsilon^m  (1-\epsilon)^{(\ntot - m)}
\end{equation}
Plugging in the observed data, her likelihood function is
\begin{equation}
\lhood(\epsilon) = \frac{99!}{9! 90!} \epsilon^{10}  (1-\epsilon)^{90}.
\end{equation}

So both you and your friend observed 10 successes out of 100 trials,
but with different {\em stopping rules}.  Your likelihood function is
based on the {\em binomial} distribution.  Your friend's is based on
the {\em negative binomial} distribution.  The two likelihoods differ
by (only!) a constant factor, so the (strong) LP says that inferences
should be {\em identical}.  In contrast, frequentist inferences that use
probabilities of data not obtained result in different confidence
intervals and $p$-values for the different stopping rules.

Amusing sidebar: The Jeffreys prior is indeed different for the two
distributions, so the use of the Jeffreys prior violates the (strong) L.P.

\subsection{Stopping rule principle}

The two efficiency measurements  have different {\em stopping rules}:
one stops after $\ntot$ events, and the other stops after $m$
events pass the trigger.  Frequentist confidence intervals depend
on the stopping rule; the likelihood function does not, except for an
overall constant.  So Bayesians will get the same answer in both
cases, unless the {\em prior} depends on the stopping rule.

The strong L.P. implies, in this example, that the inference is
independent of the stopping rule!  This {\em irrelevance} has been
elevated to the ``Stopping Rule Principle''.  (It is sometimes amusing
to ask a recent Bayesian convert if they know that they just bought
the Stopping Rule Principle.)  Concepts that average/sum over the
sample space, such as bias and tail probabilities, do not exist in the
pure Bayesian framework.

A quote by L.J. (Jimmie) Savage (Ref.~\cite{savagestopping}, p.~76), a
prominent early subjective Bayesian advocate, is widely mentioned (as
seen in Google hits, which can point you to a copy of the original
``Savage forum'' where you can read his original note):

``\dots I learned the stopping-rule principle from Professor Barnard,
in conversation in the summer of 1952. Frankly, I then thought it a
scandal that anyone in the profession could advance an idea so
patently wrong, even as today I can scarcely believe that some people
resist an idea so patently right."

\subsection{Likelihood principle discussion}
\label{lpdisc}

We will not resolve this issue, but we should be aware of it.  There
is a lot more to the Likelihood Principle than I discuss here.  See
the book by Berger and Wolpert~\cite{bergerwolpert}, but be prepared
for the Stopping Rule Principle to set your head spinning.  When
frequentist confidence intervals from a Neyman construction badly
violate the L.P., use great caution!  And when Bayesian inferences
badly violate frequentist coverage, again use great caution!

In these lectures, I omitted the important (frequentist) concept of a
``sufficient statistic'', due to Fisher.  This is a way to describe
data reduction without loss of relevant information.  E.g., for
testing a binomial parameter, one needs only the total numbers of
successes and trials, and not the information on exactly which trials
had successes.  (See Ref.~\cite{kendall1999} for math definitions.)
The ``Sufficiency Principle'' says (paraphrasing---there are strong
and weak forms) that if the observed values of the sufficient
statistic in two experiments are the same, then they constitute
equivalent evidence for use in inference.

Birnbaum famously argued (1962) that the Conditionality Principle
(Section~\ref{condLP}) and the Sufficiency Principle imply the
Likelihood Principle. Section~\ref{condLP} has a few more comments on
this point.

Controversy continues. For a recent discussion and references, see
Deborah Mayo's 2014 detailed article~\cite{mayo2014} and references
therein, with comments by six statisticians and rejoinder.

\section{Summary of three ways to make intervals}
\label{intervalsummary}

Table~\ref{threeways} summarizes properties of the three ways
discussed to make intervals.  Only frequentist confidence intervals
(Neyman's construction or equivalent) can guarantee coverage.  Only
Bayesian and likelihood intervals are guaranteed to satisfy the
likelihood principle (except when a prior such as Jeffreys's prior
violates it).

\begin{table}
\begin{center}
\caption{Summary of three ways to make intervals.  The asterisk
  reminds us that the choice of prior might violate the likelihood
  principle.}
\bigskip
\label{threeways}
\begin{tabular}{|l|c|c|c|} \hline
& Bayesian  & Frequentist    & Likelihood  \\
& credible  & confidence     & ratio \\ \hline
Requires prior pdf?           & Yes  & No & No \\
Obeys Likelihood Principle?   & Yes* & No & Yes \\
Random variable(s) in $P(\mut \in [  \mu_1,\mu_2]) = \cl$
          & $\mu_t$ & $\mu_1,\mu_2$ & $\mu_1,\mu_2$ \\
(Over)Coverage guaranteed?          & No  & Yes & No \\
Provides $P$(parameter$|$data)?   & Yes & No & No \\ 
\hline
\end{tabular}
\end{center}
\end{table}  

Table~\ref{ajpintervals}, inspired by Ref.~\cite{cousinsajp1995},
gives illustrative intervals for a specific Poisson case.  It is of
great historical and practical importance in HEP that the {\em right}
endpoint of Bayesian central intervals with a uniform prior is
mathematically identical to that of the frequentist central confidence
interval (5.92 in the example of $n=3$, but true for any $n$).  By the
reasoning of Fig.~\ref{limits_central}, this identity applies to {\em
  upper limits}.  In contrast, the prior $1/\mu$ is needed to obtain
the identity of the {\em left} endpoint of confidence intervals, and hence
of {\em lower} limits (1.37 in the example, but true for any $n$).

The fact that our field is almost always concerned with upper (rather
than lower) limits on Poisson means is responsible for the nearly
ubiquitous use of the uniform prior (instead of the Jeffreys prior,
for example); as noted in Section~\ref{fivefaces}, the main use of the
Bayesian machinery in HEP is as a technical device for generating
frequentist inference, i.e., intervals that are (at least
approximately) confidence intervals.

When a known (fixed) background is added to the problem, the Bayesian
intervals with uniform prior for $\mu$ become conservative
(over-cover), a feature that many are willing to accept in order to
obey the likelihood principle (or perhaps for less explicitly stated
reasons).  (Even without background, frequentist intervals over-cover
due to discreteness of $n$, as discussed in Section~\ref{overcov}.)

\begin{table}
\begin{center}
\caption{68 \%C.L. confidence intervals for the mean of a Poisson
  distribution, based on the single observation $n=3$, calculated by
  various methods. }
\bigskip
\label{ajpintervals}
\begin{tabular}{|l|c|c|c|c|} \hline
Method & Prior   & Interval &Length &Coverage? \\ \hline
Wald, $n \pm \sqrt{n}$  & --- &  (1.27, 4.73) & 3.36 &  no \\
Garwood, Frequentist central        & --- &  (1.37, 5.92) & 4.55 &  yes\\
Bayesian central         & 1   &  (2.09, 5.92) & 3.83 &  no \\
Bayesian central     & $1/\mu$ &  (1.37, 4.64) & 3.27 &  no \\
Bayesian central Jeffreys     & $1/\sqrt{\mu}$ &  (1.72, 5.27) & 3.55 & no \\
Likelihood ratio         & --- &  (1.58, 5.08) & 3.50 &  no \\
\hline
\end{tabular}
\end{center}
\end{table}  

\section{1D parameter space, 2D observation space}

Until now we have considered 1D parameter space and 1D observation
space.  Adding a second observation adds surprising subtleties.  As
before, $\mu$ is a 1D parameter (often called $\theta$ by
statisticians). An experiment has two observations, the set $\{x_1,
x_2\}$.  These could be:
\begin{itemize}
\item two samples from the same $p(x| \mu)$, or 
\item one sample each of two different quantities sampled from a joint
  density $p(x_1,x_2 | \mu)$.
\end{itemize}

For frequentist confidence intervals for $\mu$, we proceed with a
Neyman construction.  Prior to the experiment, for each $\mu$, one
uses an ordering principle on the sample space ($x_1,x_2$) to select
an acceptance region $\accreg(\mu)$ in the sample space ($x_1, x_2$)
such that $P((x_1,x_2) \in \accreg(\mu)) = \cl$\ (As mentioned in
Section~\ref{neymanconstruction}, this was the illustration in
Neyman's original paper.)

Upon performing the experiment and observing the values
$\{x_{1,0},x_{2,0}\}$, the confidence interval for $\mu$ at confidence
level $\cl$ is the union of all values of $\mu$ for which the
corresponding acceptance region $\accreg(\mu)$ includes the observed
data $\{x_{1,0},x_{2,0}\}$.

The problem is thus reduced to choosing an ordering of the points
$(x_1,x_2)$ in the sample space, in order to well-define
$\accreg(\mu)$, given a \cl\ This turns out to be surprisingly subtle,
exposing a further foundational issue.

\subsection{Conditioning: Restricting the sample space used \\by 
frequentists}
\label{conditioning}

We now return to the point mentioned in Section~\ref{aside} regarding
the ``whole space'' of possibilities that is considered when computing
probabilities.

In Neyman's construction in the 2D sample space ($x_1,x_2$), the
probabilities $P((x_1,x_2) \in \accreg(\mu))$ associated with each
acceptance region $\accreg(\mu)$ are {\em unconditional} probabilities
with respect to the ``whole'' sample space of all possible values of
($x_1,x_2$).  In contrast, Bayesian inference is based on a single
point in this sample space, the observed ($x_{1,0},x_{2,0}$), per the
Likelihood Principle.  There can be a middle ground in frequentist
inference, in which the probabilities $P((x_1,x_2) \in \accreg(\mu))$
are {\em conditional} probabilities conditioned on a function of
($x_1,x_2$), in effect restricting the sample space to a
``recognizable subset'' depending on the observed data.

Restricting the sample space in this way is known as {\em
  conditioning}.  Here I discuss two famous examples:

\begin{itemize}
\item A somewhat artificial example of Welch~\cite{welch1939} where
  the conditioning arises from the mathematical structure;
\item A more physical example of Cox~\cite{cox1958} where the argument
  for conditioning seems ``obvious''.
\end{itemize}

\subsubsection{Example of B.L. Welch (1939)}
\begin{figure}
\begin{center}
\includegraphics[width=0.49\textwidth]{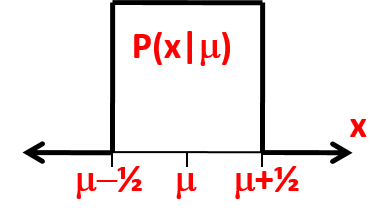}
\caption{The rectangular pdf $p(x|\mu)$ of the Welch example, centered
  on $\mu$.}
\label{welchpdf}
\end{center}
\end{figure}

In this example, $x_1$ and $x_2$ are two samples from same $p(x|
\mu)$, a rectangular pdf given by
(Fig.~\ref{welchpdf})
\begin{equation}
\label{rectangle}
p(x|\mu) = 
            \begin{cases}
             1, & \text{if~} \mu- \frac{1}{2} \le x \le \mu + \frac{1}{2}\\
             0, & \text{otherwise}.
             \end{cases}
\end{equation}
The observed data is a set of two values $\{x_1, x_2\}$ sampled from
this pdf.  From these data, the point estimate for $\mu$ is the sample mean,
$\hat\mu = \bar x = (x_1 + x_2)/2$. (Aside: if more than two samples
are observed, the point estimate is the mean of the outermost two, not the whole
sample mean; see Section~\ref{point-est}.)

\begin{figure}
\begin{center}
\includegraphics[width=0.49\textwidth]{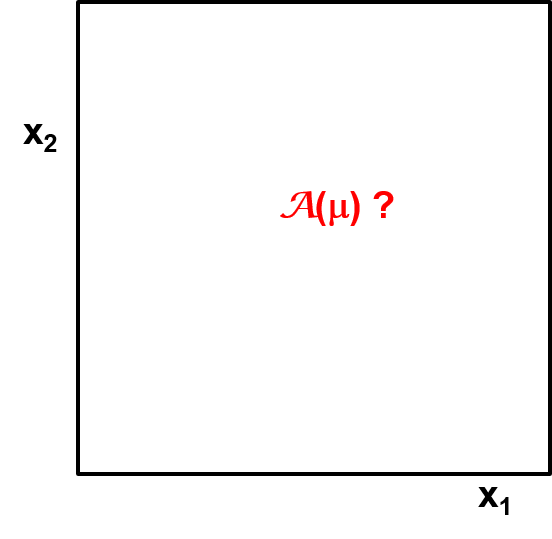}
\caption{The starting point for constructing confidence intervals in
  the Welch example: in the 2D data space centered on ($x_1=\mu, x_2=\mu$) what
  should be the acceptance region $\accreg(\mu)$ containing 68\% of
  the unit square?  One should imagine a $\mu$ axis perpendicular to the 
plane of the square, with such a square at each $\mu$.}
\label{welchA}
\end{center}
\end{figure}

What is a 68\% \cl\ central confidence interval for $\mu$?  To perform
a Neyman construction, for each $\mu$ we must define an acceptance
region $\accreg(\mu)$ containing 68\% of the unit square $(x_1, x_2)$
centered on $\mu$, as in Fig.~\ref{welchA}.  Which 68\% should one
use?  Centrality implies symmetry, but we need something else to rank
points in the plane.  The N-P Lemma suggests a likelihood ratio, but
first let's think about some examples of possible pairs $\{x_1,
x_2\}$.

\begin{figure}
\begin{center}
\includegraphics[width=0.49\textwidth]{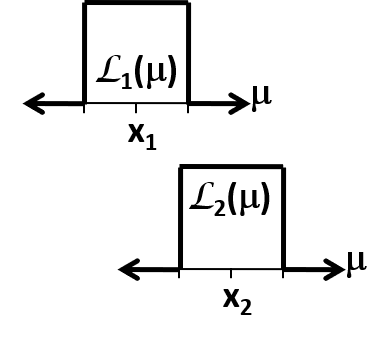}
\includegraphics[width=0.39\textwidth]{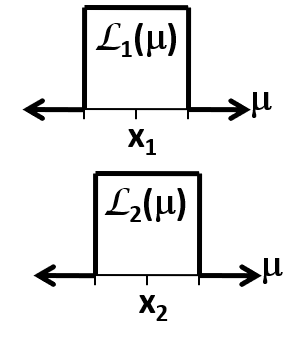}
\caption{(left) A ``lucky'' set of observations $\{x_1, x_2\}$ having
  small overlap of the likelihood functions, so that possible true
  $\mu$ is localized.  (right) an ``unlucky'' set of observations
  $\{x_1, x_2\}$, for which the second observation adds little
  additional information.}
\label{lucky}
\end{center}
\end{figure}

A ``lucky'' sample with $|x_1 - x_2|$ close to 1 is shown in
Fig.~\ref{lucky}(left): $\lhood(\mu) = \lhood_1(\mu) \times
\lhood_2(\mu)$ is very narrow.  Is it thus reasonable to expect small
uncertainty in $\hat\mu$?

An ``unlucky'' sample with $|x_1 - x_2|$ close to 0 is shown in
Fig.~\ref{lucky}(right): $\lhood(\mu)$ has full width close to 1, as
the second observation adds almost no useful information.  Should we
expect a 68\% C.L. confidence interval for $\mu$ that is the same as
for only one observation, i.e. with length 0.68?

Intuition suggests that a reasonable answer might be a confidence interval
centered on $\hat\mu$, with a length that is 68\% of the width of
$\lhood$, i.e., the interval $\hat\mu \pm 0.34(1-|x_1 - x_2|)$.

From this argument, it seems reasonable for the {\it post-data}
uncertainty to depend on $|x_1 - x_2|$, which of course cannot be
known in advance.  This quantity $|x_1 - x_2|$ is a classic example of
an {\em ancillary statistic} $A$: it has information on the {\em
  uncertainty} in the estimate of $\mu$, but no information on $\mu$
itself, because the distribution of $A$ does not depend on $\mu$. 
 An idea dating to Fisher and before is to divide the full
``unconditional'' sample space into ``recognizable subsets'' (in this
case having same or similar values of $A$), and calculate
probabilities using the ``relevant'' subset rather than the whole
space!

\begin{figure}
\begin{center}
\includegraphics[width=0.49\textwidth]{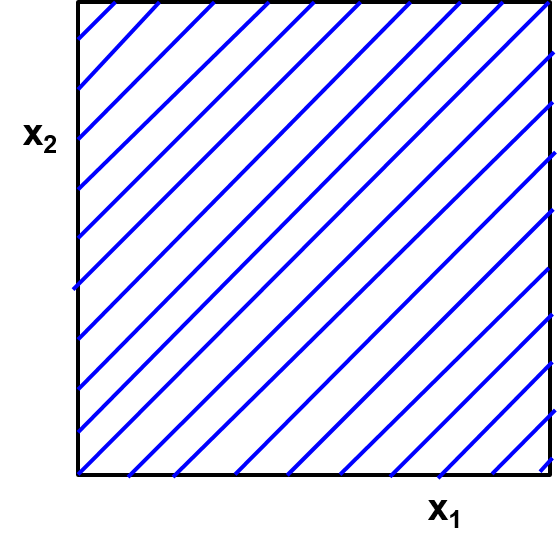}
\includegraphics[width=0.49\textwidth]{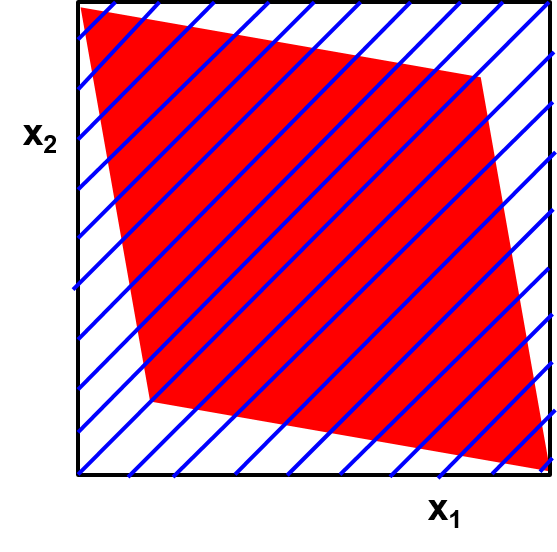}
\caption{(left) Dividing the observation space into subsets based on
  the ancillary statistic $|x_1 - x_2|$. (right), within each subset,
  acceptance regions containing the 68\% central values of $|x_1 +
  x_2|$.}
\label{neymanA}
\end{center}
\end{figure}

The (representative) diagonal lines in Fig.~\ref{neymanA}(left) show a
partition of the full sample space via the ancillary statistic $ A=
|x_1 - x_2|$.  Within each partition, in Fig.~\ref{neymanA}(right),
the shading shows a central 68\% probability acceptance region (red
fill).  We are thus using {\em conditional probabilities} (still
frequentist!)  $p(x|A,\mu)$ in this Neyman construction, with desired
probability 68\% within each partition.  (Aside: A set of measure zero
has zero probability even if non-zero pdf, so in general, care is
needed in conditioning on an exact value of continuous $A$ in
$p(x|A,\mu)$.  This is not an issue here in this example.)
The resulting $\accreg(\mu)$ for the whole square fills 68\% of the
square, so there is correct unconditional probability as well.

Imagine a plane such as this for every $\mu$, and then obtaining the
data $\{x_1, x_2\}$.  The confidence interval for $\mu$ is then the
union of all values of $\mu$ for which the observed data are in
$\accreg(\mu)$.  A moment's thought will confirm that this results in
confidence intervals centered on $(x_1 + x_2)/2$, with a length that 
is 68\% of $|x_1 - x_2|$, i.e., 
$\hat\mu \pm 0.34(1-|x_1 - x_2|)$, as intuitively
thought reasonable!

This construction is known as ``conditioning'' on the ancillary
statistic $A$.  Operationally, it can be simply stated: {\em
  post}-data, ignore the construction in the whole sample space for
values of $A$ other than that observed, and {\em proceed as if $A$ had
  been fixed in the design of the experiment, rather than randomly sampled!}

\begin{figure}
\begin{center}
\includegraphics[width=0.49\textwidth]{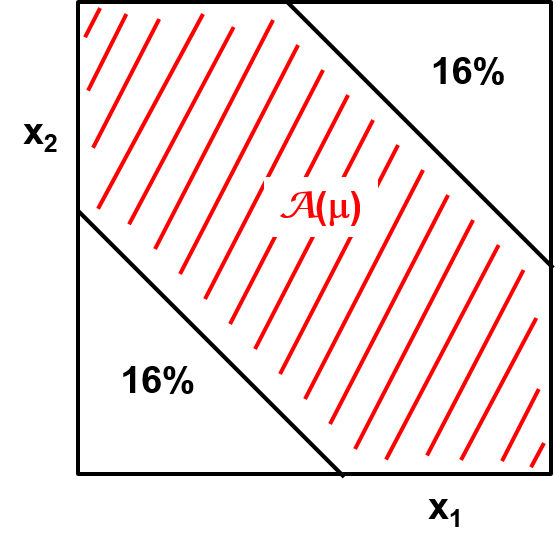}
\includegraphics[width=0.49\textwidth]{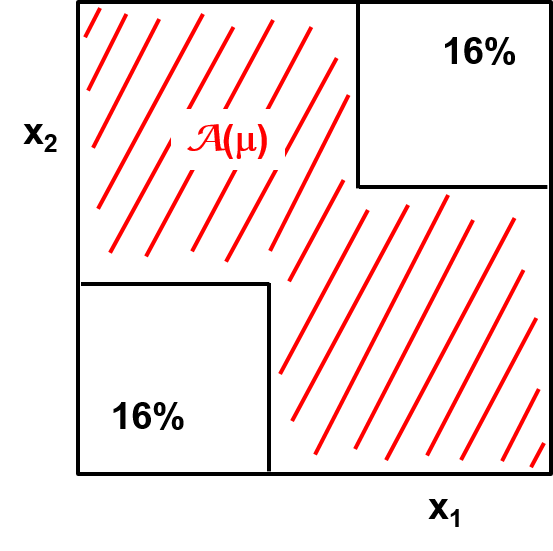}
\caption{Alternative acceptance regions having more power than that
  in Fig.~\ref{neymanA}(right).}
\label{welchpower}
\end{center}
\end{figure}

Now the catch: one can find acceptance regions $\accreg(\mu)$ that
correspond to hypothesis tests with {\em more power} (lower Type 2
error probability $\beta$) in the unconditional sample space!  A
couple of examples from the literature are shown in
Fig.~\ref{welchpower}.  The construction on the left results in 68\%
C.L. intervals with length independent of $|x_1 - x_2|$, namely
$\hat\mu \pm 0.22$ at 68\% C.L.  They obtain 68\% coverage in the
unconditional sample space by having 100\% coverage in the subspace
where $|x_1 - x_2| \approx 1$ (narrow likelihood), while badly
undercovering when $|x_1 \approx x_2|$.

Welch's 1939 paper argued {\em against} conditioning because it is
less powerful in the unconditional sample space!  Neyman's position is
not completely clear but he also seems to have been against
conditioning on ancillaries (which was Fisher's idea) when it meant an
overall loss of power~\cite{lehmann1993,berger2003}.

Most modern writers use Welch's example as an ``obvious'' argument
{\em in favor} of conditioning, unless one is in an ``industrial''
setting where the unconditional ensemble is sampled repeatedly and the
result for an individual sample is not of much interest.

\subsubsection{Efron example that is structurally similar to Welch example}

Brad Efron's talk at PhyStat-2003~\cite{efronslac} included a similar
example using the Cauchy distribution, where the ancillary statistic
is the curvature of $\lhood(\mu)$.  He called the conditional answer
``correct''.

\subsubsection{Cox two-measuring device example (1957)}

For measuring a mean $\mu$ with Gaussian resolution, one of two
devices is selected randomly with equal probability:
\begin{description}
\item{}Device \#1 with $\sigma_1$ 
\item{}Device \#2 with $\sigma_2 \ll \sigma_1$. 
\end{description}
Then a single measurement sample is made with the chosen device.

In my notation, $x_1$ is an index (1 or 2) chosen randomly and
specifying the device, and $x_2$ is the single sample from the
selected Gaussian measurement.  The total observed data is then, as
before, $\{x_1, x_2\}$.  But in this example, $x_1$ and $x_2$ are
samples of two different quantities from the joint density $p(x_1,x_2
| \mu)$.

In Ref.~\cite{behnke2013} (p.\ 112) Luc Demortier gives a nice example
in HEP: $\mu$ is the mass of a decaying particle with probability
$p_h$ to decay hadronically (mass resolution $\sigma_1$\/) and
probability $1-p_h$ to decay leptonically with mass resolution
$\sigma_2$\/).  Thus the ``measuring machine'' chosen randomly is the
detector used to measure the decay mode that is randomly chosen by
quantum mechanics.

So $\hat\mu = x_2$.  What is the confidence interval?  The index $x_1$
is an ancillary statistic, and it is reasonable (obvious?)  to
condition on it.  I.e., we report a confidence interval giving correct
coverage in the {\em subspace of measurements that used the same
  measuring device that we used}.  So the 68\% \cl\ confidence
interval is:
\begin{description}
\item{}$\hat\mu \pm \sigma_1$ if Device \#1 was randomly selected
\item{}$\hat\mu \pm \sigma_2$ if Device \#2 was randomly selected
\end{description}

Again it turns out that more powerful tests can be found.  Demortier
gives details on how the average length of intervals optimized in the
unconditional sample space is shorter in the HEP example.  Here I give
Cox's discussion.

If one is testing $\mu=0$ vs $\mu= \mu_1$, with $\mu_1$ roughly the
size of $\sigma_1$ (the larger $\sigma$\/), consider the following 
intervals: $\hat\mu \pm (0.48)\sigma_1$ if Device \#1 used (covers
true $\mu$ in 37\% of uses), and $\hat\mu \pm 5\sigma_2$ if Device \#2 used
(covers true $\mu$ nearly 100\% of uses).

Then the true $\mu$ is covered in (37/2 + 100/2)\% = 68\% of all
intervals!  The unconditional (full sample space) coverage is correct,
but conditional coverage is not.  Due to the smallness of $\sigma_2$,
the average length of all intervals {\em when averaging over the
  entire unconditional sample space} is smaller than for conditional
intervals with independent coverage.  One gives up power with Device
\#1 and uses it in Device \#2.

Cox asserts: ``If, however, our object is to say `what can we learn
from the data that we have', the unconditional test is surely no
good.''~\cite{cox1958} (p.\ 361) (See also Ref.~\cite{cox2006}
(pp.\ 47-48).)

\subsection{Conditioning in HEP}

A classic example is a {\em measurement of the branching fraction of a
  particular decay mode} when the {\em total} number of decays $N$ can
fluctuate because the experimental design is to run for a fixed length
of time.  Then $N$ is an ancillary statistic.  You perform an
experiment and obtain $N$ total decays, and then do a toy MC
simulation (Section~\ref{toymc}) of repetitions of the experiment. Do
you let $N$ fluctuate, or do you fix it to the value observed?  It may
seem that the toy MC should include your complete procedure, including
fluctuations in $N$.

However, the above arguments would point toward {\em conditioning on the
  value of the ancillary statistic actually obtained}. So your
branching fraction measurement is binomial with trials $N$.  This was
originally discussed in HEP by F. James and M. Roos~\cite{jamesroos}.
For a more complete discussion, see Ref.~\cite{cht2010}.

\subsection{Conditioning and the likelihood principle}
\label{condLP}
In summary, conditioning on an ancillary statistic $A$ means: Even
though $A$ was randomly sampled in the experimental procedure, after
data are obtained, proceed as if $A$ had been fixed to the value
observed.  Ignore the rest of the sample space with all those other
values of $A$ that you could have obtained, but did not.

{\em The Welch and Cox (and Efron) examples reveal a real conflict
  between N-P optimization for power and conditioning to optimize
  relevance}.

The assertion that inference should be conditioned on an ancillary in
the Welch example (where it comes out of the math) is often called the
``Conditionality Principle'' (CP).  Conditioning in the Cox example (a
``mixture experiment'' where the ancillary has physical meaning about
which experiment was performed) is then called the ``Weak
Conditionality Principle'' (WCP).

But note: in sufficiently complicated cases (for example if there is
more than one ancillary statistic), the procedure is less clear.  In
many situations, ancillary statistics do not exist, and it is not at
all clear how to restrict the ``whole space'' to the relevant part for
frequentist coverage.  For a comprehensive review of ancillary
statistics and their applications, see Ref.~\cite{ghosh2010}.

The pure Bayesian answer is to collapse the whole sample space to the
data observed, and refer only to the probability of the data observed,
i.e., the likelihood principle discussed in
Section~\ref{likelihoodprin}.  This is literally the ultimate extreme
in conditioning, {\em conditioning (in the continuous case) on a point
  of measure zero!}  (You can't get any more ``relevant''.)  But the
price is giving up coverage.

When there are ``recognizable subsets'' with varying coverage,
Buehler~\cite{buehler1959} has discussed how a ``conditional
frequentist'' can win bets against an ``unconditional
frequentist''. (See Refs.~\cite{cousinsvirtual,cousins2011}.)

I emphasize conditioning not only for the practical issues, but also
to explain that there are intermediate positions between the full
unconditional frequentist philosophy and the Likelihood Principle of
Section~\ref{likelihoodprin}.  A key point is that unconditional
frequentist coverage is a {\it pre-data} assessment: the entire
confidence belt is constructed independent of where the observation
lies.  Thus a big argument is whether {\em unconditional} coverage
remains appropriate {\it post-data}, after one knows where one's
observed data lies in the sample space.  When the ``measurement
uncertainty'' depends strongly on where one's data lies, then the
arguments for conditioning seem strong.  Whether or not one takes
conditioning to the extreme and considers only the (measure-zero)
subset of the sample space corresponding to the data observed is the
issue of the Likelihood Principle~\cite{bergerwolpert}.

It is not surprising that pure Bayesians argue for the importance of
{\em relevance} of the inference, and criticize frequentists for the
danger of irrelevance (and the difficulty of diagnostic of
irrelevance).  And it is not surprising that pure frequentists argue
for the importance of a useful measure of ``error rates'', in the
sense of Type 1 and Type 2 errors, coverage, etc., which may at best
be estimates if the L.P. is observed.

\section{2D parameter space, multi-D observation space}

We (finally) generalize to two parameters $\mu_1$ and $\mu_2$, with
both true values unknown.  (I hope that from context, there is no
confusion from using the subscripts 1 and 2 to indicate different
parameters, whereas in the 1D case above, they indicate the endpoints
of a confidence interval on the single parameter $\mu$.)  Let data $x$
be a multi-D vector, so the model is $p(x| \mu_1, \mu_2)$. The
observed vector value is $x_0$.

First consider the desire to obtain a 2D confidence/credible {\em
  region} in the parameter space $(\mu_1, \mu_2)$.  All three methods
discussed for intervals handle this in a straightforward (in
principle) generalization.  We mention the first two briefly and
devote a subsection to the third:

{\em Bayesian:} Put the observed data vector $x_0$ into $p(x| \mu_1,
\mu_2$) to obtain the likelihood function $\lhood(\mu_1, \mu_2$).
Multiply by the prior pdf $p(\mu_1, \mu_2$) to obtain the 2D posterior
pdf $p(\mu_1, \mu_2|x_0)$.  Use the posterior pdf to obtain credible
regions, etc., in $(\mu_1, \mu_2)$.

{\em Confidence intervals:} Perform a Neyman construction: Find
acceptance {\em regions} $\accreg(\mu_1, \mu_2)$ for $x$ as a function
of $(\mu_1, \mu_2)$.  The 2D confidence region is the union of all
$(\mu_1, \mu_2)$ for which $x_0$ is in $\accreg(\mu_1, \mu_2)$.

\subsection[Likelihood {\em regions} in $\ge 2D$ parameter 
space]{\boldmath Likelihood {\em regions} in $\ge 2D$ parameter space}
\label{rppregion}

Recall the method for 1D likelihood {\em intervals},
Eqn.~\ref{deltal}, noting that the {\em differences} in log-likelihoods correspond to {\em ratios} of likelihoods.
For a joint 2D likelihood {\em region}, first find
the global maximum of $\lhood(\mu_1, \mu_2$), yielding point estimates
$\hat\mu_1, \hat\mu_2$.  Then find the 2D contour bounded by
\begin{equation}
\label{wilks}
2\Delta
\ln \lhood = 2\ln\lhood(\hat\mu_1, \hat\mu_2) - 2\ln\lhood(\mu_1,
\mu_2) \le C,
\end{equation}
where $C$ comes from Wilks's Theorem~\cite{Wilks38}, tabulated in the PDG
RPP~\cite{pdg2024} (Table 40.2) for various C.L. and for various
values of $m$, the dimensionality of the confidence region.  Here we
have the case $m=2$, for which $C= 2.3$ for ${\approx}$68\% C.L.

\begin{figure}
\begin{center}
\includegraphics[width=0.49\textwidth]{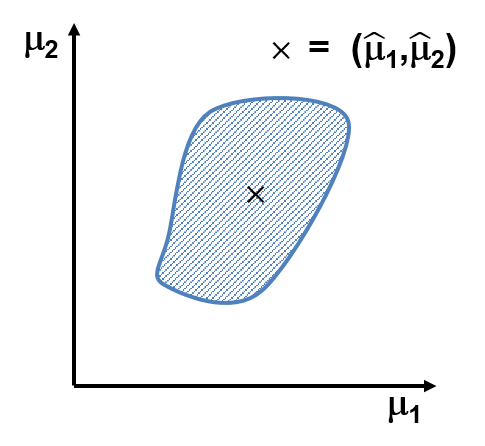}
\caption{Sketch of a 2D joint ${\approx}$68\% C.L. likelihood region for $(\mu_1,
  \mu_2)$, obtained via Eqn.~\ref{wilks}.}
\label{wilksfig}
\end{center}
\end{figure}  

This region is an approximate confidence region, as sketched in
Fig.~\ref{wilksfig}.  As in 1D, Wilks's Theorem is an asymptotic
(large $N$) result, with various ``regularity conditions'' to be
satisfied. (Again see, e.g., Ref.~\cite{ccgv2011} and references
therein.)

\subsection{Nuisance parameters}

\begin{figure}
\begin{center}
\includegraphics[width=0.49\textwidth]{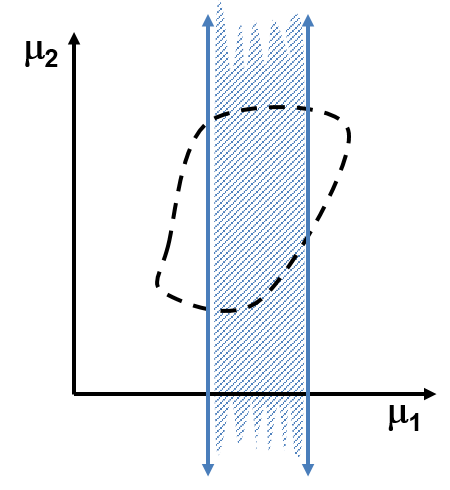}
\includegraphics[width=0.49\textwidth]{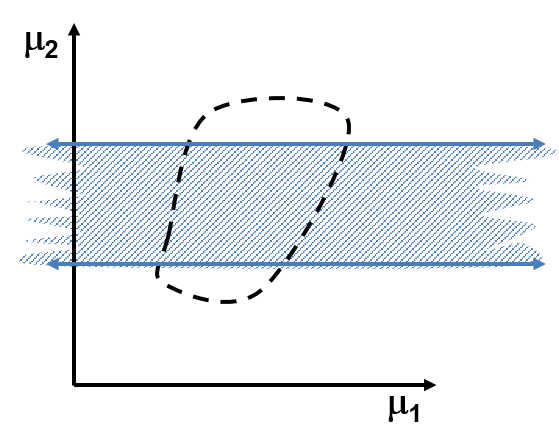}
\caption{The dashed curve is the boundary of the 2D confidence region
  when both parameters are of interest.  The shaded stripes are
  (left) 2D region to get 1D ${\approx}$68\% C.L. interval for $\mu_1$ ($\mu_2$
  is nuisance). (right) 2D region to get 1D ${\approx}$68\% C.L. interval for
  $\mu_2$ ($\mu_1$ is nuisance) }
\label{stripe}
\end{center}
\end{figure}

Frequently one is interested in considering one parameter at a time,
irrespective of the value of other parameter(s).  The parameter under
consideration at the moment is called the ``parameter of interest''
and the other parameters (at that moment) are called ``nuisance
parameters''.  E.g., if $\mu_1$ is of interest and $\mu_2$ is a
nuisance parameter, then ideally one seeks a 2D confidence region that
is a vertical ``stripe'' in the ($\mu_1, \mu_2$) plane as in
Fig.~\ref{stripe}(left); this allows the same 1D interval to be quoted
for $\mu_1$, independent of $\mu_2$.  Or, in a different moment, $\mu_2$
may be of interest and $\mu_1$ is a nuisance parameter; then one seeks
a horizontal stripe, as in Fig.~\ref{stripe}(right).  How can one
construct those stripes?

Each of the three main classes of constructing intervals (Bayesian,
Neyman confidence, likelihood ratio) has a ``native'' way to
incorporate the uncertainty on the nuisance parameters, described in
Sections~\ref{nuisbayes}--\ref{nuislikl}.  {\em But this remains a
  topic of frontier statistics research}.

\subsubsection{Systematic uncertainties as nuisance parameters}

I have begun with just one parameter of interest and one nuisance parameter.  Analyses in HEP can have hundreds or even thousands of nuisance parameters.
Systematic uncertainties provide examples of parameters that
are frequently nuisance parameters, so I mention them briefly here.  A
typical measurement in HEP has many subsidiary measurements of
quantities not of direct physics interest, but which enter into the
calculation of the physics quantity of particular interest.  E.g., if
an absolute cross section is measured, one will have uncertainty in
the integrated luminosity $L$, in the background level $b$, the
efficiency $e$ of detecting the signal, etc.  In HEP, we call these
{\em systematic uncertainties}, but statisticians (for the obvious
reason) refer to $L$, $b$, and $e$ as {\em nuisance parameters}. For
discussion of many of the issues with systematic uncertainties in HEP,
see Refs.~\cite{sinervoslac,lyonshein2007,barlowsyst}.

However, it is important to keep in mind that whether or not a
parameter is considered to be a nuisance parameter depends on context.
For example, in measurements of Higgs boson couplings, the mass of the
Higgs boson is typically regarded as a nuisance parameter.  But
clearly, the mass of the Higgs boson can itself be the primary object
of a measurement, in which case the couplings are the nuisance
parameters.

\subsection{Nuisance parameters I: Bayesian credible intervals}
\label{nuisbayes}

Construct a multi-D prior pdf $p(\textnormal{parameters})$ for the
space spanned by all parameters.  Multiply it by the likelihood function
$\lhood({\rm data}|{\rm parameters})$ for the data obtained to yield the multi-D posterior pdf.  Integrate over the full subspace
of all nuisance parameters (marginalization).  Thus obtain the 1D
posterior pdf for the parameter of interest.  Further use of the
posterior pdf is thus reduced to the case of no nuisance parameters.

{\em Problems}: The multi-D prior pdf is a problem for both subjective
and non-subjective priors.  In HEP there has been little use of the
favored non-subjective priors (reference priors of Bernardo and
Berger). The high-D integral can be a technical problem, more and more
overcome by Markov Chain Monte Carlo.

As with all Bayesian analyses, how does one interpret probability if
``default'' priors are used, so that coherent subjective probability
is not applicable?

\subsubsection{Priors for nuisance parameters}
\label{nuisprior}
It used to be (unfortunately) common practice to express, say, a 50\%
systematic uncertainty on a positive quantity as a Gaussian with 50\%
rms. Then one ``truncated'' the Gaussian by not using non-positive
values.

As mentioned in Section~\ref{pseudobayes} but worth repeating, in
Bayesian calculations, the interaction of a uniform prior for a
Poisson mean and a ``truncated Gaussian'' for systematic uncertainty
in efficiency leads to an integral that diverges if the truncation is at
origin~\cite{demortierdurham}.  In evaluating the integral
numerically, some people did not even notice!

{\em Recommendation}: Use lognormal or (certain) Gamma distributions
instead of truncated Gaussian. Recipes are in my
note~\cite{cousinslognormal}.

\subsection{Nuisance parameters II: Neyman construction}

For each point in the subspace of nuisance parameters, treat them as
fixed true values and perform a Neyman construction for multi-D
confidence regions in the full space of all parameters.  Project these
regions onto the subspace of the parameter of interest.

{\em Problems}: Typically intractable and causes overcoverage, and
therefore rarely attempted.

Tractability can sometimes be recovered by doing the construction in
the lower dimensional space of the profile likelihood function,
obtaining approximate coverage.  (This is one way to interpret the
pages in Kendall and Stuart on the likelihood ratio test with nuisance
parameters~\cite{kendall1999}. See Ref.~\cite{boblarry}) for some history.)

Typically “elimination” is done in a way technically feasible, including parametric bootstrap (Section~\ref{bootstrap}), and the coverage is studied with simulation
(Section~\ref{toymc}).

\subsection{Nuisance parameters III: Likelihood ratio intervals}
\label{nuislikl}

\begin{figure}
\begin{center}
\includegraphics[width=0.49\textwidth]{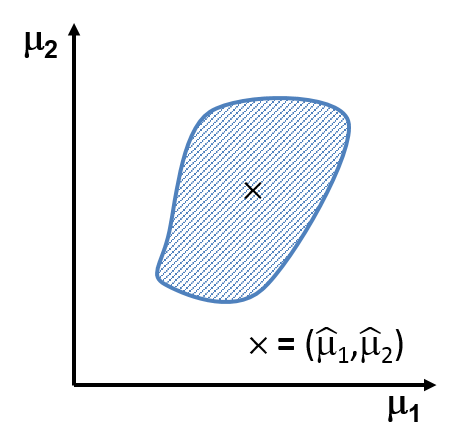}
\includegraphics[width=0.49\textwidth]{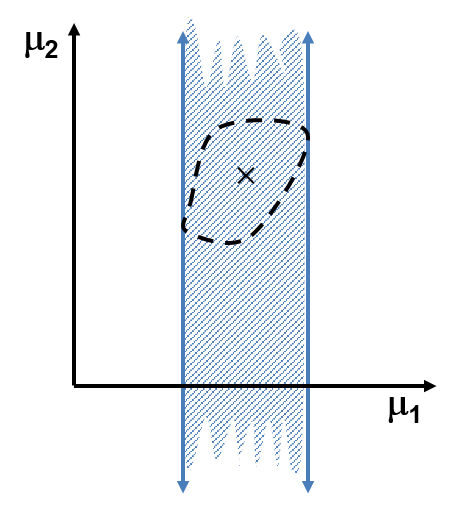}
\caption{Sketches for likelihood ratio regions. (left) The 2D confidence region from Fig.~\ref{wilksfig},
  when both parameters are of interest.  (right) On the same scale,
  the 2D confidence region if $\mu_1$ is of interest and $\mu_2$ is a
  nuisance parameter, so that effectively one obtains a confidence
  interval for $\mu_1$ that is independent of $\mu_2$.  The width of
  the stripe is smaller than the width of the extrema of the region on
  the left, since a smaller value of $C$ is used in Eqn.~\ref{wilks}:
  the dashed contour on the right is inside the solid contour on the
  left.  Both the left and the right shaded regions correspond to the
  same C.L.; the left region is relevant for joint inference on the
  pair of parameters, while the right region is relevant when only
  $\mu_1$ is of interest.}
\label{mlstripe}
\end{center}
\end{figure}

Many of us raised on MINUIT MINOS read the article by F. James,
``Interpretation of the Shape of the Likelihood Function around Its
Minimum,''~\cite{james1980}.  Whereas the 2D region in
Section~\ref{rppregion} has $m=2$ and hence $2\Delta\ln\lhood \le
2.3$, for 1D intervals on $\mu_1$, we first a make 2D {\em contour}
with the $m=1$ value, $2\Delta \ln\lhood = 1$, as shown by the black
dashed curve in Fig.~\ref{mlstripe}(right).  Then the {\em extrema} in
$\mu_1$ of this curve correspond to the endpoints of the approximate
confidence interval for $\mu_1$.

\subsubsection{{\em Profile} likelihood function}
\label{profilel}
\begin{figure}
\begin{center}
\includegraphics[width=0.49\textwidth]{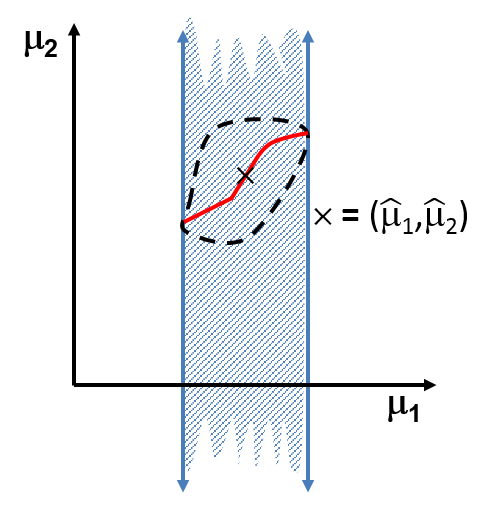}
\caption{Red curve is path $(\mu_1,\hat{\hat\mu}_2)$ along which
  profile $\lhood$ is evaluated}
\label{profile}
\end{center}
\end{figure}

At the Fermilab Confidence Limits Workshop in 2000, statistician
Wolfgang Rolke expressed the construction in a different (but {\em
  exactly equivalent}) way~\cite{rolkeclk,rolke2005}, as illustrated
in Fig.~\ref{profile} and paraphrased as follows:
\begin{itemize}
\item
For each $\mu_1$, find the value $\hat{\hat\mu}_2$ that minimizes
$-2\ln\lhood(\mu_1,\hat{\hat\mu}_2)$, shown in red.  Make a 1D plot vs $\mu_1$ of
(twice the negative log of) this ``profile likelihood function''
$-2\ln\lhood_{\rm profile}(\mu_1)$.  Use the $m=1$ threshold on
$-2\ln\lhood_{\rm profile}(\mu_1)$, i.e., Eqn.~\ref{deltal}, to obtain
intervals at the desired C.L.
\end{itemize}

The interval one obtains in Fig.~\ref{profile} is the exact {\em same}
interval as obtained by ``MINOS'' in Fig.~\ref{mlstripe}(right).  Can
you see why?  Since 2000, the ``profile'' statistical terminology has
permeated HEP. The ``hat-hat'' notation (stacked circumflex accents)
is also used, for example, by ``Kendall and Stuart''~\cite{kendall1999} in the
generalized hypothesis test that is dual to the intervals of Feldman
and Cousins.  See also Ref.~\cite{ccgv2011}.

{\em Warning:} Combining profile likelihoods from two experiments is
unreliable.  Apply profiling after combining the full 
likelihoods~\cite{lyonscombine}.

{\em Problems with profile likelihood}: 

Coverage is not guaranteed, particularly with small sample size.  By using the
best-fit value of the nuisance parameters corresponding to each value
of the parameter of interest, this has an (undeserved?) reputation for
underestimating the true uncertainties.

In Poisson problems, the profile likelihood (MINUIT MINOS) gives
surprisingly good performance in many problems. See Rolke, et
al.~\cite{rolke2005}.

In some cases (for example when there are spikes in $\lhood$),
marginalization may give better frequentist performance, I have heard.
For small sample sizes, there is no theorem to tell us whether
profiling or marginalization of nuisance parameters will give better
frequentist coverage for the parameter of interest.

\subsubsection{Aside: Profile likelihood ratio (PLR) test statistic}

In more general notation, let $\mu$ be the parameter of interest and $\nu$ be a vector of nuisance parameters, so the profile likelihood function of $\mu$ is $\lhood(\mu,\hat{\hat\nu})$.  This is also written as $\sup_\nu \lhood(\mu,\nu)$. A useful quantity for hypothesis testing (and hence for confidence interval construction) is this profile likelihood function divided by the global maximum likelihood, thus obtaining the {\em profile likelihood ratio} test statistic, 
\begin{equation}
\Lambda(\mu) = \lhood(\mu,\hat{\hat\nu}) / \lhood(\hat\mu,\hat\nu).
\end{equation}

We frequently use $-$2$\ln\Lambda$ as a test statistic.
A Bayesian-inspired alternative to the profile likelihood function, discussed below, is the “integrated” (or “marginalized”) likelihood function (also of only $\mu$),
\begin{equation}
\int \lhood(\mu,\nu) \pi(\nu)  d\nu,
\end{equation}
where $\pi(\nu)$ is a weight function in the spirit of a Bayesian prior pdf. (More generally, it could be $\pi(\nu|\mu)$.) 

\subsection{{\em Not} eliminating a nuisance parameter: the ``raster scan''}
\label{raster}

In the 2D example of Figure~\ref{mlstripe}, the right side is an
attempt to ``eliminate'' the nuisance parameter $\mu_2$ by obtaining
an interval estimate for $\mu_1$ that is independent of $\mu_2$.
However, it may be that it is preferable simply to quote a 1D interval
estimate for $\mu_1$ {\em as a function of assumed true values for
  $\mu_2$.}
In a toy neutrino oscillation example, Feldman and
Cousins~\cite{feldman1998} contrasted their unified approach with this
method, which they referred to as a ``raster scan'', in analogy with
the way old televisions drew lines across the screen.  

As an example, I am reminded that before the top quark mass was known,
other measurements (and theoretical predictions) were typically given
as a function of the unknown top quark mass, with no attempt to
``eliminate'' it (for example by putting a Bayesian prior pdf on it
and integrating it out).  In the search for the Higgs boson and
ultimately the discovery, its unknown mass was a nuisance parameter
that was also treated by raster scan: all the plots of upper limits on
cross section, as well as $p$-values testing the background-only
hypothesis, are given as a function of mass.  

When to use a raster scan is a matter of judgment; for some useful
considerations and detailed explanations, see Ref.~\cite{lyonsraster}.

\section{Evaluation of coverage with toy MC simulation}
\label{toymc}

For a single parameter of interest $\mu$, one typically reports the
confidence interval $[\mu_1,\mu_2]$ at some \cl\ after elimination of
nuisance parameters by some approximate method and construction of
intervals, perhaps involving more approximations. It is important to
check that the approximations in the whole procedure have not
materially altered the claimed coverage, defined in
Eqn.~\ref{coverage}.  Typically the performance is evaluated with a
simplified MC simulation, referred to as ``toy Monte Carlo
simulation''.  First I describe the most thorough evaluation (very CPU
intensive), and then some approximations.

In frequentist statistics, the true values of all parameters are
typically {\em fixed but unknown}. A complete, rigorous check of
coverage considers a fine multi-D grid of {\em all} parameters, and
{\em for each multi-D point in the grid}, generates an ensemble of toy
MC pseudo-experiments, runs the full analysis procedure, and finds the
fraction of intervals covering the $\mut$ of interest that was used
for that ensemble.  I.e., one calculates $P(\mut \in [ \mu_1,\mu_2])$,
and compares to \cl

Thus a thorough check of frequentist coverage includes:
\begin{enumerate}
\item Fix all parameters (of interest and nuisance) to a single set of
  true values.  For this set,
\begin{enumerate}
\item Loop over ``pseudo-experiments''
\item For each pseudo-experiment, loop over events, generating each
  event with toy data generated from the statistical model with
  parameters set equal to the fixed set.
\item Perform the same analysis on the toy events in the
  pseudo-experiment as was done for the real data.
\item Find that fraction of the pseudo-experiments for which
  parameter(s) of interest are included in stated confidence intervals
  or regions.
\end{enumerate}
\item {\em Repeat for various other fixed sets of all parameters},
  ideally a fine grid.
\end{enumerate}
{\em But}\dots the ideal of a fine grid is usually impractical.  So
the issue is what selection of ``various other fixed sets'' is
adequate.  Obviously, one should check coverage for the set of true
values set equal to the global best-fit values.  Just as obviously,
this may not be adequate.  Some exploration is needed, particularly in
directions where the uncertainty in a parameter depends strongly on
the parameter value.  One can start by varying a few critical
parameters by one or two standard deviations, trying parameters near
boundary/ies, and seeing how stable coverage is.

A Bayesian-inspired approach is to calculate a weighted average of
coverage over a neighborhood of parameter sets for the nuisance
parameters.  This requires a choice of multi-D prior. Instead of
fixing the true values of nuisance parameters during the toy MC
simulation, one samples the true parameters from the posterior
pdf of the nuisance parameters.

\subsection{Constructing the intervals with toy MC: The parametric bootstrap}

\label{bootstrap}
This approach is very common at the LHC. It generally improves on the profile likelihood ratio intervals described above. I think that it is best explained using the nested-hypothesis-test view in the duality of hypothesis tests and confidence intervals.
So, we consider the test of a particular value of the parameter of interest $\mu$, and for that fixed value (and using the data observed), we find the best-fit values of all of the nuisance parameters.
We generate toy MC  with these {\em fixed} values, and construct the distribution of the test statistic (typically a profile likelihood ratio).  For a given C.L. $= 1-\alpha$, we see if $\mu$ is rejected or not. By trying various values of $\mu$, we construct the confidence interval for $\mu$.

\subsection{Hybrid techniques: Introduction to pragmatism}
\label{pragmatism}

Given the difficulties with all three classes of interval estimation,
especially when incorporating nuisance parameters, it can be useful to
relax foundational rigor and:
\begin{itemize}
\item Treat nuisance parameters in a Bayesian way (marginalization)
  while treating the parameter of interest in a frequentist way.
  Virgil Highland and I were early advocates of this for the
  luminosity uncertainty in upper limit
  calculation~\cite{cousinshighland1992}.  At PhyStat 2005 at Oxford,
  Kyle Cranmer revealed problems when used for the background mean in a
  5$\sigma$ discovery context~\cite{cranmeroxford2005}.  For a review
  of the background case and connection to George Box's semi-Bayesian
  ``prior predictive $p$-value'', see Cousins, Linnemann, and
  Tucker~\cite{clt2008}.
\item Or, use the Bayesian framework (even without the priors
  recommended by statisticians), but evaluate the frequentist
  performance~\cite{bergerclk}.  In effect (as in profile likelihood)
  one gets approximate coverage while respecting the L.P. In fact, the
  statistics literature has attempts to find prior pdfs that lead to
  posterior pdfs with good frequentist coverage: {\em probability
    matching priors. At lowest order in 1D, the matching prior is the Jeffreys
    prior!}~\cite{welchpeers1963}.
\end{itemize}

\subsection{A few looks at the literature on nuisance 
parameters, and studies of coverage}

In the mid-2000s, Luc Demortier and I both looked in the statistics
literature regarding nuisance parameters. I thought that my note was
fairly thorough until I read his!  Our writeups:
\begin{itemize}
\item R.D. Cousins, ``Treatment of Nuisance Parameters in High Energy
  Physics, and Possible Justifications and Improvements in the
  Statistics Literature,'' presented at the PhyStat 2005 at
  Oxford~\cite{cousinsoxford2005} with response by statistician Nancy
  Reid~\cite{reidoxford2005}.
\end{itemize}
\begin{itemize}
\item Luc Demortier, ``P Values: What They Are and How to Use
  Them,''~\cite{lucpvalue}. See also Luc's scholarly Chapter 4 on
  interval estimation in Ref.~\cite{behnke2013}.
\end{itemize}

Recently, statistician Larry Wasserman and I performed an ``informal review"   of marginalization versus profiling at a dedicated PhyStat meeting, and posted our writeups on the arxiv~\cite{boblarry}, including a discussion of the parametric bootstrap.

Numerous studies have been done for elimination of nuisance parameters
in the test statistic (typically a likelihood ratio), many concluding
that results are relatively insensitive to profiling vs
marginalization, so that the choice can be made based on CPU time.  See
for example John Conway's talk and writeup at
PhyStat-2011~\cite{conwayphystat}.  It seems that the method for treating
nuisance parameters in the toy MC generation of events may be more
important than the treatment chosen in the test statistic: with poor treatment in
the test statistic, one may lose statistical power but still calculate
coverage correctly, while poor treatment in the toy MC generation may
lead to incorrect coverage calculation.

\subsection{``State of the art'' 
for dealing with nuisance parameters}

In HEP, all three main classes of methods are used for the parameter of interest.  In addition:
\begin{itemize}
\item Both marginalization and profiling are used to treat
  nuisance parameters. At present, I think that profiling is (much) more common at the LHC. 
\item Many people have the good practice of checking coverage.
\item The parametric bootstrap is increasingly used to improve coverage.
\item Too little attention is given to priors, in my opinion.  But the uniform prior for Poisson mean is ``safe'' (from frequentist point of
  view) for {\em upper} limits (only!).
\end{itemize}

A serious analysis using any of the main methods requires coding up
the model $p(x|\mu)$. (It is needed at $x=x_0$ to obtain the
likelihood function, and at other $x$ as well for confidence
intervals.)  Doing this (once!) with the RooFit~\cite{roofit} modeling
language gives access to RooStats~\cite{roostats} techniques for all
three classes of calculations, and one can mix/match nuisance
parameter treatments.  Software such as the Combine suite of tools~\cite{combine} is available.

\section{Downward fluctuations in searches for excesses}
\label{downward}

As mentioned in Section~\ref{centralproblems} and discussed in more
detail in Section~\ref{negativex}, a key problem that has been a
driver of the development of special methods for upper limits in HEP
is the situation where there is Gaussian measurement resolution near a
physical boundary.  Specifically, in the Gaussian model $p(x|
\mu,\sigma)$ in Eqn.~\ref{eqn-gaussian}, $\mu$ may be a quantity that
is physically non-negative, e.g., a mass, a mass-squared, or an
interaction cross section.  Recall (Section~\ref{negativex}) that in
this case negative values of the parameter $\mu$ {\em do not exist in
  the model}, but that negative values of the observation $x$ {\em do
  exist} in the sample space.

\begin{figure}
\begin{center}
\includegraphics[width=0.6\textwidth]{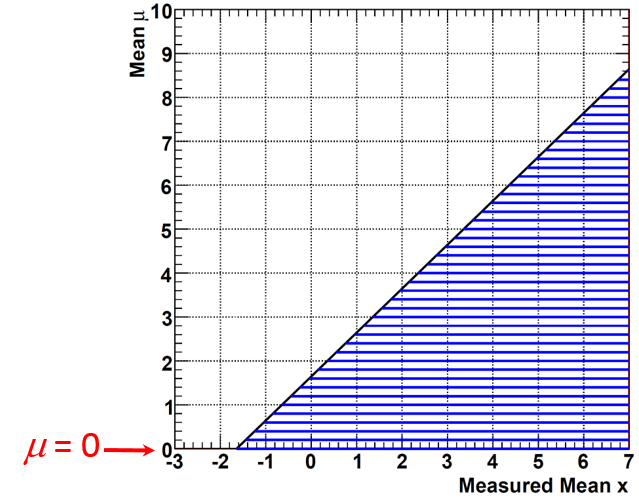}
\caption{The traditional frequentist construction for one-sided upper
  limits at 95\% C.L., for a Gaussian measurement with unit rms. 
A vertical line drawn through observed $x_0$,
  for $x_0 < -1.64 \sigma$, intersects no acceptance intervals,
  resulting in an empty-set confidence interval.}
\label{ULgauss}
\end{center}
\end{figure}

The traditional Neyman construction of frequentist one-sided 95\%
C.L. upper limits, for $\alpha = 1 - $C.L. $= 5$\%, is shown in
Fig.~\ref{ULgauss}.  As the observation $x_0$ becomes increasingly
negative, the standard frequentist upper limit (obtained by drawing a
vertical line at observed $x_0$) becomes small, and then for $x_0 <
-1.64 \sigma$, the upper limit is the {\em null} set!

Some people prefer to extend the construction to negative $\mu$,
which is a bad idea in my opinion, since the model {\em does not exist}
there (!); this leads to a different description of the issue,
so-called ``unphysical'' upper limits.

In any case, one should report enough information so that the consumer
can make interval estimates by any desired method.  This would include
the observed $x$ (not constrained to positive values) and the model $p(x|\mu)$.
Such information is also essential for combining results of different 
experiments.

This was an acute issue 20--30 years ago in experiments to measure the
$\bar\nu_e$ mass (actually the mass-squared) in tritium $\beta$ decay:
several observed $x_0 < 0$.  (With neutrino mixing, $\bar\nu_e$ is presumably not
a mass eigenstate, but one still speaks loosely of its mass, which is actually an
expectation value.)
At the time (and still today,
unfortunately), $x_0$ was referred to as the ``measured value'', or
point estimate $\hat m_\nu^2$.  The resulting confusion and
resolutions make a very long story; see my ``virtual talk'', ``Bayes,
Fisher, Neyman, Neutrino Masses, and the LHC''~\cite{cousinsvirtual}
and arXiv post~\cite{cousins2011}.  (These also contain an
introduction to ``Buehler's betting game'', related to conditioning.)

The confidence intervals proposed by F-C and described in
Section~\ref{negativex} are just one of the three proposed ways to
deal with this issue that have been widely adopted.  The other two are
Bayesian with a uniform prior (for the quantity with Gaussian
resolution), and the \cls\ criterion of the next subsection.

\subsection{\cls}
\label{cls}
The unfortunately named \cls\ is the traditional frequentist
one-tailed $p$-value for upper limits divided by another tail
probability (associated with the alternative hypothesis), i.e., by a
number less than 1. The limits are thus (intentionally) conservative.
A definition and discussion in good notation is in the PDG RPP~\cite{pdg2024}
(Eqn. 40.84, Section 40.4.2.4). \cls\ is a generalization of an
earlier result by G\"unter Zech~\cite{zech1988} that considered the
case of Poisson distribution with background and applied a
non-standard form of conditioning on an inequality. The rationale is
further described by Alex Read in the early
papers~\cite{readclw,readjphysg}; an implementation with further
commentary is described by Tom Junk~\cite{junkcls}.

What is new (non-standard statistics) in \cls\ is combining two
$p$-values into one quantity.  (This is referred to as a
``modification'' to the usual frequentist $p$-value.)  The combination
(equivalent to a non-standard form of conditioning) is designed to
avoid rejecting H$_0$ when the data are similarly incompatible with
H$_1$.  This situation can arise when an experiment has little sensitivity
for distinguishing H$_0$ from H$_1$.

There is no established foundation in the statistics
literature for this as far as I know.  In fact, the two {\em
  un}combined $p$-values are considered to be the (irreducible)
post-data ``evidence'' for simple-vs-simple hypothesis testing in a
philosophical monograph by Bill Thompson~\cite{thompson2007}
(p.\ 108). Clearly, with both $p$-values at hand, as with the Higgs
spin-parity example in Section~\ref{higgsCP}, the consumer has the
complete information regarding post-data tail probabilities.

(Ofer Vitells has unearthed a suggestion in a 1961 paper by
Birnbaum~\cite{birnbaum1961} (p.\ 434) that combines the pre-data Type
I and Type II errors rates $\alpha$ and $\beta$ into one quantity in
the identical manner, i.e., $\alpha/(1-\beta)$; Birnbaum's motivation
was that this quantity equals the likelihood ratio in a very special
case where the test statistic is the single binary digit ``reject
H$_0$'' or ``do not reject H$_0$'' (rather than the full likelihood
ratio).  But this seems to be obscure to modern statisticians, and
Birnbaum's last paper~\cite{birnbaum1977} examined both $\alpha$ and
$1-\beta$ (the usual ``power''), not just the combination
$\alpha/(1-\beta)$, in defining his ``confidence concept''.)

In any case, the \cls\ quantity does have properties that many find to
be attractive. In particular, for the two simple prototype problems
(Poisson with known mean background, and the bounded Gaussian
problem), the results are numerically identical to the Bayesian
answers with uniform prior, and hence the likelihood principle is respected. 
These Bayesian interval estimates over-cover from a frequentist point,
which is not considered to be as bad as under-coverage.
(Regarding the uniform priors, for both these models 
negative $\mu$ {\em does not exist} in the model, so it is incorrect to 
speak of the prior being zero for negative $\mu$, 
as in Section~\ref{negativex}) 

The step of combining the two $p$-values into one quantity is called
``the \cls\ criterion'' in (most) CMS papers.  Unfortunately, the
calculation of the $p$-values themselves, typically using a
likelihood-ratio test statistic, is sometimes called the
``\cls\ method'' in HEP.  But these $p$-values themselves have long
existed in the statistics literature and should be designated that way
(and not denoted with equally unfortunate names that the original
papers of \cls\ use for them).

The many issues of $p$-values are of course inherited by \cls, namely:

\begin{itemize}
\item What specific likelihood ratio is used in the test statistic;
\item How nuisance parameters are treated (marginalization, profiling);
\item What ensembles are used for ``toy MC'' simulation used to get the
  distribution of the test statistic under H$_0$ (e.g. no Higgs) and H$_1$
  (e.g. SM with Higgs).
\end{itemize}
LEP, Tevatron, and LHC Higgs experimenters differed in the choices
made (!).

\section{ATLAS and CMS Conventions}

For many years, ATLAS and CMS physicists have collaborated on
statistics tools (the RooStats~\cite{roostats} software), and
attempted to have some coherence in methods, so that results could be
compared, and (when worth the effort) combined.

A key development was the paper by Cowan, Cranmer, Gross, and Vitells
(CCGV)~\cite{ccgv2011} that extends asymptotic formulas to various
cases where Wilks's theorem is not valid.

As the CCGV asymptotic formulas correspond to the ``fully
frequentist'' treatment of nuisance parameters, for consistency we
tended to use that treatment in many cases at small $N$ as well.  Toy
MC simulation is thus performed in an approximate frequentist manner
in which the underlying parametric distribution of the data is known
up to one or more parameters, and sampling is performed from that
distribution after substituting estimates for the parameter(s).  (This
is known as the {\em parametric bootstrap}~\cite{luc2012slac}.)

For upper limits, there was a lot of discussion among CMS and ATLAS
physicists in the early LHC days without convergence, with the result
that the two experiments' physics coordinators in 2010 decreed that
\cls\ (Section~\ref{cls}) be used in most cases.

The ATLAS/CMS Higgs boson results followed these trends. A jointly
written description is, ``Procedure for the LHC Higgs boson search
combination in Summer 2011''~\cite{LHC-HCG}.

Many issues were further discussed and described in the ATLAS-CMS
combination papers for mass~\cite{atlascmsmass} and
couplings~\cite{atlascmscouplings}.  In particular, a lot of attention was paid to correlations.  

In the last few years,
Feldman-Cousins~\cite{feldman1998} starts to be used in some analyses,
without my pushing.  (Initially, some at the LHC were very opposed,
evidently because it could return a two-sided interval not including
zero when they insisted on a strict upper limit.)

The CMS Collaboration maintains a large suite of software called ``Combine"~\cite{combine}, building on techniques developed by the CMS and ATLAS Collaborations for the Higgs boson measurements and combinations, as well as numerous other applications. These tools continue to evolve rapidly.
Individual tools currently used by ATLAS (in addition to Roofit and Roostats mentioned above) include(pointers courtesy Kyle Cranmer)
HistFactory~\cite{histfactory} for specific modelling in histogram based analyses); 
HistFitter~\cite{histfitter}, which sits on top of HistFactory and offers top-level steering functionality more like CMS's Combine);
TRexFitter, that is like HistFitter and widely used, but no paper yet; pyhf~\cite{pyhf}, a Python-based implementation of HistFactory, which also has a different format for saving results, which is being used by HEPData; and cabinetry~\cite{cabinetry}, a Python-based approach to tools like HistFitter/TRexFitter.
 
\section{My advocacy for $>$15 years}

Have in place tools to allow computation of results using a variety of
recipes, for problems up to intermediate complexity:
\begin{itemize}
\item Bayesian with analysis of sensitivity to prior
\item Profile likelihood ratio (MINUIT MINOS)
\item Frequentist construction with approximate treatment of nuisance
  parameters
\item Other ``favorites'' such as \cls\  (an HEP invention).
\end{itemize}

The community can (and should) then demand that a result shown with
one's preferred method also be shown with the other methods, {\em and
  sampling properties studied}.

When the methods all agree, we are in asymptotic Nirvana (idyllic
state).  When methods disagree, we are reminded that the results are
answers to different questions, and we learn something! E.g.:
\begin{itemize}
\item Bayesian methods can have poor frequentist properties
\item Frequentist methods can badly violate the Likelihood Principle.
\end{itemize}

In fact, the community reached the point of having the tools in place
(RooStats~\cite{roostats}) by the time of the Higgs boson discovery,
and they have continued to be improved.  What is not as far along is
the ``demands'' of the community, in my opinion.  I would prefer that
it be more common for papers to compare explicitly the chosen method
with other methods, as is sometimes done.

\section {Unsound statements you can now avoid}

\begin{itemize}

\item ``It makes no sense to talk about the probability density of a
  constant of nature.''

\item ``Frequentist confidence intervals for efficiency measurements
  don't work when all trials give successes.''

\item ``We used a uniform prior because this introduces the least
  bias.''

\item ``We used a uniform positive prior as a function of
the parameter of interest.''

\item ``A noninformative prior probability density does not contain
  any information.''

\item ``The total number of events could fluctuate in our experiment,
  so {\em obviously} our toy Monte Carlo simulation should let the
  number of events fluctuate.''

\item ``We used Delta-likelihood contours so there was no Gaussian
  approximation.''

\item ``A five-sigma departure from the SM constitutes a discovery.''

\item ``The confidence level tells you how much confidence one has
  that the true value is in the confidence interval.''

\item ``We used the tail area under the likelihood function to measure
  the significance.''

\item ``Statistics is obvious, so I prefer not to read the literature
  and just figure it out for myself!''

\end{itemize}

\appendix

\section{Details of spin discrimination example of simple-vs-simple hypothesis test}
\label{secjhep}

\newcommand\zp{\rm Z^{\prime}}
\newcommand\invmm{M_{\ell\ell}}
\newcommand\scrl{{\cal L}} 
\newcommand\costh{\cos\theta^{\ast}}
\newcommand\thcs{\theta^{\ast}_{\rm CS}}
\newcommand\thcsi{\theta^{\ast}_{{\rm CS},i}}
\newcommand\G{\rm G^*}

\newcommand\costhcs{\cos\thcs}
\newcommand\costhcsi{\cos\thcsi}
\newcommand\LR{\lambda}
\newcommand\LRr{\LR^{rec}}
\newcommand\LRcut{\LR^{\rm cut}}

\newcommand\HA{{\rm H}_{\rm A}}
\newcommand\HB{{\rm H}_{\rm B}}

Ref.~\cite{cousinsspin} describes an enlightening toy example with
simple hypotheses.  The setup is the observation of a new resonance
with a mass of 1.5 TeV in the dilepton final state at the LHC
(unfortunately optimistic thus far).  We consider here the case of
just two simple hypotheses for the spin of the new resonance, either a
spin-1 vector boson or a spin-2 graviton.  In order to avoid confusion
between subscripts and spin values, the hypotheses are called $\HA$ and
$\HB$.

\begin{figure}
\begin{center}
\includegraphics[width=0.9\textwidth]{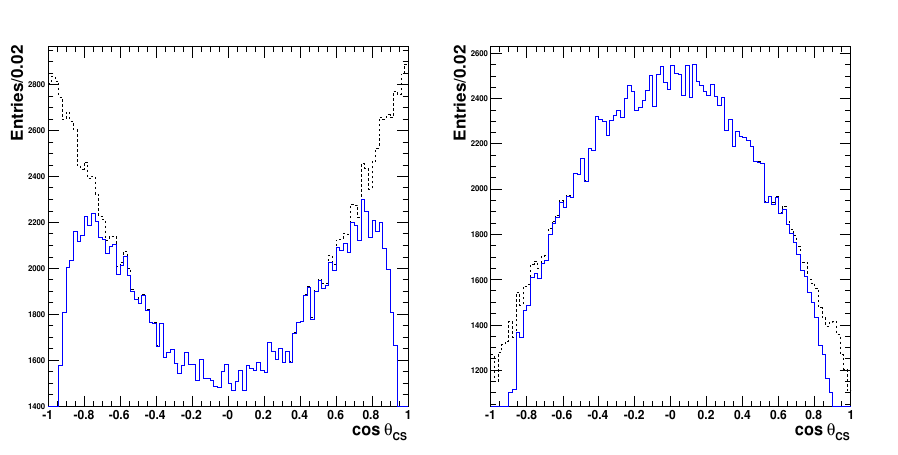}
\caption{Histograms of $\costhcs$ of individual events 
from the generated (dashed) and accepted (solid) samples of 
spin-1 $\zp$ (left) and spin-2 $\G$ (right) for 1.5 TeV mass.
From Ref.~\cite{cousinsspin}.}
\label{fig:angdists}
\end{center}
\end{figure}

The distinguishing observable is the so-called Collins-Soper angle
$\thcs$, which is a useful approximation to the angle in the CM frame
between the incoming quark and the outgoing $\mu^-$.
Figure~\ref{fig:angdists} shows the simulated distributions of
$\costhcs$ for the two spin hypotheses.  The events in the detector's
geometrical acceptance (solid histograms) are used in the analysis.
The histograms numerically define the models $p(\costhcs|\HA)$ and  
$p(\costhcs|\HB)$.

As dictated by the Neyman--Pearson lemma (Section~\ref{secNP}), 
the optimal test statistic
for distinguishing the two hypotheses is the likelihood ratio
\begin{equation}
\lambda = \lhood(\HA) / \lhood(\HB),
\end{equation} 
from which one usually considers the monotonic function $-2\ln\lambda$.
For a data set with $N$ events indexed by $i$,
$\lambda$ is the product of event likelihood
ratios, so that $-2\ln\lambda$ is the sum of the individual event quantities,
\begin{equation}
\label{sumlambda}
-2\ln\lambda =\sum_i^N -2\ln\left (\frac{p(\costhcsi|\HA)}{ p(\costhcsi|\HB)}\right).
\end{equation}

\begin{figure}
\begin{center}
\includegraphics[width=0.9\textwidth]{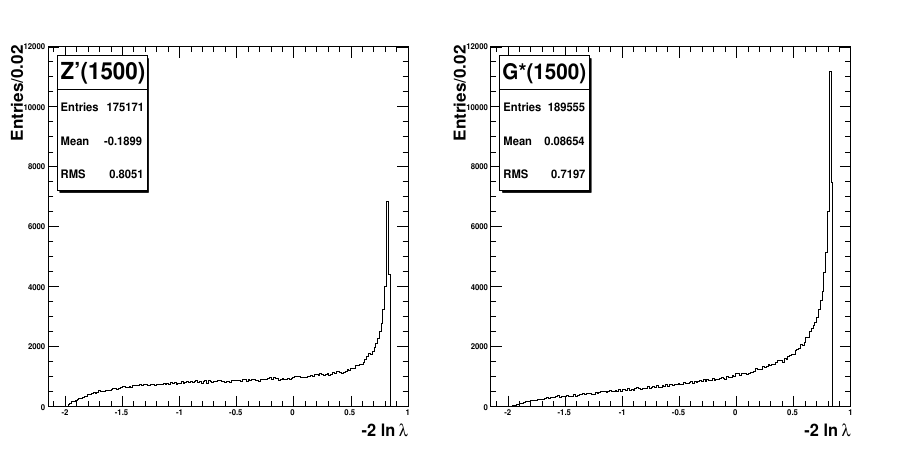}
\caption{Histograms of $-2\ln\LR = 2\ln\scrl(\HA) - 2\ln\scrl(\HB)$
for individual events from the accepted samples generated 
according to $\HA$ (spin-1) and $\HB$
(spin-2) for a 1.5 TeV resonance. From Ref.~\cite{cousinsspin}.}
\label{fig:lrdists}
\end{center}
\end{figure}

Figure~\ref{fig:lrdists}(left) and right are histograms of the
individual terms in the sum, with events on the left simulated
according to $\HA$ as in Figure~\ref{fig:angdists}(left), and events on the
right simulated according to $\HB$, as in Figure~\ref{fig:angdists}(right).

\begin{figure}
\begin{center}
\includegraphics[width=0.9\textwidth]{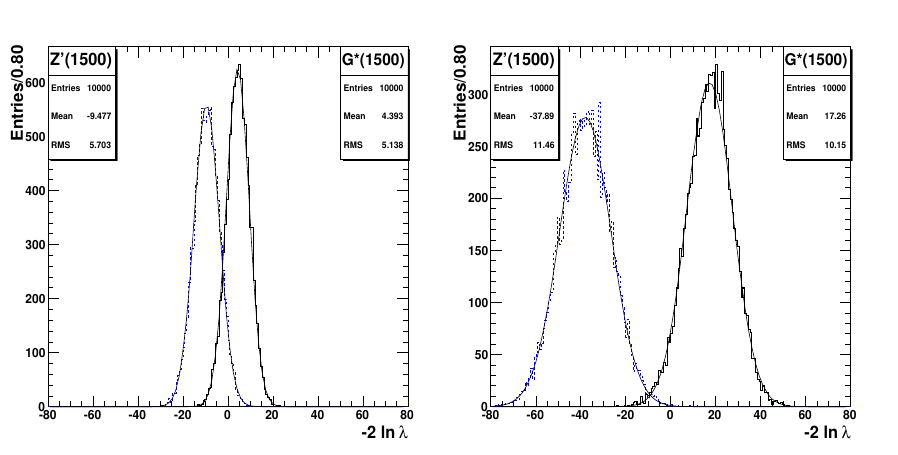}
\caption{Histograms of $-2\ln\LR$ for simulated
experiments with 1.5 TeV dilepton mass, each of which is a sum of 
$N$ values of $-2\ln\LR$ from spin-1 and spin-2 events
in Fig.~\protect \ref{fig:lrdists}, for $N=50$ (left) and $N=200$ (right) 
events per experiment.  The superimposed Gaussian curves have
means fixed at a factor of $N$ 
greater than that of the events in Fig.~\protect \ref{fig:lrdists}, and
rms deviations a factor of $\sqrt{N}$ greater. From Ref.~\cite{cousinsspin}.}
\label{fig:lrdists_n50}
\end{center}
\end{figure}

We then consider data sets (``experiments'') that each contain a
sample of $N$ events from the distributions in Figure~\ref{fig:lrdists},
with the values summed as in Eqn.~\ref{sumlambda}.
Figure~\ref{fig:lrdists_n50} shows on the left the distribution for
10,000 simulated experiments, each with $N=50$, and on the right for
$N=200$.  We see the dramatic effect of the central limit theorem,
which says that each histogram of sums tends to Gaussian despite
the highly non-Gaussian distribution of the addends from
Fig.~\ref{fig:lrdists}; and that furthermore the mean and rms of each are
correctly related to those in Fig.~\ref{fig:lrdists}.

\begin{figure}
\begin{center}
\includegraphics[width=0.57\textwidth]{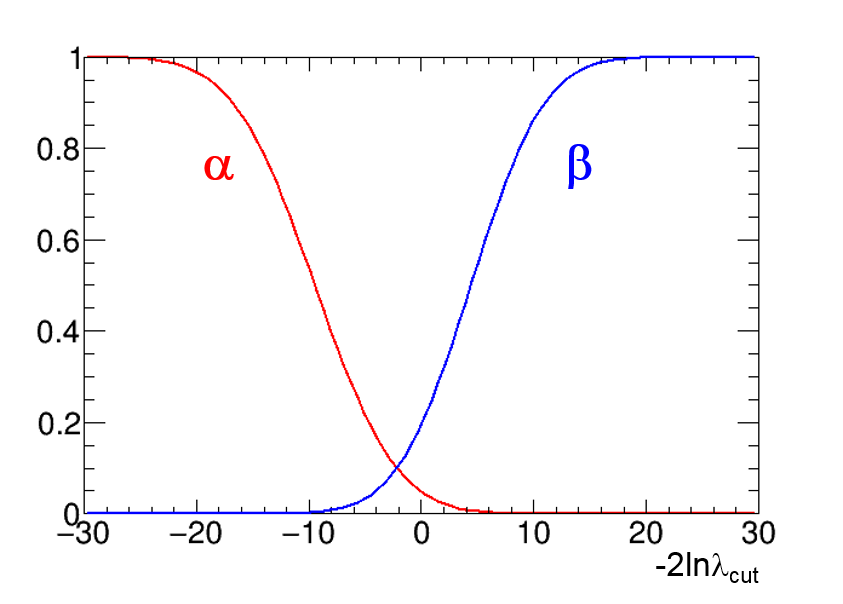}
\includegraphics[width=0.42\textwidth]{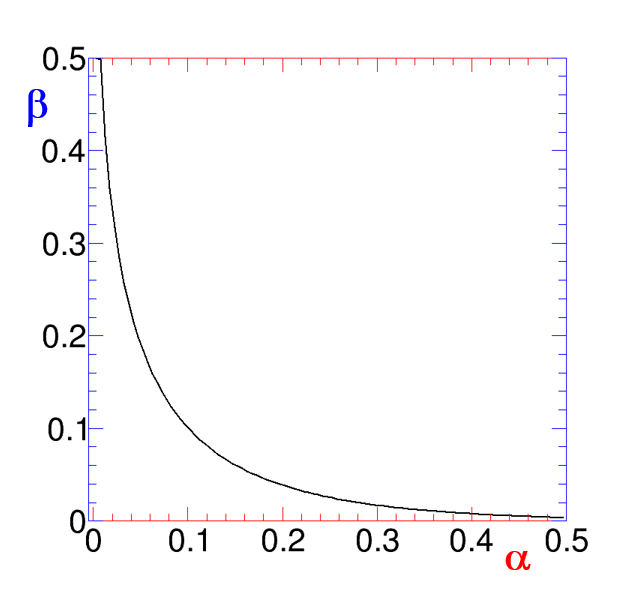}
\caption{On the left are plots of $\alpha$ and $\beta$ as a function of the cut
$-2\ln \lambda_{{\rm cut,}\alpha}$ that is applied to the histograms in 
Fig.~\ref{fig:lrdists_n50}(left).  On the right is a plot of $\beta$ vs $\alpha$
derived from the curves on the left. (Reverse the vertical axis to get the ROC curve of power $1-\beta$ vs $\alpha$).}
\label{alphabeta}
\end{center}
\end{figure}

Recalling the prescriptions in Sections~\ref{hypotest} and
\ref{secNP}, to perform a N-P hypothesis test of $\HA$, one finds the
cutoff $\lambda_{{\rm cut,}\alpha}$ for the desired Type I error probability $\alpha$
and rejects $\HA$ if $\lambda \le \lambda_{{\rm cut,}\alpha}$. The Type II error
probability $\beta$ then follows.  Both of these error probabilities are easily
obtained from the histograms in Fig.~\ref{fig:lrdists_n50} and plotted as a
function of $-2\ln \lambda_{{\rm cut,}\alpha}$, in
Fig.\ref{alphabeta}(left), for $N=50$.  One can also plot $\beta$ vs
$\alpha$ as in Fig.~\ref{alphabeta}(right) for $N=50$.  As noted in Section~\ref{alphachoice},
the choice of operating point on the $\beta$ vs $\alpha$ curve (or equivalent ROC
curve of power $1-\beta$ vs $\alpha$) requires multiple considerations.

\section{Further discussion of goodness of fit}
\label{app-gof}

As introduced in Section~\ref{goodness},
a goodness-of-fit (g.o.f.) test is generally defined as a
test of the null hypothesis 
(typically composite, i.e., having adjustable parameters) when an
alternative hypothesis has not been explicitly specified.  In a typical example,
one has observed data $\vec x$, and the null hypothesis $H_0$ is that
$\vec x$ is a random sample from a specific probability density
function (pdf) $p_0(\vec x; \vec\mu)$; here $\vec\mu$ indicates parameters
that are typically not specified in advance, but rather set to their
``best-fit'' values. A g.o.f.\ test is then used to test
$H_0$.  As noted in Section~\ref{probpdf}, the given pdf $p_0$ is often called ``the model''
(short for ``the statistical model'').  Despite the issues raised in this
note, g.o.f.\ tests constitute an important step in data analysis;
in fact, the discussion here indicates that using more than one g.o.f.
test is advisable.  

The most common g.o.f.\ test is surely the chisquare g.o.f.\ test used in
introductory lab classes, either for measurements of dependent
variables $x_i$ as a function of an independent variable (say current
vs.\ voltage), or for binned data.  As analyses with unbinned
likelihood functions have become commonplace in HEP, the use of
unbinned g.o.f.\ tests has increased as well. For unbinned measurements
in one dimension (1D), the Kolmogorov-Smirnov (K-S) test is commonly
used, probably because it has been readily available in HEP software
packages (CERNLib and ROOT) for decades, and (like the chisquare test)
the interpretation of the test statistic is asymptotically independent
of the null hypothesis model $p_0$.  However, as this note emphasizes,
other tests may be more appropriate than the K-S test (for example
when one is interested in departures from $p_0$ in the tails).

\begin{figure}
  \centering
  \includegraphics*[width=0.7\textwidth]{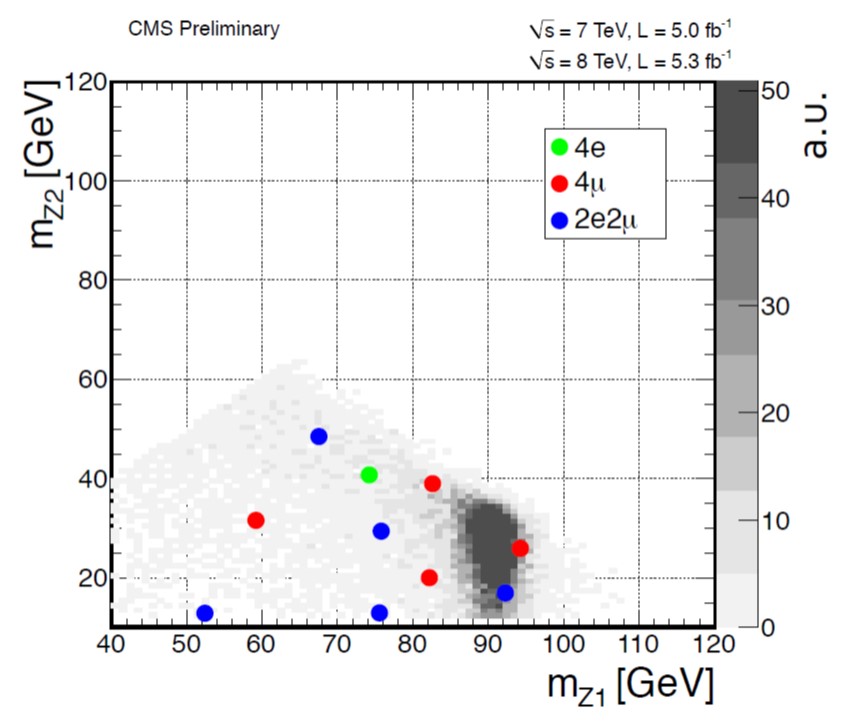}
  \caption{Scatter plot shown by Joe Incandela at the CMS talk on the 
discovery of the Higgs-like boson~\cite{july4}. For Higgs to $ZZ^\ast$ candidates
in the four-lepton final state, the horizontal axis is the mass of the
higher-mass pair ($Z_1$), while the vertical axis is the mass of the
lower-mass pair ($Z_2$). The large dots are the 10 events in data.
The gray shading, peaking in the region around (90 GeV,30 GeV), is the
expected pdf for an SM Higgs with mass of 126 GeV.}
  \label{fig:z1z2}
\end{figure}

For g.o.f.\ tests of unbinned data in more than one dimension, there are
no conventions in HEP, despite sporadic work in the last 25 years
or more.  An example of a potential application arose in the $ZZ^\ast$
Higgs discovery channel, in which the events were expected to have one
$Z$ nearly on mass shell and the other off-shell.  The CMS data
presented at the discovery talk on 4 July 2012 included
Figure~\ref{fig:z1z2}, a scatter plot of the invariant mass of the
lower-mass pair vs the higher-mass pair.
In this case, the null hypothesis $p_0$ could be taken as SM Higgs
production and decay, for which the pdf is shown in gray shading. To
the eye, there seemed to be fewer on-shell Z's than expected.  This was
noted and presumed to be a statistical fluctuation (borne out when
more data was obtained).  As far as I know, no unbinned 2D g.o.f.\ test
of $p_0$ was attempted.  One can imagine that such tests might be
useful for future discovery plots in low-statistics regions.

As noted above, in common g.o.f.\ tests the model $p_0$ has parameters that are
adjusted to obtain the ``best fit'', in which case the g.o.f.\ test is
performed using the best-fit parameters.  In such a case, it is
important keep in mind that the g.o.f.\ test is a test of the complete
model including the best-fit adjustable parameters.  This is distinct
from the separate inference problems of ``measuring'' the parameters
and uncertainties on them (known to statisticians as point estimation
and interval estimation, respectively).

\subsection{Relevance of Neyman--Pearson Lemma to g.o.f.\ tests}

In approaching the theory of g.o.f.\ tests, it is useful to recall
the Neyman--Pearson (N-P) theory of hypothesis testing, as discussed in
Section~\ref{hypotest}.  The N-P theory provides the
language in which the limitations of g.o.f.\ tests become clear.
Recalling in particular the testing of simple hypothesis $H_0$ vs simple hypothesis $H_1$ in Section~\ref{secNP}, for a given $\alpha$, $\beta$ is minimized      
if the test statistic $T$ is the likelihood ratio $\LR$ in Eqn.~\ref{NPLR} 
that depends explicitly on $H_1$. By definition, in a g.o.f.\ test $H_1$ 
has not been specified. Thus the problem of choosing a ``best'' $T$ 
for a g.o.f.\ test of $H_0$ is not well-posed!  There typically exist many
possible choices for $T$ in a g.o.f.\ test.  Some may be particularly
popular (e.g. the ubiquitous chi-square g.o.f.\ test), but that does not
mean that they are ``best'' in any general sense.

In fact, it follows from the Neyman--Pearson Lemma that for any choice
of $T$ used for a g.o.f.\ test, that choice will tend to have high power
against some alternative $H_1$ if it happens that $T$ is approximately
monotonic with the likelihood ratio $\LR=\lhood(H_0)/\lhood(H_1)$,
while the choice is vulnerable to having poor power against
alternative hypotheses from which $T$ bears no relationship to
$\lambda$.  Thus, a choice of g.o.f.\ test statistic $T$ picks out
alternatives to $H_0$ (sometimes called {\em directions} of deviations
from $H_0$) for which $T$ has higher discrimination power.  This is
true whether or not one is aware of it!

Although the strict superiority of $\LR$ as test statistic $T$ no
longer holds with composite hypotheses, the lesson remains that the
choice of $T$ defines directions of high discrimination power.

\subsection{Prototype problem: test of uniform density on (0,1)}

A prototype g.o.f.\ test is the following: suppose that $\vec
y$ consists of a set of $N$ numbers $\{y_i\}$ between 0 and 1, and you
want to test the hypothesis that they were obtained by sampling $N$
times from a uniform density on (0,1).  While this may seem to be an
artificial special case (useful for testing the validity of a
pseudorandom number generator), in fact many models can be re-cast
into this form without loss of generality by using the 
probability integral transform of Section~\ref{PIT}. That is, 
we can consider the more general null hypothesis $H_0$ as the model in
which each number $x_i$ in the set of $N$ numbers $\{x_i\}$ is an
independent random sample from a general pdf $p(x)$ defined in the
domain $a < x < b$, with $x$ continuous.  Then the
probability integral transform of Eqn.~\ref{eqpit} defines $y(x)$ such
that $p(y)$ is
uniform on (0,1). Thus without loss of generality, the g.o.f.\ test of
the null hypothesis $H_0$ for the model $p(x)$ is re-cast as the
hypothesis that the set $\{y_i\} = \{y(x_i)\}$ is a random sample from
the uniform density on (0,1).  (If $x$ is discrete, there are
complications which are discussed in some of the cited papers.)

Since $1${}$-${}$y$ is also uniform on (0,1), typically some other
consideration (such as the distribution of $x$ under a different $p$)
dictates if one end of the interval is of more interest than the
other.

Over the last century, a plethora of tests have been invented to test
for uniformity of $\vec y$ on (0,1), and hence test for $\vec x$ drawn
from any given continuous $p_0(x)$.  The book by D'Agostino and
Stephens~\cite{dagostino} contains a comprehensive discussion.  
The more recent article by Marhuenda et al.~\cite{marhuenda} defines and
compares a plethora of tests against standard sets of parameterized
alternatives (higher or lower density near 0 or 1, or both, etc.).
Subsets of tests in
common use in HEP are discussed by F.~James~\cite{james2006}, and a few
are implemented in ROOT.  

\subsubsection{Tests based on the empirical distribution function}

The most widely used methods in HEP are based on the {\em cumulative
distribution function} (CDF) and {\em empirical distribution function}
(EDF).  While in science ``distribution'' is often used synonymously with
probability density function (pdf), in statistics ``distribution
function'' is often short for {\em cumulative distribution function}, 
which is an {\em integral} of a pdf.  Often $f(x)$ is used for
a pdf, and upper case $F(x)$ is used for its CDF:
\begin{equation}
\label{cdf}
F(x)=\int_{-\infty}^x f(x^\prime)dx^\prime.
\end{equation}
Given $N$ observed values $x_i$ as above, the EDF is
\begin{equation}
\label{edf}
F_N(x)= \frac{{\rm number~of~observed~values~}\le x}{N}.
\end{equation}
Thus $F_N(x)$ is an increasing piecewise-constant function, starting
from 0 for $x$ less than the smallest observed $x_i$, increasing by
$1/N$ at every observed value, and obtaining unity above the highest
observed value of $x_i$.

If the observed $\vec x$ is drawn from $p_0(x)$, then we expect the
EDF $F_N(x)$ to be similar to the CDF calculated from $p_0(x)$, which
we call $F_0(x)$.  A variety of g.o.f.\ tests are thus based on various
definitions of the ``distance'' between $F_0(x)$ and $F_N(x)$.  These
tests include, for example, the Kolmogorov-Smirnov test (based on the extremum of
$|F_0(x)- F_N(x)|$), and the Cram\'er-von Mises family based on integrals weighted
by a function $w(x)$ of the squared difference:
\begin{equation}
\label{cvm}
N \int_{-\infty}^{+\infty} \left(F_0(x)- F_N(x)\right)^2 w(x) dF(x).
\end{equation}
The unweighted case (i.e. $w(x) = 1$) corresponds to the classic 
Cram\'er-von Mises statistic. 
The weight function $w(x)$ can be chosen to give more power against certain
deviations.  For example, the Anderson-Darling (AD) test is designed to
have more power against deviations at both ends of the distribution, 
with a weight function 
\begin{equation}
w(x) = \frac{1}{F(x)(1-F(x)}.
\end{equation}
Since departures from assumed Gaussianity are often in the tails, AD
is reputed to be useful for testing Gaussianity, as well as being
generally useful.  There are also versions that 
emphasize only one tail~\cite{dagostino}.
On the other hand, both the Kolmogorov-Smirnov family and the Cram\'er-von Mises family have 
variations that make the endpoints {\em not} special by posing the
problem on a circle.

Keeping in mind the probability integral transform and the fact that
Eqn.~\ref{cvm} is defined in terms of $dF(x)$ rather than $dx$, one
can see that the distribution of such test statistics under the null
hypothesis does not depend on the specific form $p_0$ in the null.
Such tests are called distribution-free and are popular since standard
tables can be computed and used.  Nowadays, with the ability to
simulate data sets and obtain the null distribution directly, it can
be worth exploring the use of more powerful tests that do not have
this property.

\subsubsection{Other families of tests of uniformity on (0,1)}
Refs.~\cite{dagostino,marhuenda}
describe multiple classes of tests in addition to those based on the EDF.
These include tests based on the ``ordering statistics'', i.e.,  
on the ordered set of the observed points ${y_i}$
(testing mean values of $i$th points, moments of differences $y_j-y_i$, etc.).
Among the many tests, one can attempt to choose ``omnibus'' tests that
perform reasonably well against a number of different alternatives.  The
conclusion of Ref.~\cite{marhuenda} is that a member of the class of tests called
``Neyman smooth tests'' is unique in being in the top-10 most powerful
tests for all the alternatives that they considered.  

Neyman smooth tests and more general ``smooth'' tests seem to have had
little use in HEP thus far, despite popularity in the statistics
literature~\cite{raynor,thas}.  They are tests where the alternative
hypothesis is constructed by fitting the data to a sum of Legendre
polynomials.  (The variant that Ref.~\cite{marhuenda} studies uses
Schwarz's Bayesian information criterion~\cite{schwarz1978} to choose
the degree of polynomials.)  This spirit of using the data to construct
the alternative hypothesis used in a likelihood ratio (or asymptotic
equivalent) is similar to that of the saturated model discussed below
for binned data. Recently, a real-world example using the Neyman
smooth test in X-ray astronomy was published by astrophysicists
collaborating with statistician Algeri and her student
Zhang~\cite{algerismooth}. Grosso et al.~\cite{grosso} have
investigated a further generalization that uses machine learning to
construct an alternative model that is used in a likelihood ratio.

\subsection{Chisquare g.o.f.\ and variants}
As noted above, Neyman and Pearson taught us that (even for simple
hypotheses) the best test of the null hypothesis depends on the
alternative, and hence there is no universally best g.o.f.\ test.
Nonetheless, the ubiquity of the chisquare g.o.f.\ test attests to its
utility, at least for picking up certain departures from the null.  In
its usual form for uncorrelated Gaussian (normal) distributed data,
one has
\begin{equation}
 \chi^2 = \sum_i \frac{(d_i - f_i)^2}{\sigma_i^2},
 \label{chisq}
\end{equation}
where $d_i \pm \sigma_i$ is the $i$th measured data point with rms
deviation $\sigma_i$ (each assumed to be a known constant), and $f_i$
is the model prediction (perhaps with parameters) to be compared with
$d_i$.  (If $\sigma_i$ is not a known constant at each $i$, but
depends on the unknown true value of the model at $i$, then there are
subtleties beyond the scope of this note.) Since the test statistic
$\chi^2$ is a function of the random data, it is itself a random
variable, and in unbounded Gaussian applications it has a probability
density function \cite{pdg2024} which is itself also frequently called
chisquare.  The potentially confusing ambiguity in multiple
meanings is usually resolvable by context. This could also be
avoided by using another name, such as $S^2$ or $Q^2$ for the left
hand side of Eqn.~\ref{chisq}, but I yield to common practice in this
note.

In HEP, we can also encounter situations in which the so-called
``regularity conditions'' are not met, so that the distribution of the
test statistic in Eqn.~\ref{chisq} is not a chisquare function; two
common cases are when the true or best-fit values of parameters are on
the boundary (physical constraint such as non-negativity), and when
there are issues with degrees of freedom not being well-defined.
Again, these issues are beyond the scope of this note.

More information regarding the chisquare test, including the
generalization of Eqn.~\ref{chisq} to include correlations, is in the
PDG RPP~\cite{pdg2024}.

\subsubsection{Gaussian chisquare g.o.f.\ is a likelihood ratio using the saturated model}

For the same data and model as above, the likelihood for the null
hypothesis $H_0$ is:
\begin{equation}
 \lhood(H_0)  =  \prod_i \frac{1}{\sqrt{2\pi\sigma_i^2}}
       \exp\left(-(d_i - f_i)^2/2\sigma_i^2\right).
 \label{like}
\end{equation}
It is sometimes said that $-2\ln{\lhood}$ is equal to $\chi^2$; this
is clearly not correct since the former quantity has extra constant
terms.  Understanding how the extra terms ``disappear'' in the
g.o.f.\ test (from the point of view of N-P testing) is enlightening.

Given only the null hypothesis and the data, one can use the data to
invent an alternative hypothesis for which the model $f_i$ is equal to
the data $d_i$ at every measured value!  Such a model, which typically
needs as many parameters are there are data points, is called a {\it
saturated model} \cite{lindsey}.

For the Gaussian data above, the saturated model sets $f_i = d_i$, so
that the likelihood of the alternative hypothesis $H_1$ (the saturated
model) is (since $\exp(0) = 1$)
\begin{equation}
 \lhood(H_1)  = \prod_i \frac{1}{\sqrt{2\pi\sigma_i^2}}. 
 \label{likesat}
\end{equation}
Inspired by Eqn.~\ref{NPLR}, we consider the {\em ratio} of the two likelihoods 
above, i.e, 
\begin{equation}
 \LR = \lhood(H_0)/\lhood(H_1)  =  \prod_i \exp\left(-(d_i - f_i)^2/2\sigma_i^2\right),
 \label{ratio}
\end{equation}
and thus
\begin{equation}
 \chi^2 = -2\ln\lambda.
 \label{chilike}
\end{equation}
From this point of view, the constants in $-2\ln\lhood$ were not just
ignored; they were canceled when a ratio was formed.  (There are other
paths to the chisquare expression not as relevant to this note.) Since the
saturated model does not depend on the parameters of the original
model, the maximum of $\lambda$ is of course at those parameters that
maximize the original $\lhood(H_0)$.
\subsubsection{Pearson's chisquare for binned histograms}
The original chisquare variable is that of Karl Pearson in 1900,
designed for histograms (``frequency tables'') with
multinomial-distributed bin contents, and defined as
\begin{equation}
 \chi^2 = \sum_i \frac{(d_i-f_i)^2}{f_i}.
\label{chip}
\end{equation}
(In a multinomial histogram, the total number of events is fixed by
design, so there is one fewer independent bin content than in Poisson
data.)  This is very similar to Eqn.~\ref{chisq}, since in multinomial
and Poisson data, $f_i$ can be a proxy for $\sigma_i^2$.

\subsubsection{Use of saturated model to construct improved 
chisquare g.o.f.\ for binned histograms} 

In 1983, Baker and Cousins \cite{bakercousins1984} reviewed construction
of likelihood ratios as in Eqn.~\ref{NPLR} using saturated models
for testing g.o.f.\ for Poisson and multinomial data in histograms.
(They did not call them saturated models, instead citing a 1928 paper
by Neyman and Pearson.)  By that time, it had been widely noted that
using a Poisson likelihood model for fitting typical HEP histograms cured a 
defect of fits using Pearson's chisquare, namely that the area under
the fitted curve was not equal to the observed number of events.  It was less well
known how to construct a g.o.f.\ test consistent with such a likelihood fit.
For histograms with independent
Poisson-distributed bin contents $d_i$, the result for the g.o.f.\ test 
statistic,
\begin{equation}
-2\ln\lambda = 2\sum_i f_i - d_i +d_i\ln(d_i/f_i),
\end{equation}
was called $\chi^2_{\lambda,p}$ since asymptotically its pdf tends to
the chisquare distribution.  This Poisson form is mentioned in the
PDG's Review of Particle Properties \cite{pdg2024}; some time ago it was
decided that it is best just to denote it by $-2\ln\lambda$ as some
felt that calling it $\chi^2$ might encourage people to forget that it
only asymptotically follows the $\chi^2$ distribution (and only if
conditions are satisfied).  As noted in the Baker-Cousins paper,
probably the safest thing to do is to study the distribution by Monte
Carlo.

Heinrich~\cite{heinrich} has studied the distribution and moments of
$\lambda$ for small statistics, and makes the point that for the
asymptotic formulas to be valid, the contents of all bins must each be
large.  More generally, how to bin data is another complex problem
since binning discards information on the location of events within
the bin, and suppresses the ability to observe high-frequency
components.  Thus the robustness of results (or lack thereof) to the
binning chosen should be understood and communicated.

\subsubsection{Tests complementary to the chisquare g.o.f.\ test}
The tests based on chisquare or asymptotic equivalents discard
all or most of the information on the ordering of the deviations as 
well as the signs of the deviations.  Hence they can easily miss a trend
in the deviations that is obvious to the eye (e.g., all deviations of the same sign,
or a trend due to a slope not present in the null hypothesis).
Thus, authors such James~\cite{james2006} suggest complementing the chisquare
test with a ``runs test'', often called the Wald-Wolfowitz runs test.
Performing more than one g.o.f.\ test raises the issue of whether or not
the results can be combined into one overall g.o.f.\ summary. Doing so
might be useful in some contexts, but one should be aware that
combining p-values is itself fraught with ambiguity, as discussed in
Ref.~\cite{cousinscomb}.

\subsection{Caution against absolute likelihood as g.o.f.\ test statistic}
Occasionally one finds the recommendation to use as a g.o.f.\
test statistic the absolute likelihood at its maximum, as opposed to a
{\it ratio} of likelihood maxima; typically Monte Carlo simulation is
recommended as the way to get the null distribution of the test
statistic.  Although this might at first seem plausible, it is a
flawed concept: unlike the likelihood {\em ratio}, such a g.o.f.\ statistic is
without foundation, and power can vary arbitrarily with the metric.
Simple examples can make clear that the value of the likelihood at its
maximum must be compared to something.  For example, for Poisson
counts, the probability of observing 100 events when $\mu=100.0$ is
much less than the probability of observing 1 event when $\mu=1.0$,
even though in both cases the data perfectly fit the theory.  For binned
data, the
saturated model provides the reference of the largest value that the
likelihood can be for that data (for {\it any} model), and hence
provides a reasonable normalization for the maximum observed for a
more constraining model.

Heinrich~\cite{heinrichunbinned} discusses the pitfalls of using the
absolute likelihood as a g.o.f.\ statistic.  For unbinned likelihoods,
which are common in HEP, the problem is exacerbated.  One might hope
that one could just take binned g.o.f.\ in the limit of small bins, but
the answer depends on the way the limit is taken, i.e., in which
metric the bin boundaries are equally spaced.  

\subsection{Discussion}
From the above, it is clear that, for ``g.o.f.\ test'' defined in its
purest form of no specified alternative, there is no unique ``best'' g.o.f.
test.  Therefore trying several can give some indication of how well the
chosen statistics model approximates reality, and in which directions departures
might be suspected (heavier tails, skewness, etc.).  It is therefore useful to be 
aware of directions against which one would like to have power, and to choose g.o.f.\ tests
appropriately; simple toy MC simulations, such as those in the cited references,
can help in this regard. For a g.o.f.\ test for
unbinned data in one dimension, one has a variety of tests to choose among
\cite{dagostino,marhuenda}, depending on roughly what sort of alternatives one
wants power against.  A recent monograph is by
Thas~\cite{thas}.   If, on the other hand, there is a specific alternative of interest, then 
typically one leaves the world of generic g.o.f.\ tests and gains power by constructing 
alternative-specific likelihood ratio tests (as in the Higgs boson searches).

Generalization of these g.o.f.\ tests to higher
dimensions, as in Fig.~\ref{fig:z1z2}, remains a topic of research.
Some comparisons were made by Aslan and Zech at Durham in 
2002 \cite{aslanzechdurham},
including their proposed {\em energy test} \cite{aslanzechenergy}.
Williams~\cite{williamsgoodness} has provided a more recent review, including work by Cuadras and Fortiana (first pointed out to me by Ilya Narsky) that is closely related to the energy test.  Williams also discusses the close relationship between g.o.f.\ tests and two-sample tests, which are beyond the scope of this note. One can of course
resort to binning the data for g.o.f.\ tests even if the parameters are fit with
an unbinned likelihood; in this case as noted above the robustness with
respect to the binning chosen needs to be understood.

\section{Look-elsewhere effect}

In these lectures, I did not have time for the important (and
increasingly mandatory) calculation of the LEE.  A starting point for
self-study is the discussion by Louis Lyons, ``Comments on `Look
Elsewhere Effect' ''~\cite{lyonslee2010}.  See also Section 9.2 of my
paper on the Jeffreys-Lindley Paradox~\cite{cousinsJL}.

An important paper for practical calculations, and also for
qualitative insight, is by Eilam Gross and Ofer Vitells, ``Trial
factors for the look elsewhere effect in high energy
physics,''~\cite{grossvitells2010}.

\section{Bayesian hypothesis testing  (model selection)}
\label{modelselection}

As mentioned very briefly in Sections~\ref{bayesintro},
\ref{pseudobayes}, and \ref{intervalsummary}, the duality used in
frequentist hypothesis testing (Section~\ref{duality}) is not used in
Bayesian statistics.  The usual methods follow Chapter 5 of Harold
Jeffreys's book~\cite{jeffreys1961}: Bayes's Theorem is applied to the
models themselves after integrating out all parameters, including the
parameter of interest!  This is typically presented by Bayesian
advocates as ``logical'' and therefore simple to use, with great
benefits such as automatic ``Occam's razor'' penalizing less
predictive models, etc.

In fact, Bayesian model selection is full of subtleties, and even for
the experts, it can be a ``can of worms'' (James
Berger~\cite{berger2006}, Rejoinder, p.\ 459).  As just one
indication, Jeffreys and followers use {\em different priors} for
integrating out parameter(s) in model selection than for the same
parameter(s) in parameter estimation.  Here I mainly just say: Beware!
There are posted/published applications in HEP that lack foundation, in particular by
Bayesian standards.  As mentioned in Section~\ref{pseudobayes}, a
pseudo-Bayes example in PRL provoked me to write a
Comment~\cite{cousins2008} that has some references to useful Bayesian
literature.
 
The dependence on prior probabilities for the {\em models themselves}
can be factored out, leading to a ``Bayes factor'' that is the ratio
of posterior odds to prior odds.  However, the Bayes factor still
depends on prior pdfs for {\em parameters} in the models, and this
leads to direct sensitivity to the prior pdf for a parameter that is
in one model but not in the other.  For testing H$_0$:$\mu=\mu_0$ vs
H$_1$:$\mu\ne\mu_0$, improper priors for $\mu$ that work fine for
estimation become a disaster. Adding a cut-off to make them proper
just gives (typically arbitrary) cutoff dependence.

In the asymptotic limit of lots of data, your answers in a test of a
point null vs a continuous alternative (either the probability H$_0$
is true, or also the Bayes factor) {\em remain directly proportional
  to the prior pdf for the parameter of interest}.  This is {\em
  totally different} behavior compared to Bayesian interval
estimation, where the effect of a prior typically becomes negligible
as the likelihood function becomes narrowly peaked at large $N$.

For a review and comparison to $p$-values in the discovery of Higgs
boson, see my paper, ``The Jeffreys-Lindley Paradox and Discovery
Criteria in High Energy Physics''~\cite{cousinsJL}.  As mentioned in
Section~\ref{pseudobayes}, see also Chapter 12 by Harrison Prosper in
Ref.~\cite{behnke2013}.

\section{Point estimation}
\label{pointest}

Most of this paper is about intervals; I do not say much about
what to quote as the ``measured value'' (typically using the MLE
by default).  Statisticians call this the
``point estimate''.
There is a huge literature on point estimation; see e.g. Chapters 7
and 8 in Ref.~\cite{james2006}.

If you are well-grounded in interval estimation, one approach is to
use that machinery to get a point estimate.  E.g., one might take the
mid-point of (say) your 68\% \cl\ central interval. But a better
approach is probably to let the \cl\ go to 0, so that your interval
gets shorter and shorter, and use the limiting point.  For both the LR
ordering for confidence intervals (F-C), and for likelihood ratio
intervals, this results in the maximum likelihood estimate.

To give an idea of how rich the subject is, I show a few interesting
things from Ref.~\cite{james2006}.

\subsubsection{Point estimation: Traditional desiderata}
\label{point-est}
\begin{itemize}
\item Consistency: Estimate converges toward true value as number of
  observations $N$ increases
\item Unbiasedness: Expectation value of estimate is equal to the true
  value.  (Bias and consistency are independent properties; see
  Fig. 7.2 in Ref.~\cite{james2006}.)
\item Efficiency: Estimate has minimum variance
\item Minimum loss of information: (technical definition)
\item Robustness: Insensitivity to departures from the assumed
  distribution
\end{itemize}
One can add:
\begin{itemize}
\item Simplicity: transparent and understandable
\item Minimum computer time: still relevant in online applications,
  less relevant otherwise
\item Minimum loss of physicist's time (how much weight to put on
  this?)
\end{itemize}

These desired properties can be impossible to achieve simultaneously.
How to choose?  A thorough analysis requires further input: what are
the costs of not incorporating various desiderata?  Then formal {\em
  decision theory} can be used to choose an estimator.

In practice in HEP, maximum likelihood estimates are often used (even
though they are typically not unbiased).  Typically the MLE is
consistent and has other excellent asymptotic properties (e.g.,
estimate is asymptotically normal).  For finite $N$, the MLE works
well in the so-called exponential family that includes binomial,
Poisson, and Gaussian.  As noted in Section~\ref{invariantl}, the MLE
is invariant under reparameterization. (This means that if it is
unbiased in one metric, it is typically biased in other metrics, as
discussed in Ref.~\cite{barlow}, Section 5.3.1.)

\subsubsection{Alternatives to the arithmetic mean when model is non-Gaussian}

Fred James~\cite{james2006} (pp.\ 209 ff) has a very illuminating
discussion of the MLEs for a diverse set of pdfs, representing long
tails, short tails, and in between.  If $p(x| \mu) = f(x- \mu)$, where
$f$ is some pdf, then $\mu$ is called a {\it location parameter}.
Some common examples of pdfs where $\mu$ is a location parameter are:
\begin{description}
\item{Normal:} $p \sim \exp(-(x- \mu)^2/2\sigma^2)$

\item{Uniform:}  p = constant for $|x- \mu|<a$; $p=0$ otherwise

\item{Cauchy:} $p \sim 1/(a^2 + (x- \mu)^2)$

\item{Double exponential:} $p \sim \exp( -a |x- \mu| )$
\end{description}
These examples are all symmetric about  $\mu$:     $p(\mu+y) = p(\mu-y)$

Suppose you are given N=11 values of $x$ randomly sampled from $p(x|
\mu)$.  What estimator (function of the 11 values) gives you the
``best'' estimate of $\mu$ ?  If by ``best'' you mean minimum
variance, it is the M.L. estimate, resulting in a different formula
for each of the above cases!  Three of the four are special cases of
$L^p$, the estimator that minimizes the sum over the observations of
$|x_i - \mu|^p$.  Different values of $p$ put different emphasis on
observations in the tails.

{\em Only for the normal} pdf is the MLE equal to the arithmetic mean,
obtained by the familiar least-squares technique ($p=2$).  Can you
guess what the MLE is for the uniform and the exponential cases?  (For
the Cauchy pdf, with its long tails, the arithmetic mean is
particularly useless; see also the discussion by Efron~\cite{efronslac}).

{\em If the true distribution departs from that assumed, then the
  estimate of location is no longer optimal. Sensitivity is in the
  tails!}  See the nice discussion of asymptotic variance and
robustness in Ref.~\cite{james2006} (pp.\ 211 ff).

\section*{Acknowledgments}
Thanks to many in HEP (Frederick James, Gary Feldman, Louis Lyons, Luc
Demortier, and numerous others) from whom I learned\dots and to many
statisticians that Louis invited to PhyStat meetings.  For Bayesian
statistics, that was especially Jim Berger (multiple times) and Michael
Goldstein, and more recently, David van Dyk (multiple times). Thanks
also to the CMS Statistics Committee (Olaf Behnke et al., including Igor Volobouev, who 
pointed to Refs.~\cite{marhuenda,thas}) for many
discussions and comments on earlier versions of the slides and
writeup\dots and to the authors of numerous papers from which I
learned, including early (1980s) Bayesian papers by Harrison
Prosper\dots.  Thanks also to Diego Tonelli of the LHCb experiment for
encouragement to update the slides in 2017 and for comments on an
earlier version.

This work was partially supported by the U.S.\ Department of Energy
under Award Number {DE}--{SC}0009937.

\bibliography{statstheory}{}
\bibliographystyle{cms}
 
\end{document}